\input harvmac 
\input epsf 

\vskip 1cm

 \Title{ \vbox{\baselineskip12pt\hbox{  Brown Het-1305 }}}
 {\vbox{
\centerline{ Finite Factorization equations  and   }
\vskip.08in
\centerline{ Sum Rules  for  }
\vskip.08in
\centerline{  BPS correlators in  $N=4$ SYM theory }  }}

\centerline{$\quad$ { Steve Corley, Sanjaye Ramgoolam   } }
\smallskip
\centerline{{\sl Department of Physics}}
\centerline{{\sl Brown  University}}
\centerline{{\sl Providence, RI 02912 }}
\smallskip
\centerline{{\tt scorley@het.brown.edu }}
\centerline{{ \tt ramgosk@het.brown.edu }}

\vskip .3in 

A class of exact non-renormalized extremal correlators of half-BPS operators 
in $N=4$ SYM, with $U(N)$ gauge group, is shown to satisfy 
finite factorization equations reminiscent of 
topological gauge  theories.   The finite 
factorization equations can be generalized, beyond the extremal case, 
to a class of correlators involving observables with 
a simple pattern of $SO(6)$ charges.
The simple group theoretic form of the correlators 
allows equalities between ratios of 
correlators in $N=4$ SYM and Wilson loops in Chern-Simons theories at  
$k= \infty$,  correlators of appropriate observables 
in topological $G/G$ models  and Wilson loops in  two-dimensional 
Yang-Mills theories.  The correlators  also obey sum rules
which can be generalized to off-extremal correlators. The simplest 
sum rules can be viewed as large $k$ limits of the Verlinde formula 
using the Chern-Simons correspondence. For special classes of correlators, 
the saturation of the factorization equations by a small subset of 
the operators in the large $N$ theory
is related to the emergence of semiclassical objects 
like KK modes and giant gravitons in the dual $ ADS \times S$ 
background.  We comment on an intriguing symmetry between
KK modes and giant gravitons.

\Date{ May  2002 }

\lref\cjr{
S.~Corley, A.~Jevicki and S.~Ramgoolam,
``Exact correlators of giant gravitons from dual N = 4 SYM theory,''
arXiv:hep-th/0111222, to appear in Adv. Theor. Math. Phys.
}
\lref\rus{ 
B.~E.~Rusakov,
``Loop Averages And Partition Functions In U(N) Gauge Theory On Two-Dimensional Manifolds,''
Mod.\ Phys.\ Lett.\ A {\bf 5}, 693 (1990).
}
\lref\witymii{ 
E.~Witten,
``On Quantum Gauge Theories In Two-Dimensions,''
Commun.\ Math.\ Phys.\  {\bf 141}, 153 (1991).
}
\lref\bk{ 
D.~Berenstein, C.~P.~Herzog and I.~R.~Klebanov,
``Baryon spectra and AdS/CFT correspondence,''
arXiv:hep-th/0202150.
} 
\lref\mikh{ 
A.~Mikhailov,
``Giant gravitons from holomorphic surfaces,''
JHEP {\bf 0011}, 027 (2000)
[arXiv:hep-th/0010206].
} 
\lref\fulhar{ W. Fulton and J. Harris, ``Representation Theory, ''
Springer Verlag, 1991  } 
\lref\witint{ 
E.~Witten,
``Gauge Theories And Integrable Lattice Models,''
Nucl.\ Phys.\ B {\bf 322}, 629 (1989).
} 
\lref\witsyfr{
E.~Witten,
``Supersymmetric Yang-Mills theory on a four manifold,''
J.\ Math.\ Phys.\  {\bf 35}, 5101 (1994)
[arXiv:hep-th/9403195].
}
\lref\blauthom{ 
M.~Blau and G.~Thompson,
``Derivation of the Verlinde formula from Chern-Simons theory 
and the G/G model,''
Nucl.\ Phys.\ B {\bf 408}, 345 (1993)
[arXiv:hep-th/9305010].
}
\lref\vafmarin{
J.~M.~Labastida, M.~Marino and C.~Vafa,
``Knots, links and branes at large N,''
JHEP {\bf 0011}, 007 (2000)
[arXiv:hep-th/0010102].
}
\lref\gopvaf{ 
R.~Gopakumar and C.~Vafa,
``On the gauge theory/geometry correspondence,''
Adv.\ Theor.\ Math.\ Phys.\  {\bf 3}, 1415 (1999)
[arXiv:hep-th/9811131].
} 
\lref\vafoog{ 
H.~Ooguri and C.~Vafa,
``Knot invariants and topological strings,''
Nucl.\ Phys.\ B {\bf 577}, 419 (2000)
[arXiv:hep-th/9912123].
}
\lref\lomonesh{ 
A.~Losev, G.~W.~Moore, N.~Nekrasov and S.~Shatashvili,
``Four-dimensional avatars of two-dimensional RCFT,''
Nucl.\ Phys.\ Proc.\ Suppl.\  {\bf 46}, 130 (1996)
[arXiv:hep-th/9509151].
}
\lref\grotayii{
D.~J.~Gross and W.~I.~Taylor,
``Twists and Wilson loops in the string theory of two-dimensional QCD,''
Nucl.\ Phys.\ B {\bf 403}, 395 (1993)
[arXiv:hep-th/9303046].
}
\lref\samuel{ 
S.~Samuel,
``U(N) Integrals, 1/N, And The Dewit-'T Hooft Anomalies,''
J.\ Math.\ Phys.\  {\bf 21}, 2695 (1980).
}
\lref\obzub{ 
K.~H.~O'Brien and J.~B.~Zuber,
``Strong Coupling Expansion Of Large N QCD And Surfaces,''
Nucl.\ Phys.\ B {\bf 253}, 621 (1985).
}
\lref\eulwil{ 
S.~Ramgoolam,
``Wilson loops in 2-D Yang-Mills: Euler characters and loop equations,''
Int.\ J.\ Mod.\ Phys.\ A {\bf 11}, 3885 (1996)
[arXiv:hep-th/9412110].
}
\lref\mig{ 
A.~A.~Migdal,
``Recursion Equations In Gauge Field Theories,''
Sov.\ Phys.\ JETP {\bf 42}, 413 (1975)
[Zh.\ Eksp.\ Teor.\ Fiz.\  {\bf 69}, 810 (1975)].
}
\lref\rus{
B.~E.~Rusakov,
``Loop Averages And Partition Functions In U(N) Gauge Theory On Two-Dimensional Manifolds,''
Mod.\ Phys.\ Lett.\ A {\bf 5}, 693 (1990).
}
\lref\cmrrev{
S.~Cordes, G.~W.~Moore and S.~Ramgoolam,
``Lectures on 2-d Yang-Mills theory, equivariant 
cohomology and topological field theories,''
Nucl.\ Phys.\ Proc.\ Suppl.\  {\bf 41}, 184 (1995)
[arXiv:hep-th/9411210].
}
\lref\witjones{ 
E.~Witten,
``Quantum Field Theory And The Jones Polynomial,''
Commun.\ Math.\ Phys.\  {\bf 121}, 351 (1989).
}
\lref\friedkras{ 
L.~Freidel and K.~Krasnov,
``2D conformal field theories and holography,''
arXiv:hep-th/0205091.
}
\lref\tuvi{ 
V.G. Turaev, O.Yu.Viro, ``State sum invariants and quantum
6J-symbols,''
 Topology, 31, 865-902 (1982) } 
\lref\lewtye{
H.~Kawai, D.~C.~Lewellen and S.~H.~Tye,
``A Relation Between Tree Amplitudes Of Closed And Open Strings,''
Nucl.\ Phys.\ B {\bf 269}, 1 (1986).
}
\lref\moseg{G.Moore, ``Lectures on Branes, K-theory and RR charges,'' 
based on work of G. Moore and G. Segal, given at the Newton Institute 
Cambridge, April 2002 } 
\lref\bmn{
D.~Berenstein, J.~M.~Maldacena and H.~Nastase,
``Strings in flat space and pp waves from N = 4 super Yang Mills,''
JHEP {\bf 0204}, 013 (2002)
[arXiv:hep-th/0202021].
} 
\lref\cpss{
C.~Kristjansen, J.~Plefka, G.~W.~Semenoff and M.~Staudacher,
``A new double-scaling limit of N = 4 super Yang-Mills theory and PP-wave  strings,''
arXiv:hep-th/0205033.
} 
\lref\zelo{ D. P. Zelobenko, ``Compact Lie groups and their
representations, '' Translations of Mathematical monographs, Vol 40, 
American Mathematical Society,  1973} 
\lref\barac{ A.O.Barut and R.Raczka, ``Theory of Group
Representations and Applications,'' World Scientific, 1986 } 
\lref\dhokfree{ 
E.~D'Hoker and D.~Z.~Freedman,
``Supersymmetric gauge theories and the AdS/CFT correspondence,''
arXiv:hep-th/0201253.
}
\lref\af{ 
L.~Andrianopoli and S.~Ferrara,
``Short and long SU(2,2/4) multiplets in the AdS/CFT correspondence,''
Lett.\ Math.\ Phys.\  {\bf 48}, 145 (1999)
[arXiv:hep-th/9812067].
} 
\lref\afsz{ 
L.~Andrianopoli, S.~Ferrara, E.~Sokatchev and B.~Zupnik,
``Shortening of primary operators in N-extended SCFT(4) and  
harmonic-superspace analyticity,''
Adv.\ Theor.\ Math.\ Phys.\  {\bf 3}, 1149 (1999)
[arXiv:hep-th/9912007].
}
\lref\agmo{
O.~Aharony, S.~S.~Gubser, J.~M.~Maldacena, H.~Ooguri and Y.~Oz,
``Large N field theories, string theory and gravity,''
Phys.\ Rept.\  {\bf 323}, 183 (2000)
[arXiv:hep-th/9905111].
}
\lref\guma{ 
M.~Gunaydin and N.~Marcus,
``The Spectrum Of The S**5 Compactification Of The Chiral N=2, D = 10 Supergravity 
And The Unitary Supermultiplets Of U(2, 2/4),''
Class.\ Quant.\ Grav.\  {\bf 2}, L11 (1985).
}
\lref\gumz{
M.~Gunaydin, D.~Minic and M.~Zagermann,
``Novel supermultiplets of SU(2,2$|$4) and the AdS(5)/CFT(4) duality,''
Nucl.\ Phys.\ B {\bf 544}, 737 (1999)
[arXiv:hep-th/9810226].
}
\lref\ski{ 
W.~Skiba,
``Correlators of short multi-trace operators in N = 4 supersymmetric  Yang-Mills,''
Phys.\ Rev.\ D {\bf 60}, 105038 (1999)
[arXiv:hep-th/9907088].
}
\lref\lmrs{ 
S.~M.~Lee, S.~Minwalla, M.~Rangamani and N.~Seiberg,
``Three-point functions of chiral operators in D = 4, N = 4 SYM at  large N,''
Adv.\ Theor.\ Math.\ Phys.\  {\bf 2}, 697 (1998)
[arXiv:hep-th/9806074].
}
\lref\mooreseib{ 
G.~W.~Moore and N.~Seiberg,
``Classical And Quantum Conformal Field Theory,''
Commun.\ Math.\ Phys.\  {\bf 123}, 177 (1989).
}
\lref\alvgaum{
L.~Alvarez-Gaume, C.~Gomez and G.~Sierra,
``Duality And Quantum Groups,''
Nucl.\ Phys.\ B {\bf 330}, 347 (1990).
}
\lref\schweig{J. Fuchs, I. Runkel, C. Schweigert, 
``TFT construction of RCFT correlators I: Partition functions,''
 hep-th/0204148 } 
\lref\pasqsal{ 
V.~Pasquier and H.~Saleur,
``Common Structures Between Finite Systems 
And Conformal Field Theories Through Quantum Groups,''
Nucl.\ Phys.\ B {\bf 330}, 523 (1990).
}
\lref\prud{A.P. Prudnikov, Yu. A. Brychokov, and O. I. Marichev, 
``Integrals and Series, Volume I'', Gordon and Breach Science
Publishers, 1986, page 628.}
\lref\intril{
K.~A.~Intriligator,
``Bonus symmetries of N = 4 super-Yang-Mills correlation 
functions via  AdS duality,''
Nucl.\ Phys.\ B {\bf 551}, 575 (1999)
[arXiv:hep-th/9811047].
}
\lref\harvmit{ 
N.~R.~Constable, D.~Z.~Freedman, M.~Headrick, S.~Minwalla, L.~Motl, A.~Postnikov and W.~Skiba,
``PP-wave string interactions from perturbative Yang-Mills theory,''
arXiv:hep-th/0205089.
}
\lref\ofer{ 
O.~Aharony and S.~S.~Razamat,
``Exactly marginal deformations of N = 4 SYM and 
of its supersymmetric  orbifold descendants,''
arXiv:hep-th/0204045.
}
\lref\oferi{
O.~Aharony, B.~Kol and S.~Yankielowicz,
``On exactly marginal deformations of N = 4 SYM and 
type IIB  supergravity on AdS(5) x S**5,''
arXiv:hep-th/0205090.
}
\lref\ehsw{
B.~U.~Eden, P.~S.~Howe, E.~Sokatchev and P.~C.~West,
``Extremal and next-to-extremal n-point correlators in four-dimensional  SCFT,''
Phys.\ Lett.\ B {\bf 494}, 141 (2000)
[arXiv:hep-th/0004102].
}
\lref\ehpsw{
B.~U.~Eden, P.~S.~Howe, A.~Pickering, E.~Sokatchev and P.~C.~West,
``Four-point functions in N = 2 superconformal field theories,''
Nucl.\ Phys.\ B {\bf 581}, 523 (2000)
[arXiv:hep-th/0001138].
}
\lref\bernas{
D.~Berenstein and H.~Nastase,
``On lightcone string field theory from super Yang-Mills and holography,''
arXiv:hep-th/0205048.
}
\lref\wittqft{
E.~Witten,
``Topological Quantum Field Theory,''
Commun.\ Math.\ Phys.\  {\bf 117}, 353 (1988).
}
\lref\malda{
J.~M.~Maldacena,
``The large $N$ limit of superconformal field theories and supergravity,''
Adv.\ Theor.\ Math.\ Phys.\  {\bf 2}, 231 (1998)
[Int.\ J.\ Theor.\ Phys.\  {\bf 38}, 1113 (1999)]
[arXiv:hep-th/9711200].
}
\lref\GKP{
S.~S.~Gubser, I.~R.~Klebanov and A.~M.~Polyakov,
``Gauge theory correlators from non-critical string theory,''
Phys.\ Lett.\ B {\bf 428}, 105 (1998)
[arXiv:hep-th/9802109].
}
\lref\WittenQJ{
E.~Witten,
``Anti-de Sitter space and holography,''
Adv.\ Theor.\ Math.\ Phys.\  {\bf 2}, 253 (1998)
[arXiv:hep-th/9802150].
}
\lref\bala{
V.~Balasubramanian, M.~Berkooz, A.~Naqvi and M.~J.~Strassler,
``Giant gravitons in conformal field theory,''
JHEP {\bf 0204}, 034 (2002)
[arXiv:hep-th/0107119].
}
\lref\verlinde{
E.~Verlinde,
``Fusion Rules And Modular Transformations In 2-D Conformal Field Theory,''
Nucl.\ Phys.\ B {\bf 300}, 360 (1988).
}
\lref\McGreevyCW{
J.~McGreevy, L.~Susskind and N.~Toumbas,
``Invasion of the giant gravitons from anti-de Sitter space,''
JHEP {\bf 0006}, 008 (2000)
[arXiv:hep-th/0003075].
}
\lref\HashimotoZP{
A.~Hashimoto, S.~Hirano and N.~Itzhaki,
``Large branes in AdS and their field theory dual,''
JHEP {\bf 0008}, 051 (2000)
[arXiv:hep-th/0008016].
}
\lref\DasST{
S.~R.~Das, A.~Jevicki and S.~D.~Mathur,
``Vibration modes of giant gravitons,''
Phys.\ Rev.\ D {\bf 63}, 024013 (2001)
[arXiv:hep-th/0009019].
}
\lref\GrisaruZN{
M.~T.~Grisaru, R.~C.~Myers and O.~Tafjord,
``SUSY and Goliath,''
JHEP {\bf 0008}, 040 (2000)
[arXiv:hep-th/0008015].
}
\lref\HokerDM{
E.~D'Hoker, J.~Erdmenger, D.~Z.~Freedman and M.~Perez-Victoria,
``Near-extremal correlators and vanishing supergravity couplings in  AdS/CFT,''
Nucl.\ Phys.\ B {\bf 589}, 3 (2000)
[arXiv:hep-th/0003218].
}
\lref\ErdmengerPZ{
J.~Erdmenger and M.~Perez-Victoria,
``Non-renormalization of next-to-extremal correlators in N = 4 SYM 
and  the AdS/CFT correspondence,''
Phys.\ Rev.\ D {\bf 62}, 045008 (2000)
[arXiv:hep-th/9912250].
}

\def\Phid{\Phi^{\dagger}}
\def\cO{ {\cal{O}}  } 
\def\di{ \partial } 
\def\diz#1{ \partial z_{#1 }} 
\def\dizb#1{ \partial \bar z_{#1} } 
\def\zb#1{ \bar z_{#1} } 

\noblackbox

\newsec{ Introduction } 

 In \cjr\ we gave a systematic study 
 of extremal correlators of the most general half BPS 
 operators of the $U(N)$ theory.  These include single trace as well as 
 multiple trace operators. Young Diagrams 
 associated with $U(N)$ representations are useful 
 in characterizing the operators. When we use a Young Diagram 
 basis, the correlators take a very simple form. 
 The two point functions are diagonal in this basis 
 and the three point functions are proportional to 
 fusion coefficients of the unitary groups, i.e Littlewood Richardson 
 coefficients. Analogous group theoretic quantities 
 enter the higher point correlators. 
 Extremal correlators are also distinguished in that 
they obey non-renormalization theorems \refs{ \ehpsw,\ehsw}. 
Half-BPS operators 
and their non-renormalization theorems 
 have been the subject of many papers \refs{ \intril, 
\af,\afsz,\guma,\gumz,\lmrs,
 \ski, \ehpsw,\ErdmengerPZ,\HokerDM,\ehsw}. 
 A more complete list of references can be found \agmo\dhokfree.

 In this paper we will explore several 
 consequences of these results. 
 In the first place the extremal correlators 
 obey finite { \it factorization }   and { \it fusion } 
 equations which are 
 of a form similar to the ones that appear in 
 topological gauge theories, for example three-dimensional Chern-Simons 
 theory, two-dimensional $G/G$ models and two dimensional Yang-Mills. 
 The factorization equations imply bounds 
 on  the large $N$ growth of the correlators which guarantee that 
 a probabilistic interpretation of certain correlators 
 in terms of overlaps of incoming and outgoing states 
 is sensible.  
 The group theoretic nature of the correlators 
 also implies that certain { \it sum rules } can be 
 written down relating weighted sums of 
 higher point functions to 
 lower point functions.

 The similarity of the factorization equations 
 to those of topological gauge theories (TGT's)  
 leads us to investigate more detailed connections 
 between observables in TGT's and the SYM4 correlators. 
 We find that there are simple identities between 
 ratios of extremal correlators and appropriate 
 observables. In the SYM4-Chern- Simons correspondence, 
 there are natural relations involving both 
 observables on $S^3$ and observables on $S^2 \times S^1 $. 
After making these identifications, the 
 factorization equations for correlators 
 indeed map to factorization equations in 
 TGT's.  The sum rules we found above turn out 
 to reduce, in simple cases,  to relations between 
 Chern-Simons on $S^2 \times S^1$ and on $S^3 $. 
They are also related to the large $k$ limit of the Verlinde formula
\verlinde.  These results are section 7. 

 Many of these basic remarks extend beyond 
 extremal correlators. The basic technical tool 
 here is the fact that perturbative calculations 
 based on free fields involving arbitrary composites 
 are naturally described in terms of projection 
 operators in tensor spaces, which can be represented 
 in an economical fashion in  diagrammatic form.   
 We will call these { \bf projector diagrams.} Examples of 
  projector diagrams are Figures 7, 10, 12, 13.  
 These projectors 
 in tensor spaces can be related to projectors 
 acting in products of irreducible $U(N)$ 
 representations. The latter diagrams in turn can be related 
 to graphs where edges are labelled by irreps and vertices
 correspond to Clebsch-Gordan coefficients.
 We will call these { \bf projector graphs  }.
 Examples of such projector graphs are in Figures 11, 18 and 19. 
  Such diagrams, or slight variations thereof, 
 are familiar from various related contexts : 
 two dimensional Rational Conformal field Theory, 
 three-dimensional Chern-Simons theory, integrable lattice models, 
 and two dimensional Yang-Mills theory. 
 Thus we are able to map correlators in SYM4 to observables in 
Chern-Simons theory, in a very general manner. This map is particularly 
 simple in the cases where the correlators in SYM4
 are chosen to have simple spacetime dependence.

 The factorization equations 
 can be generalized to an equation we will call 
 {\bf staggered factorization }. The terminology 
 is based on the structure of the  projector diagrams
 or projector graphs which 
 allow such factorization. We will describe the 
 SYM4 correlators which admit such a factorization. 
 We observe that staggered factorization is related 
 to simple operations on Wilson loop expectation values in Chern-Simons 
 or in two-dimensional Yang-Mills theory. In the context of 
Chern-Simons theory, it is related to connected sums of links ( Wilson
 loops).     
In the context of two dimensional Yang-Mills, it is related to 
 Wilson loops with tangential intersections.  These points
are developed in section 8.
 
 The extremal correlators, which were first 
 identified as an interesting set of correlators 
in the context of studies of non-renormalization theorems, 
 were shown in \cjr\ to have  simple dependence on the 
$U(N)$ representation content of the operators involved, 
that is, the dependence is entirely in terms of 
dimensions and fusion coefficients rather than $6J$ symbols 
or such.  
This simplicity, given the results of this paper, 
can be traced to the factorization equations. 
Since these factorization equations have been generalized to staggered 
factorization,   
 it is tempting to conjecture that the 
 correlators which can be simplified using 
 staggered factorization, into dimensions and fusion coefficients, 
 are non-renormalized. 
 We will give some very heuristic 
 arguments in favour of such a conjecture in section 8 
 and leave a  detailed exploration of the question to the future. 

 In section 9 we will argue that the basic factorization equations, 
 derived above using explicit knowledge of the 
 complete space of half-BPS representations and their extremal 
correlators, actually follow from general considerations based 
 on the  OPE and superalgebra. As such they should work, 
 not just for $U(N)$ but also for the $SU(N)$ theory.  
 We do not have a direct proof in the case of $SU(N)$, 
 but we describe, in section 10, the complete set of half-BPS 
 representations in that case and write down some formulae
 for the relevant correlators, which should be constrained 
 by the basic factorization. We predict that some 
 clever manipulation of the relevant quantities, which 
 involve sums of products of Littlewood-Richardson 
 coefficients will give a direct proof of the factorization equations
 here.

 In section 11 we make some comments on giant gravitons \McGreevyCW. 
 These were the original motivation for the current study. 
 The first remark is that 
 the group theoretic formulae for three point functions imply 
 a symmetry between Kaluza-Klein (KK) correlators and giant graviton 
 correlators -- it would be fascinating to understand this symmetry 
 from the spacetime side.  The second, not unrelated,  remark
 concerns the special properties that the factorization equations have  
 when the extremal correlators involve giant graviton states. 

In appendix 1 we derive, starting from some 
group integrals familiar from lattice gauge 
theory, some basic properties of 
projection operators in tensor space, in particular the 
relation between unitary group projectors and symmetric group 
projectors. In appendix 2 we derive the relation between 
fusion multiplicities of unitary groups and branching multiplicities
of Symmetric groups from the same point of view. 
 In appendix 3 we give a general argument underlying staggered 
factorization, which allows a trace in tensor space
to be related to a product of traces in tensor space.  
In terms of projector diagrams, it relates one projector 
diagram to a product of projector diagrams. 
appendix 4 gives some steps in the diagrammatic derivation of 
correlators of descendants, which are discussed further in section 5. 
appendix 5 reviews
some $SO(6)$ group theory which is useful in section  8 
in describing the class of correlators which obey staggered 
factorization. 
In appendix 6, we translate the results of \cjr\ where general 
extremal correlators were given in the Schur Polynomial basis, 
to get some extremal correlators of traces. These have  appeared recently
in the context of pp-waves \refs{\bmn,  \cpss, \bernas, \harvmit }. 
The perspective of this paper suggests the existence of
interesting connections 
between pp-wave backgrounds and Chern-Simons theory. 
  
\newsec{ Factorization and Unitarity  for extremal correlators } 

 In \cjr\ a $1-1$ correspondence was established 
between half-BPS representations in $U(N)$
maximally supersymmetric Yang-Mills theory and 
symmetric polynomials in the eigenvalues of 
a complex matrix $\Phi$ or Schur Polynomials.  $\Phi$
is one combination of the six Hermition scalars
of ${\cal N}=4$ SYM given by $\Phi = \phi_1 + i \phi_4$. 
 We recall from \cjr\ that $k \rightarrow l $ 
 multi-point extremal correlation functions are given by  
\eqn\nrhtmx{\eqalign{ 
&   \langle ~~ \chi_{ R_1 } \bigl( \Phi  (x_1) \bigr)
 \chi_{R_2}  \bigl( \Phi (x_2) \bigr) \cdots 
  \chi_{R_k}  \bigl( \Phi (x_k ) \bigr)   ~  
\chi_{S_1} \bigl( \Phi^{\dagger}   (0) \bigr) \cdots
\chi_{S_l} \bigl( \Phi^{\dagger } (0) \bigr)   ~~\rangle  \cr 
   &= \sum_{S} g(R_1, R_2 \cdots R_k ; S ) 
 {  n(S)! Dim S \over d_S } g(S_1, S_2,
\cdots S_l ; S )   { 1 \over 
 ( x_1)^{2n ( R_1 ) } \cdots  (x_k)^{2n(R_k)  } }      \cr }}
where 
\eqn\defgrrr{ g(R_1, R_2, \cdots R_k ; S ) = 
\int dU \chi_{R_1} ( U ) \chi_{R_2} ( U ) \cdots \chi_{R_k} ( U )
 \chi_S ( U^{\dagger} ) } 
where the characters and measure for $U(N)$  are normalized so that 
$ \int dU \chi_R ( U ) \chi_S ( U^{\dagger}) = \delta_{RS } $. 
The sum runs over all irreps of $U(N)$ having 
$n( S)  = n( R_1 ) + \cdots + n(R_k) $ boxes.  
$Dim S $ is the dimension of the $U(N)$ irrep. associated
with the Young Diagram $S$ and $d_S$ is the dimension of 
the symmetric group irrep. associated with the same Young diagram. 
The $g$ factors can be expressed in terms 
of the basic fusion coefficient $g(R_1,R_2; S)$ 
which can be calculated using the Littlewood-Richardson 
combinatoric rule for combining Young Diagrams
\refs{\barac, \fulhar, \zelo }.

The four-point extremal correlators satisfy the following 
factorization condition. 
\eqn\fact{\eqalign{ 
&  \langle ~ \chi_{R_1} \bigl( \Phi  (x_1) \bigr) ~
 \chi_{R_2} \bigl( \Phi  (x_2) \bigr) ~ \chi_{S_1}
  \bigl( \Phi^{\dagger}  (0 ) \bigr) \chi_{S_2} 
\bigl( \Phi^{\dagger} (0) \bigr)  ~\rangle
\cr 
 & = \sum_{S} { \langle ~
 \chi_{R_1} \bigl( \Phi  (x_1) \bigr)  \chi_{R_2} \bigl( \Phi  (x_2)
\bigr) \chi_{S}
  \bigl( \Phi^{\dagger} (0) \bigr)  ~\rangle  ~
 \langle ~\chi_{S}  \bigl( \Phi ( y ) \bigr)  
\chi_{S_1} \bigl( \Phi^{\dagger}  (0) \bigr) \chi_{S_2}  \bigl( \Phi^{\dagger} 
(0 ) \bigr)  ~\rangle 
\over   \langle~  \chi_{S} \bigl( \Phi (0) \bigr) \chi_S \bigl(
 \Phi^{\dagger}  (y ) \bigr)  
~\rangle  }
\cr } }
where $y \neq 0$ but is otherwise arbitrary.  

There is an obvious generalization to 
higher point functions 
\eqn\facti{\eqalign{ 
& \langle ~
 \prod_{i} \chi_{R_i} \bigl( \Phi  (x_i) \bigr)  \prod_{j} 
\chi_{S_j} \bigl( \Phi^{\dagger} (0) \bigr)  ~\rangle \cr 
& = \sum_{S} { \langle ~  \prod_{i} \chi_{R_i} \bigl( \Phi  (x_i)
\bigr) ~~ \chi_S
\bigl( \Phi^{\dagger}  (0) \bigr)  ~ \rangle  ~
\langle ~ \chi_{S} \bigl( \Phi  (y )\bigr) \prod_j  
\chi_{S_j} \bigl( \Phi^{\dagger} (0) \bigr) 
~ \rangle 
\over    \langle ~
 \chi_S \bigl( \Phi  (0 ) \bigr) \chi_{S} \bigl( \Phi^{\dagger} (y  )
\bigr)  ~\rangle 
} \cr
}}

Taking equation \fact\ with $R_1 = S_1$ and $R_2 = S_2$ we see that 
it can be expressed as saying that the 
sum of  normalized two-point functions is equal to $1$. 
\eqn\sumtw{ 
 \sum_{S} { \langle ~ \chi_{R_1} ( \Phi )  \chi_{R_2}( \Phi ) \chi_{S}
  ( \Phi^{\dagger} ) ~ \rangle  
\over \sqrt {   
\langle ~
\chi_{R_1}  ( \Phi ) \chi_{R_2}  ( \Phi ) \chi_{R_1} ( \Phi^{\dagger}
)  \chi_{R_2} ( \Phi^{\dagger} ) ~\rangle  \langle~ \chi_{S} ( \Phi ) 
\chi_S ( \Phi^{\dagger} ) ~ \rangle  }}
= 1 }

This equation guarantees that the normalized 
two-point functions can be interpreted as amplitudes 
for transitions between states described by the wavefunctions
$ \chi_{R_1 } ( \Phi )  \chi_{R_2} ( \Phi ) $ on the one hand and 
$ \chi_S ( \Phi ) $ on the other. It shows that 
when the incoming state is a product of highest weight 
states of half-BPS representations of the superalgebra,  
the probability for the outgoing state to be the dual to a highest weight 
of a half-BPS representation is $1$.

\subsec{ Fusion identities } 

 The result \nrhtmx\ also implies that 
 higher point functions can be expressed in 
 terms of lower point functions. Consider, for example 
the four-point function 
\eqn\fusion{\eqalign{   
& \langle  ~~  \chi_{R_1} ( \Phi  ( x_1 ))  ~ \chi_{R_2} ( \Phi  ( x_2) )
~ \chi_{R_3} ( \Phi  ( x_3  )) ~  \chi_S ( \Phi^{\dagger}  (0)) ~~ 
 \rangle  \cr 
& = \sum_{S_1} ~~ g(R_1, R_2, S_1 )~  g(S_1, R_3, S ) ~ { n(S) !  Dim (  S )
 \over d_S}    ~  x_1^{-2n_1}  x_2^{-2n_2} x_3^{-2n_3}  \cr 
& = \sum_{S_1}  
\langle~~ \chi_{R_1} ( \Phi  ( x_1))  ~ \chi_{R_2} ( \Phi  ( x_2 )) ~ 
 \chi_{S_1}  ( \Phi^{\dagger}  ( 0 )) ~~ \rangle \cr 
& {  1 \over \langle ~~ \chi_{S_1} ( \Phi ( y )) ~
\chi_{ S_1 } ( \Phi^{\dagger}  
  ( 0)) ~~ \rangle }   ~~~~~
  \langle ~~ \chi_{S_1} ( \Phi  ( y ))  ~ \chi_{R_3} ( \Phi  ( x_3 ))
 ~ \chi_{S}  ( \Phi^{\dagger}   ( 0 )) ~~ \rangle
\cr }} 
To obtain the second line we have 
used  \defgrrr\ for $g(R_1,R_2,R_3 ; S) $ in terms
of group integrals and expanded  the class function 
$\chi_{R_1} ( U ) \chi_{R_2} ( U) $ 
into a sum of irreducible characters of $U(N)$. 
This yields the coefficients $g(R_1,R_2; S_1) $ which can in turn 
be re-expressed in terms of three-point functions. 
The simple spacetime dependences of these extremal 
correlators allow the functions of $x_i$ on the left and 
right to match. 

 The generalization to higher point functions is 
\eqn\fusioni{\eqalign{
& \langle  \prod_{i=1}^{k}  \chi_{R_i } ( \Phi  ( x_i  )) \chi_S ( 
\Phi^{\dagger}
 (0)) \rangle 
= \sum_{ S_1 \cdots S_{k-2 } }   
 \langle \chi_{R_{1 }}  ( \Phi  ( x_{1})   ) 
\chi_{R_{2}  } ( \Phi  (x_{2} )) 
 \chi_{S_{1}  }  ( \Phi^{\dagger}  ( 0)) \rangle \cr 
 & \prod_{i=1}^{k-2 }
{ 1 \over \langle \chi_{S_{ i } }   ( \Phi  ( 0)) \chi_{S_{i} }
( \Phi^{\dagger}   ( y_i )) \rangle } 
 \langle \chi_{S_{i} } ( \Phi  ( y_i )) \chi_{ R_{ i+2 }}
(\Phi^{\dagger} 
(x_{i+2}  ))
    \chi_{S_{ i+1 }} ( \Phi (0)) ~~ \rangle \cr }}
 Here $ S_{k-1} \equiv S $ and $y_{k-1} = 0$.

In section 9, we outline  a general derivation of these
equations based on properties of the superalgebra
symmetry and the OPE. 
From the OPE we expect 
such factorization equations to be true, but typically 
we would expect infinitely many operators to be involved 
in the intermediate sums. In two-dimensional 
CFTs finite sums are possible, for general 
correlators,  because there is 
an infinite conformal algebra symmetry. In the
case at hand, such equations would typically involve infinite 
sums, but the sums are finite when we work with
extremal correlators of half-BPS operators.

\newsec{ Elements of diagrammatic derivations } 

 The calculation of the finite $N$ extremal correlators 
 in \cjr\ as well as further calculations of finite $N$ 
 non-extremal correlators can be simplified 
 by using a diagrammatic method. 

 The matrix $ \Phi $ is an operator which transforms  
 states of an $N$ dimensional vector space $V$. 
 The matrix elements of $\Phi $ are $ \Phi^{i}_{j}$. 
\eqn\matel{ 
\langle ~ e^i | ~\Phi ~  |e_j \rangle = \Phi^i_j }  


  We can naturally extend the action of $ \Phi $ 
 to the n-fold tensor product $V^{\otimes n } $,  by considering 
 the operator $ \Phi \otimes \Phi \cdots \Phi $. 
 The matrix elements are now $ \Phi^{i_1}_{j_1} \Phi^{i_2}_{j_2}
 \cdots \Phi^{i_n}_{j_n}$.  
\eqn\mateltns{ 
\langle ~ e^{i_1} \otimes e^{i_2} \otimes \cdots  \otimes e^{i_n} | 
\Phi \otimes \Phi \otimes \cdots \otimes \Phi  
| e_{j_1} \otimes e_{j_2} \otimes \cdots \otimes e_{j_n} \rangle 
= \Phi^{i_1}_{j_1} \Phi^{i_2}_{j_2}
 \cdots \Phi^{i_n}_{j_n}
}

\ifig\phiinvn{Diagram for $\Phi$ operator in $V^{\otimes n }$ }
{\epsfxsize3.0in\epsfbox{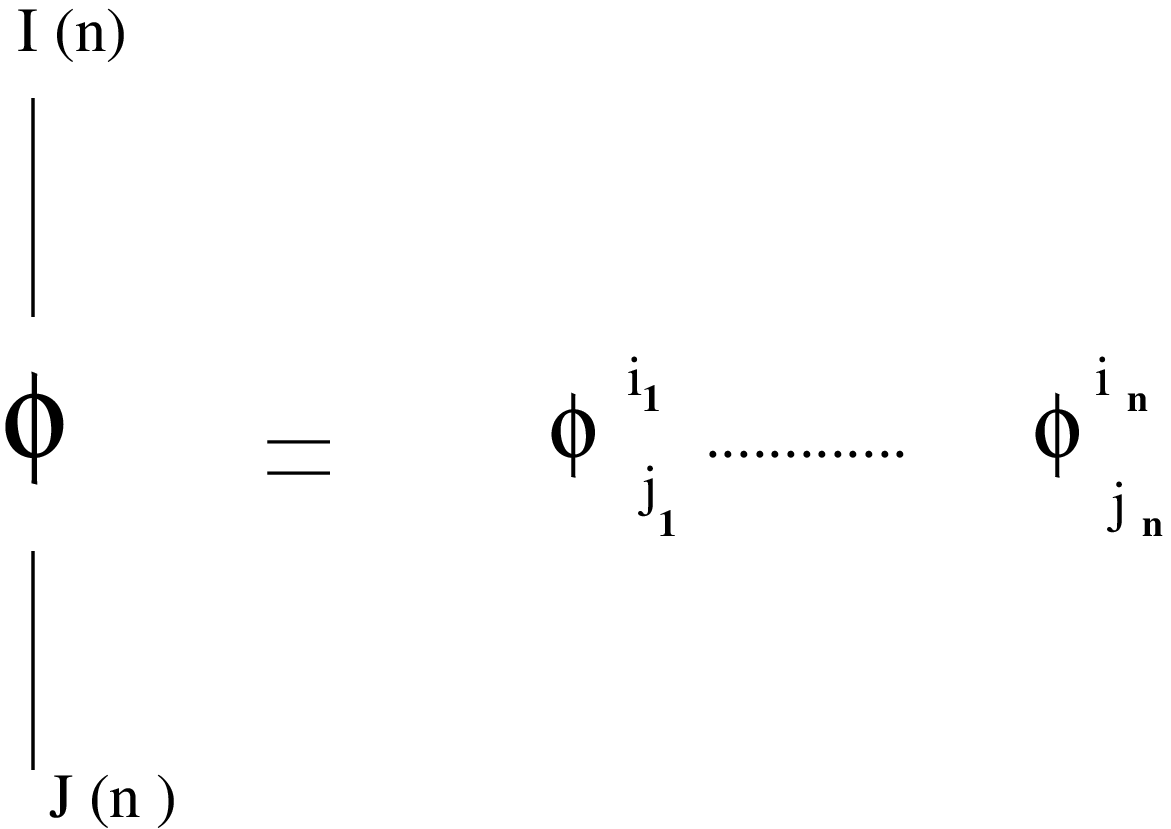} }

Sometimes it is convenient to use a multi-index $I(n)$ 
for an index set $(i_1, i_2, \cdots i_n ) $. The equation 
\mateltns\ is presented diagrammatically in \phiinvn.

\ifig\tracephi{ Diagram for Trace of $\Phi$ }
{\epsfxsize2.5in\epsfbox{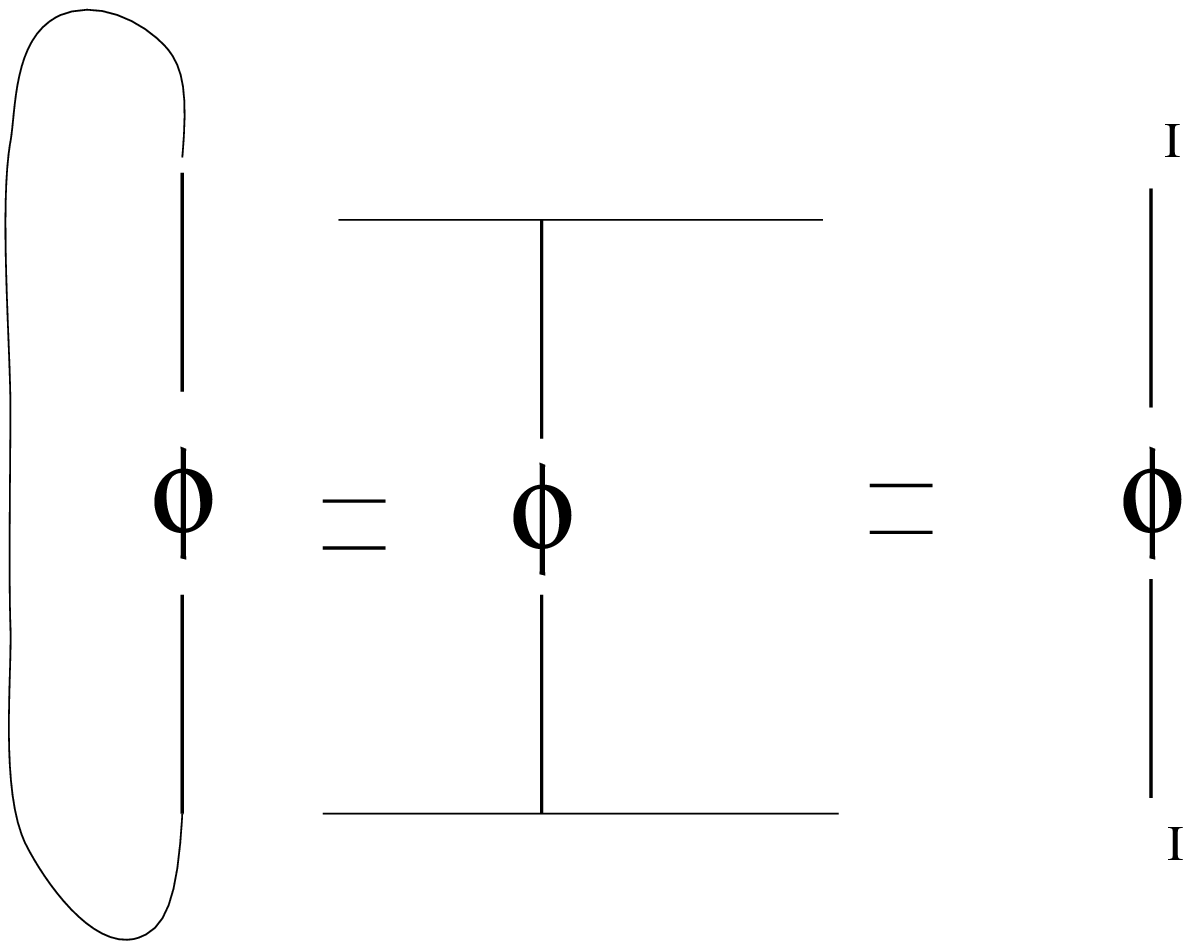} }

The trace of the operator $\Phi $ acting on 
$ V^{\otimes n } $ is denoted by the diagram in \tracephi. 
When we do free field contractions, we 
end up summing over different permutations
which describe how we are contracting. 
It is useful to recall that the 
matrix element of the permutation $\gamma $ 
acting on $V^{\otimes n }$ is 
\eqn\matsig{ ( \gamma )^{i_1 ~ i_2 \cdots i_n }_{j_1 ~ j_2 \cdots j_n } 
   = \delta^{i_1}_{j_{\gamma(1) } }
  \delta^{i_2}_{j_{\gamma(2) }}  \cdots  \delta^{i_n}_{j_{\gamma(n) } } } 

The basic correlator can be written as 
\eqn\bascr{\eqalign{  
& \langle ~~ \Phi^{i_1}_{j_1} (x_1) \cdots \Phi^{i_n}_{j_n} (x_1) ~~
       (\Phi^{\dagger})^{k_1}_{l_1} (x_2) \cdots 
   (\Phi^{\dagger})^{k_n}_{l_n} (x_2) ~~ \rangle
\cr 
& = (x_1-x_2)^{- 2 n } ~~ \sum_{\gamma } 
\delta^{i_1}_{l_{\gamma(1)} } \delta^{i_2}_{l_{\gamma(2)} } \cdots 
\delta^{i_n}_{l_{\gamma(n)} } ~~
\delta^{k_1}_{j_{ \gamma^{-1} (1) }} \delta^{k_2}_{j_{\gamma^{-1} (2) }}
\cdots \delta^{k_n}_{j_{ \gamma^{-1} (n) }} \cr }}

A slightly more compact way of writing this 
is 
\eqn\bascri{ \langle ~~\Phi^{I(n)}_{J(n)} (x_1)~~ 
( \Phi^{\dagger}  )^{K(n)}_{L(n)} (x_2) ~~\rangle 
= (x_1-x_2)^{- 2 n }  ~~ \sum_{\gamma}   ( \gamma )^{I(n)}_{L(n)  } 
~~(  \gamma^{-1} )^{K(n)}_{J(  n ) } }  
 
\ifig\twoptdiag{Diagram for two-point function  }
{\epsfxsize3.0in\epsfbox{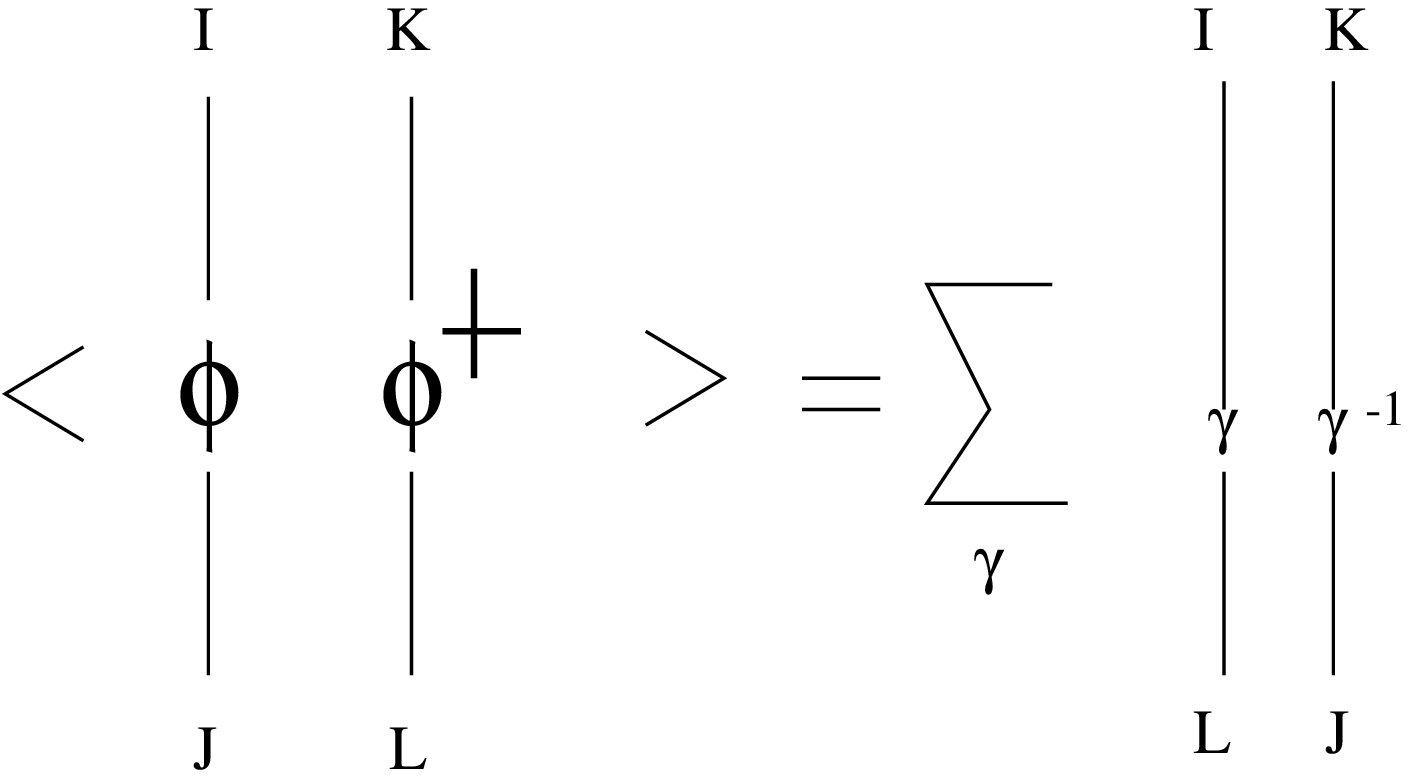} } 

The notation $I(n)$ indicates that $I$ is a 
 multi-index involving $n$ indices $i_1, \cdots i_n$. 
When the number of indices involved is 
clear from the context, we will not need to make it explicit, 
and we can write : 
\eqn\bascrii{ \langle  ~~\Phi^{I}_{J} (x_1) ~~
( \Phi^{\dagger}  )^{K}_{L} (x_2)  ~~\rangle 
= (x_1-x_2)^{- 2 n } ~~\sum_{\gamma}   ( \gamma )^{I}_{L } ~~
( \gamma^{-1} )^{K}_{J }. } 
This can be expressed in diagrammatic form as in \twoptdiag.

In the diagrams, we will not make the $x$ dependences
explicit. 
We can also keep  the positions 
of the indices unchanged with respect to the first 
term, at the cost of introducing a twist, that
is, a permutation acting on $ V^{\otimes n } \otimes V^{\otimes n }$ 
which switches the first $n$ copies with the last $n$ 
copies, compare figures 3 and 4.

\ifig\nexti{ Alternative diagram for two-point function   }
{\epsfxsize3.0in\epsfbox{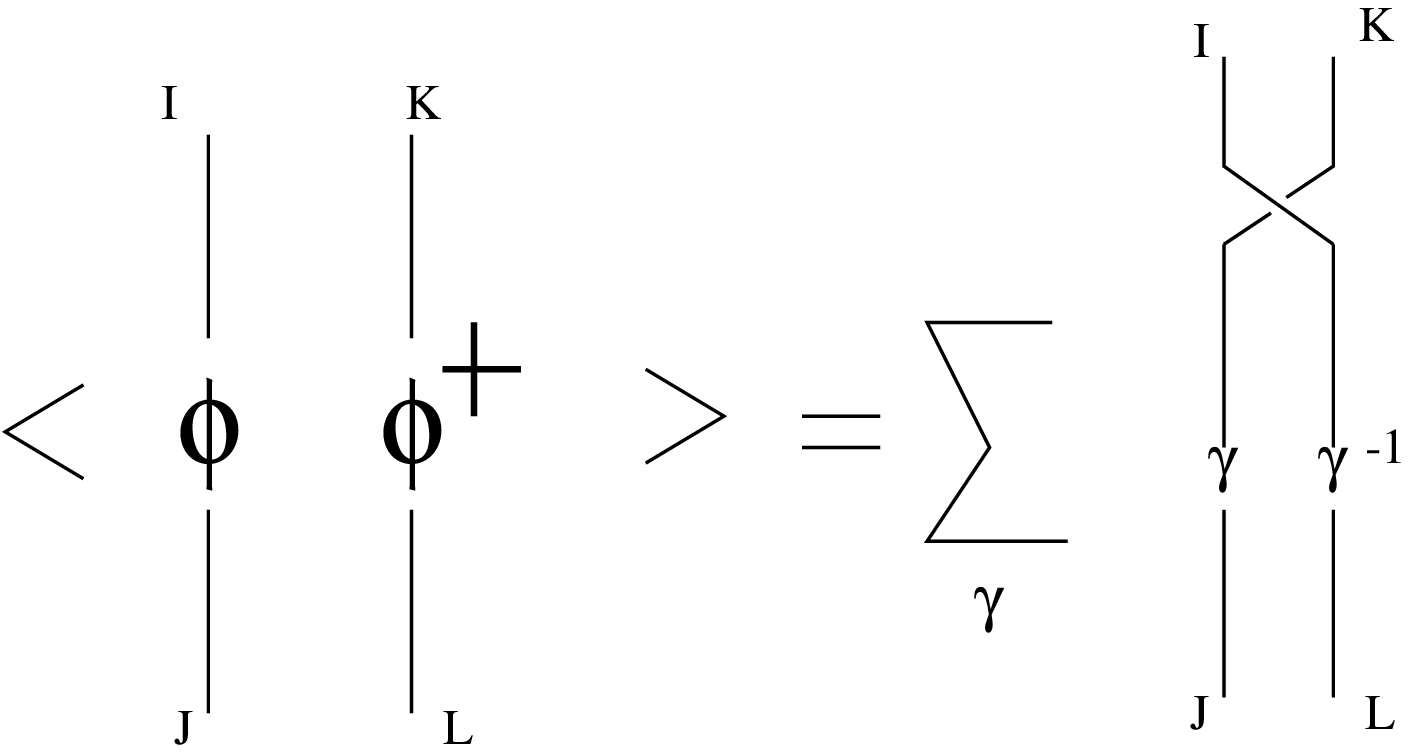} }

The advantage of the last step is that we 
can express the key correlator \bascri\ in a completely index-free 
diagrammatic way as in figure 5, with the understanding 
that when we put back the indices to recover the more 
familiar formula, there is no reshuffling of multi-indices 
between left and right hand side of the equations.  

\ifig\nexti{ Index-free diagram for two-point function }
{\epsfxsize3.0in\epsfbox{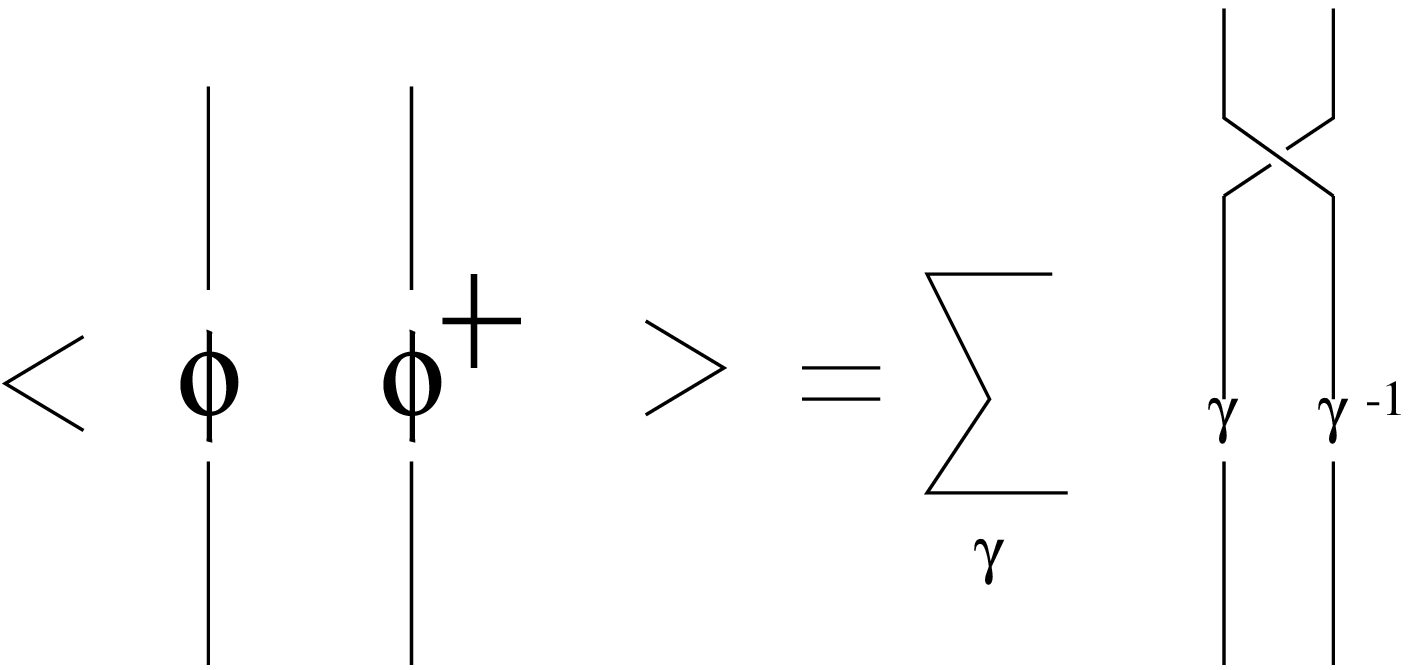} }

Now we express some of the correlators evaluated
in \cjr\ using the diagrammatic notation.
Gauge invariant correlators are obtained
by contracting the free indices with
the $U(N)$ invariant metric $\delta^{i}_{j}$.
In \cjr\ these contractions were taken in the
Schur polynomial basis, that is one contracts
the $\Phi (\Phi^{\dagger})$ indices with
operators 
\eqn\proj{\eqalign{{1 \over d_R} (P_{R})^{J(n)}_{I(n)} & =
{1 \over n !} \sum_{\sigma \in S_n} \chi_{R}(\sigma)~~
 ( \sigma )^{J(n)}_{ I (n))} \cr
& = {Dim R \over d_R} \int dU \chi_R (U) ~~ (U^{\dagger})^{J(n)}_{I(n)}
}}
where in the first line we have given the symmetric group
form of the projector and in the second line the
unitary group integral form of the projector. The equality 
of these two projectors when acting on tensor 
space is proved in appendix 1.
Indeed it is straightforward to show that $P_R$ is
a projection operator satisfying
\eqn\projident{P_{R} P_{S} = \delta_{R,S} P_S}
and the trace condition
\eqn\projtrace{tr(P_{R}) = d_R ~ Dim R.}
To calculate the two-point function $d_R d_S \langle \chi_R (\Phi)
\chi_S (\Phi^{\dagger}) \rangle$ one 
has to evaluate 
$ \langle tr ( P_R \Phi ) tr ( P_S \Phi^{\dagger} )\rangle$. 
This is illustrated in the left of figure 6, with 
$R,S$ drawn for $P_R, P_S$. The correlator of 
$\Phi$'s  is evaluated using figure 4, to give 
the middle diagram of figure 6, a sum over $\gamma$ being understood. 
An obvious diagrammatic manipulation or equivalently 
an identity relating traces in $ V^{\otimes n } \otimes V^{\otimes n
}$
to traces in $  V^{\otimes n }$ leads to the final diagram in figure 6.

\ifig\untwopoint{Diagram of the $U(N)$ two point function. }
{\epsfxsize3.0in\epsfbox{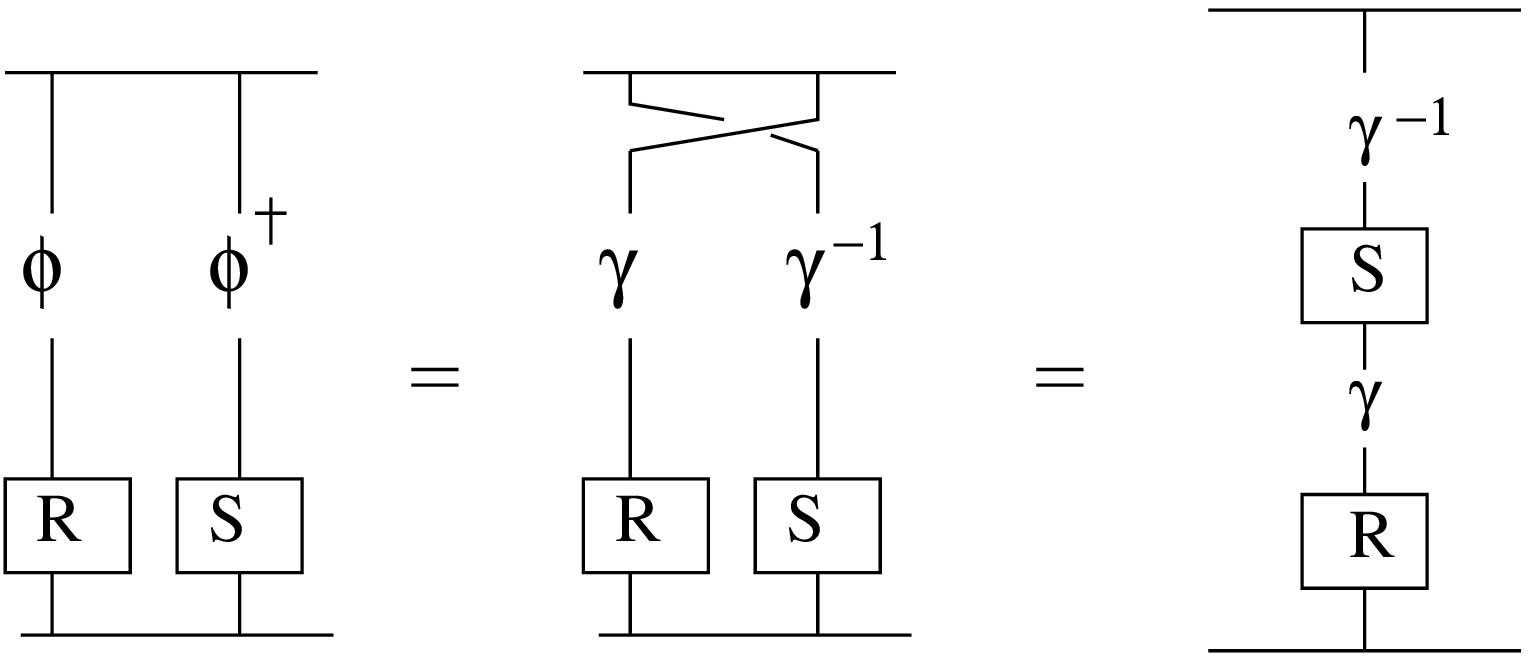} }

Explicitly one has
\eqn\untwopoint{\eqalign{
\langle \chi_R (\Phi) \chi_S (\Phi^{\dagger}) \rangle
& = {1 \over d_R d_S} \sum_{\gamma \in S_n} tr(P_R(\gamma^{-1} P_S
\gamma))
\cr
& = {n! Dim R \over d_R} \delta_{R,S}
}}
where we have used the properties \projident\ and \projtrace\ and
the fact that $P_S$ commutes with $\gamma$
to evaluate \untwopoint.

More generally for the multi-point correlator
we have 
\eqn\mulpt{\eqalign{  
&  \langle \chi_{R_1} (\Phi) \cdots \chi_{R_n} (\Phi)
\chi_{S_1} ( \Phi^{\dagger}) \cdots 
\chi_{S_m} ( \Phi^{\dagger})  \rangle \cr 
& = { 1 \over  \prod_i d_{R_i} } { 1 \over \prod_i d_{S_i}} 
  \langle tr (P_{R_1} \Phi ) \cdots tr ( P_{R_k}  \Phi ) 
           ~~ tr ( P_{S_1} \Phi^{\dagger } ) \cdots tr ( P_{S_l}
 \Phi^{\dagger } ) \rangle \cr }}
As in the case of the two point function, 
we can express this in a diagram
and then use figure 4 to simplify. The resulting diagram 
can be manipulated into figure 7.   To avoid clutter
we denote the projection operators $P_{R}$ in 
the projector diagrams by just $R$.

\ifig\unmultifig{Diagram of a $U(N)$ multi-point correlator. }
{\epsfxsize3.0in\epsfbox{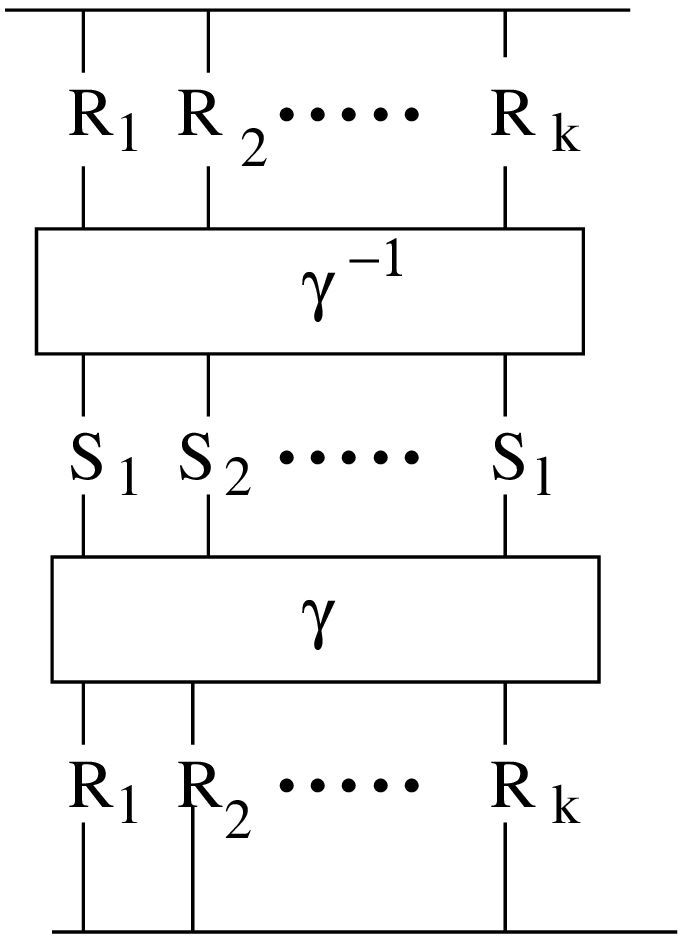} }

Written out this is
\eqn\unmulti{\eqalign{
\langle \chi_{R_1} (\Phi) & \cdots \chi_{R_k} (\Phi)
\chi_{S_1} (\Phi^{\dagger}) \cdots 
\chi_{S_l} (\Phi^{\dagger})  \rangle
= {1 \over d_{R_1} \cdots d_{R_k} d_{S_1} \cdots d_{S_l}} \cr
& \times \sum_{\gamma \in S_n} tr(
(P_{R_1} \otimes \cdots \otimes P_{R_k}  ) 
\gamma^{-1}( P_{S_1} \otimes \cdots \otimes P_{S_l}) \gamma) 
}}
The integer $n$ is given by
 $n= n(R_1) + \cdots n(R_k ) = n(S_1) + \cdots n(S_l)$. 
To evaluate the trace we proceed as follows.  The projection
operator $( P_{S_1} \otimes \cdots \otimes P_{S_l})$ projects
onto a subspace of $V^{\otimes n}$ 
corresponding to a reducible representation of $U(N)$.
To decompose this reducible representation into
irreducible representations we recall 
the theorem, see eg. \fulhar\zelo, that $V^{\otimes n }$
can be decomposed as
\eqn\decomp{V^{\otimes n} \cong \oplus_S S \otimes s  }
where the sum is over all  Young diagrams 
corresponding to 
irreducible representations
$S$  of $U(N)$ with dimension $Dim S$
and irreps $s$ of $S_n $ with dimension $d_S$.
 When $n$ is smaller than $N$ \decomp\  includes 
all irreps $s$  of $S_n$, but when
$n \ge N$, it only includes those
irreps of $S_n$ which correspond to Young Diagrams with 
no column of length larger than $N$. 
As a representation
of $U(N) \times  S_n$, $V^{\otimes n}$ decomposes
into irreps $S \otimes s $ with unit multiplicity. 
Equivalently, 
$V^{\otimes n}$ consists of $d_S$ copies
of each irrep $S$ of $U(N)$. 
This multiplicity of irreps of $U(N) $ in 
tensor space has the consequence that Schur Polynomials 
are related to traces of projectors by a factor $ {1 \over d_S }$. 
\eqn\schupro{ 
\chi_S ( \Phi ) = { 1 \over d_S } tr ( P_S \Phi ) }

We can compute the trace by inserting a 
complete set of projectors $P_S$ acting in 
$V^{\otimes n}$. 
\eqn\com{\eqalign{ 
&  tr( (P_{R_1} \otimes \cdots \otimes P_{R_k}  ) 
\gamma^{-1}( P_{S_1} \otimes \cdots \otimes P_{S_l}) \gamma)
\cr 
& = \sum_S  
tr( 
(P_{R_1} \otimes \cdots \otimes P_{R_k}) P_S P_S 
(\gamma^{-1}( P_{S_1} \otimes \cdots \otimes P_{S_l}) \gamma)) \cr }}
The $P_S$ can be expressed in 
terms of a unitary group integral or a symmetric 
group sum as in \proj. Now $ \sum_{\gamma } \gamma^{-1}
 (P_{S_1} \otimes \cdots P_{S_l}  )
\gamma $ commutes with the action of $ U(N) \otimes S_n$ 
in $V^{\otimes n} $. If we write the projectors $P_{S_i}$ in terms 
of symmetric groups, it is clear that this operator commutes with 
$U(N)$. The averaging by $\gamma$ guarantees that it also commutes
with $S_n$.  By Schur's Lemma, and taking advantage 
of the fact that $P_S$ projects onto a single irreducible
representation of $U(N) \times S_M  $ 
 we can factor the trace 
to get 
\eqn\factrd{\eqalign{ 
&   \sum_{ \gamma} tr \left( (\tau^{-1}
(P_{R_1} \otimes \cdots \otimes P_{R_k}) \tau)
(\gamma^{-1}( P_{S_1} \otimes \cdots \otimes P_{S_l}) \gamma) \right) \cr 
&  =  \sum_{ \gamma } 
 tr( ( P_{R_1} \otimes \cdots \otimes P_{R_k})  ~ P_S )
 {1 \over d_S Dim S }  
tr  ( P_S \gamma^{-1}( P_{S_1} \otimes \cdots \otimes P_{S_l}) \gamma)  }}  
The manipulation has a simple diagrammatic meaning. 
We have factored the diagram in figure 7 to a pair of diagrams 
as in figure 8. The factors can now be evaluated. 

\ifig\firstfacdiag{Diagram illustrating the factored
form of \unmultifig.}
{\epsfxsize4.0in\epsfbox{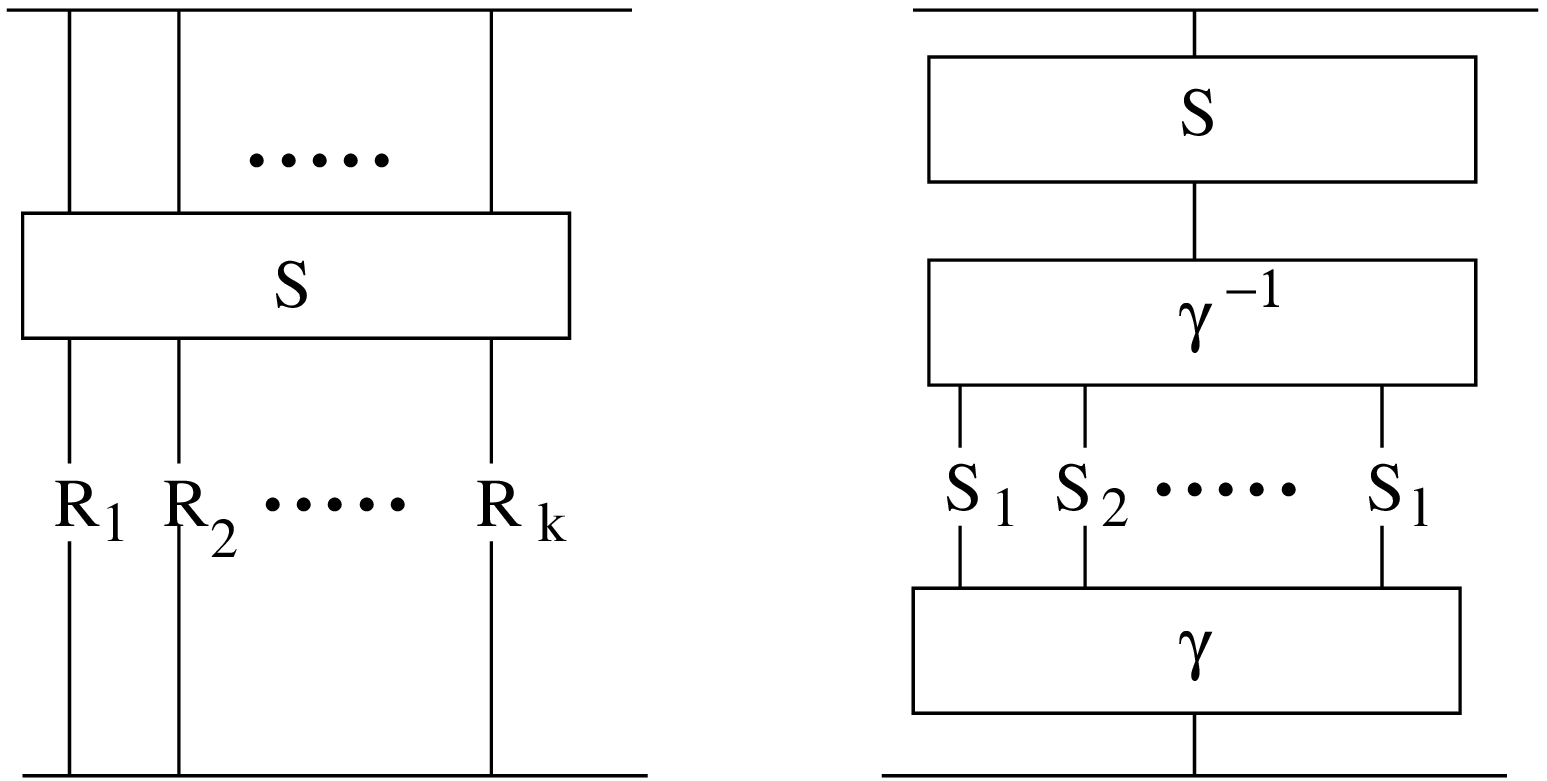} }

Equivalently we state a relation between 
tensor products of projection operators
in $ V^{\otimes n_1} \otimes \cdots V^{\otimes n_l}$
and projectors in $ V^{\otimes ( n_1 + n_2 + \cdots + n_l )}$.   
Schur's lemma
implies the decomposition
\eqn\firstmultidecomp{\sum_{\gamma} \gamma^{-1}
( P_{S_1} \otimes \cdots \otimes P_{S_l}) \gamma
=\sum_{S}
\alpha(S_1, \cdots, S_l;S) P_S
}
To fix the coefficient $\alpha(S_1, \cdots, S_l;S)$ we
multiply the expression by a projection
operator $P_{S'}$ and trace.  The result is 
\eqn\multidecomp{\sum_{\gamma} \gamma^{-1}
( P_{S_1} \otimes \cdots \otimes P_{S_l}) \gamma
=d_{S_1} \cdots d_{S_l} \sum_{S} {n_S ! \over d_S}
g(S_1, \cdots, S_l;S) P_S
}
where the trace has been evaluated to be
\eqn\simp{\sum_{\gamma } tr(\gamma^{-1} 
(P_{S_1} \otimes \cdots \otimes P_{S_l}) \gamma ~ P_S )
= n! { d_{S_1}  \cdots d_{S_l}  } g(S_1, \cdots S_l ;  S ) 
    Dim ~ S.}   
The trace can be evaluated either
directly using the $U(N)$ form of the projection
operators given in \proj\ or just by using the 
fact that we can convert the trace 
in tensor spaces with projectors $ P_{S_1} \otimes \cdots  P_{S_l} $ 
into a trace in the tensor product of irreducible 
$U(N)$ representations with a factor $d_{S_1} \cdots d_{S_l}$. 
Further the tensor product $ S_1 \otimes  \cdots S_l$ contains the 
representation $S$ with a multiplicity 
$ g(S_1, S_2 \cdots S_l ;  S )$. 
Taking the trace gives a factor of $Dim S $.

Collecting all the factors we get 
\eqn\unmultifin{\eqalign{
\langle \chi_{R_1} (\Phi) & \cdots \chi_{R_k} (\Phi)
\chi_{S_1} (\Phi^{\dagger}) \cdots 
\chi_{S_l} (\Phi^{\dagger})  \rangle \cr
& = \sum_S g(R_1,\cdots,R_k;S) {n(S) ! ~ Dim S \over d_S}
g(S_1, \cdots, S_l;S).}}
which was derived in \cjr\ by using characters rather 
than projectors.

\newsec{ Sum rules for three-point funtions } 

The basic idea for sum rules
is that sums of projectors onto Young diagrams
in tensor space gives $1$, as in \decomp. 
In terms of diagrams we have figure 9. 
\ifig\bassmrl{ Basic identity leading to sum rules } 
{\epsfxsize3.0in\epsfbox{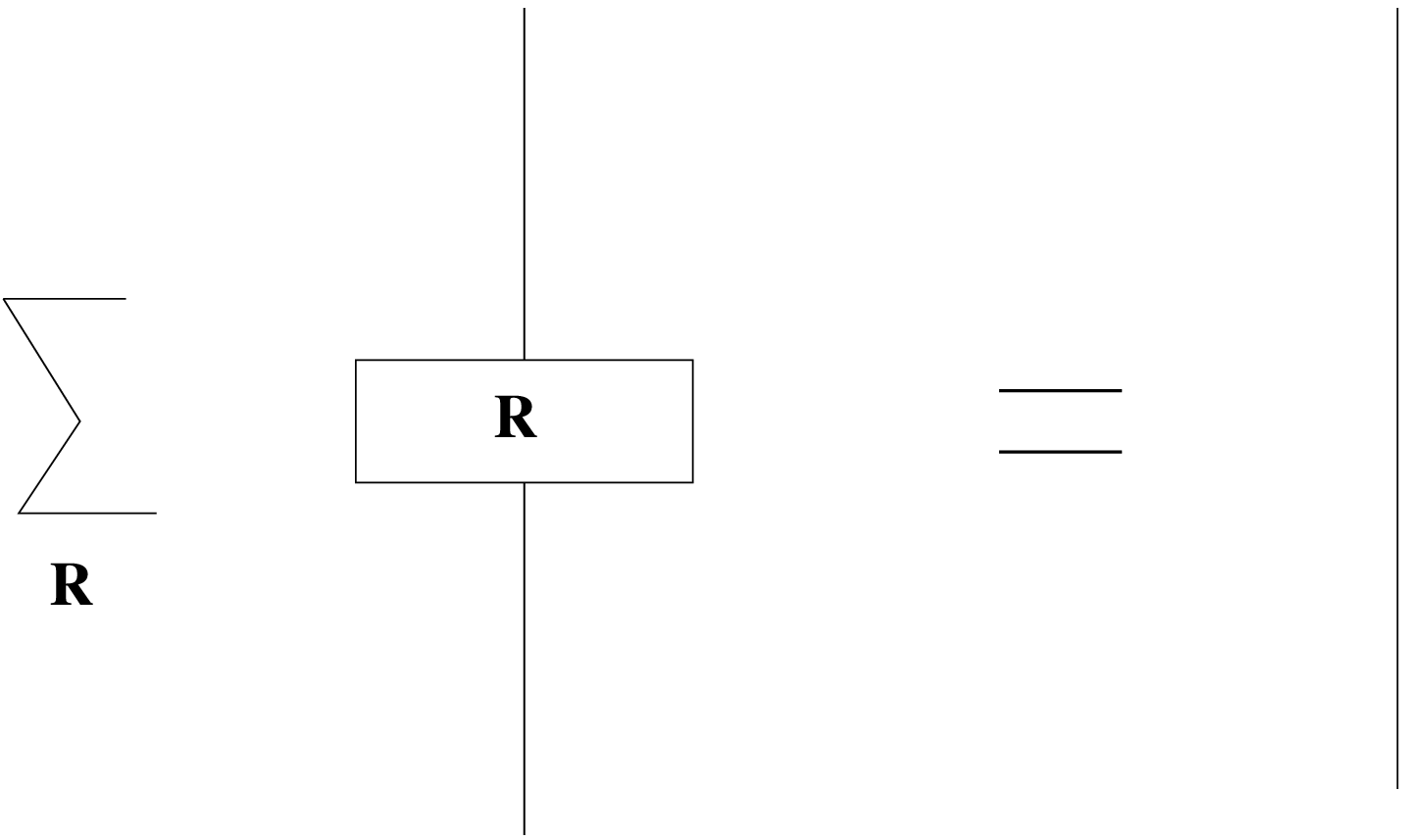} }

Consider the three-point function 
\eqn\thrpt{\eqalign{ 
& \langle \chi_{R_1} \bigl( \Phi (x) \bigr) \chi_{R_2} \bigl( \Phi(x)
\bigr) \chi_S ( \Phi^{\dagger} ( 0) ) ~ \rangle \cr 
& = {1 \over x^{2 n(S)}} g(R_1, R_2 ; S) { n(S) ! Dim S \over d_S} \cr 
& =  {1 \over x^{2 n(S)}} 
{ n(S) ! \over d_{R_1} d_{R_2} d_S } tr \bigl( ( P_{R_1} \otimes P_{R_2} )
P_S \bigr) \cr 
}} 
The integer $n(S)$, the number of boxes in
the Young Diagram of $S$, is related to 
the numbers of boxes $R_1$ and $R_2$ 
by $n(S) = n(R_1) + n(R_2) $. 
When we sum over $R_1$ we get the identity operator 
in $V^{\otimes n(R_1)} $. Summing over $R_2$ we get the 
identity in $V^{\otimes n(R_2) }$. This leads to 
$tr_{ n(S) } ( P_{S} ) $, which is just the dimension 
of the irrep of $U(N) \times S_n $ associated with the 
Young Diagram of $S$, i.e $d_{S} Dim ( S ) $.  
\eqn\smrlxt{\eqalign{&
 \sum_{R_1 R_2} 
d_{R_1} d_{R_2} \langle
 ~\chi_{R_1} \bigl( \Phi (x)  \bigr) \chi_{R_2} \bigl( \Phi (x)  \bigr) 
 \chi_{S} \bigl( \Phi^{\dagger} (0)  \bigr) ~\rangle \cr 
& ={1 \over x^{2 n(S)}} Dim ( S ) n(S)! 
 = d_S \langle  \chi_{S} (  \Phi(x) ) \chi_{S} \bigl( \Phi^{\dagger}(0)
\bigr)
 \rangle  \cr }}

Expressed in terms of 
$ \hat \chi_R = d_R \chi_R $ we have 
\eqn\smrlext{\eqalign{&
 \sum_{R_1 R_2} 
 \langle {  \hat \chi_{R_1}}  \bigl( \Phi(x)  \bigr) { \hat \chi_{R_2} 
\bigl( \Phi (x)  \bigr) } 
 { \hat \chi_{S}}  \bigl( \Phi^{\dagger} (0) \bigr) \rangle  \cr 
& =  \langle {\hat \chi_{S} \bigl(  \Phi (x)  \bigr)} { \hat  \chi_{S} } 
\bigl( \Phi^{\dagger} (0) \bigr) ~ \rangle  \cr }}

Another sum rule is obtained from just summing over 
the  single representation $S$. 
\eqn\smrlxti{\eqalign{
& \sum_{S } d_{S} \langle \chi_{R_1} \bigl( \Phi (x)  \bigr)
 \chi_{R_2} \bigl( \Phi(x)  \bigr) 
 \chi_{S} \bigl( \Phi^{\dagger} (0) \bigr) \rangle \cr 
& = d_{ R_1} d_{R_2} 
{ n(S)  ! \over n(R_1)! n(R_2) !} \langle \chi_{R_1} \bigl( \Phi
(x) \bigr) 
\chi_{R_1} \bigl( \Phi^{\dagger} (0) \bigr)  ~ \rangle  
\langle ~ \chi_{R_2} \bigl( \Phi(x)  \bigr) 
\chi_{R_2} \bigl( \Phi^{\dagger} (0)  \bigr)  \rangle  \cr }}

In the above we presented the sum rules 
using  properties of projectors. This is indeed the
approach that works in  general. In the
simple extremal case considered above, 
we may also write down the above equations
by recalling the fact that 
$g(R_1,R_2 ; S ) $ is a fusion coefficient for 
$U(N) $ so that 
\eqn\fusid{ 
 \sum_S Dim S ~ g(R_1,R_2 ; S ) = Dim R_1 ~ Dim R_2 } 
which leads to \smrlxti. 
Since it also has an interpretation 
as a branching coefficient for $ S_{n(R_1) } \times S_{n(R_2)}
\rightarrow  S_{n(S) } $, we immediately write down  
\eqn\bchid{ 
\sum_{R_1,R_2} d_{R_1} d_{R_2} g(R_1,R_2; S) = d_{S} }
which leads to \smrlxt.
The equality between the fusion coefficient 
and branching coefficient follows from Schur 
duality \fulhar. For completeness a proof is given 
in appendix 2.

The argument based on projectors generalizes 
to cases where the original correlator 
cannot be written in terms of fusion coefficients.  
These sum rules can also be used to simplify 
non-extremal three-point functions. 
The basic fact that leads to sum rules 
is that once we have expressed the correlator in terms 
of a trace in tensor space of a  sequence of projectors, 
each of which acts in some subspace of the tensor space, 
it is natural to expect that the trace will simplify 
if some projectors are summed 
to leave the identity as in figure 9. 

\newsec{ Formulae for non-extremal three-point functions }

 We will consider two types of departure from extremality. 
 One involves descendants. Another involves generalizing 
 position dependencies while leaving the operators 
 to be highest weights and their duals. 
 A new kind of sum rule will be possible here -- where 
 we sum over one projector and then simplify using 
 the staggered factorization of appendix 3.

\subsec{ Correlators of descendants : $ \sum_{i} \Delta_i \ne \sum_{j}
\bar \Delta_j $} 

As a specific example of a non-extremal correlator we consider
the descendant 
$\langle E_{21}^k \chi_{R_1} ( \Phi_1 )  Q_{12}^k  \chi_{R_2} ( \Phi_1 ) 
\chi_{R_3} ( \Phi_1^{\dagger} ) \rangle$
of the extremal correlator
$\langle \chi_{R_1} (\Phi_1) \chi_{R_2} (\Phi_1) 
\chi_{R_3} (\Phi^{\dagger}_1) \rangle$ evaluated in 
section 2.
The operators $E_{21}$ and $Q_{12}$ are elements of
the Lie algebra of $SO(6)$ and are defined in appendix 5.
For the
case at hand $E_{21}$ converts $\Phi_1$ scalars to
$\Phi_2$ scalars and $Q_{12}$ converts $\Phi_1$ scalars
to $\Phi^{\dagger}_2$ scalars.  Explicitly we have
\eqn\eacts{E_{21}^k \chi_{R_1} (\Phi_1) = {(n_1 + k)! \over n_1 !}
{Dim R_1 \over d_{R_1}} \int dU_1 \chi_{R_1} (U_1)
(tr(U^{\dagger}_1 \Phi_2))^k 
(tr(U^{\dagger}_1 \Phi_1))^{n_1}}
in terms of unitary group integrals, with a similar expression
for $Q_{12}^k \chi_{R_2} (\Phi_1)$.  By the diagrammatic rules we 
can show that the above non-extremal correlator is equal to a trace
of a sequence of projectors acting on tensor space, up to a factor
$  { 1\over d_{R_1} d_{R_2} d_{R_3} }
(n_1 + n_2)! k!  { ( n_1+k )! ( n_2+k)!
\over n_1! n_2! } $. The factor  $(n_1 + n_2)! k!$ arises from 
all possible ways of contracting the $\Phi_1$ and $\Phi_2$
fields respectively.  Equivalently this factor arises from
summing over the permutations $\gamma_1$ and $\gamma_2$ appearing
in appendix 4 where the derivation of the correlator
is explained diagrammatically.  The factors $(n_1 + k)! / n_1 !$
and $(n_2 + k)! / n_2 !$ arise from the number of ways
of converting $k$ $\Phi_1$'s into $\Phi_2$'s and
$\Phi^{\dagger}_2$'s respectively as in \eacts.
The factor of  ${ 1\over d_{R_1} d_{R_2} d_{R_3} } $ 
comes from the relation between projectors and Schur 
Polynomials given in \schupro.  

\ifig\projdesc{ Projector diagram for a correlator of descendants  }
{\epsfxsize1.6in\epsfbox{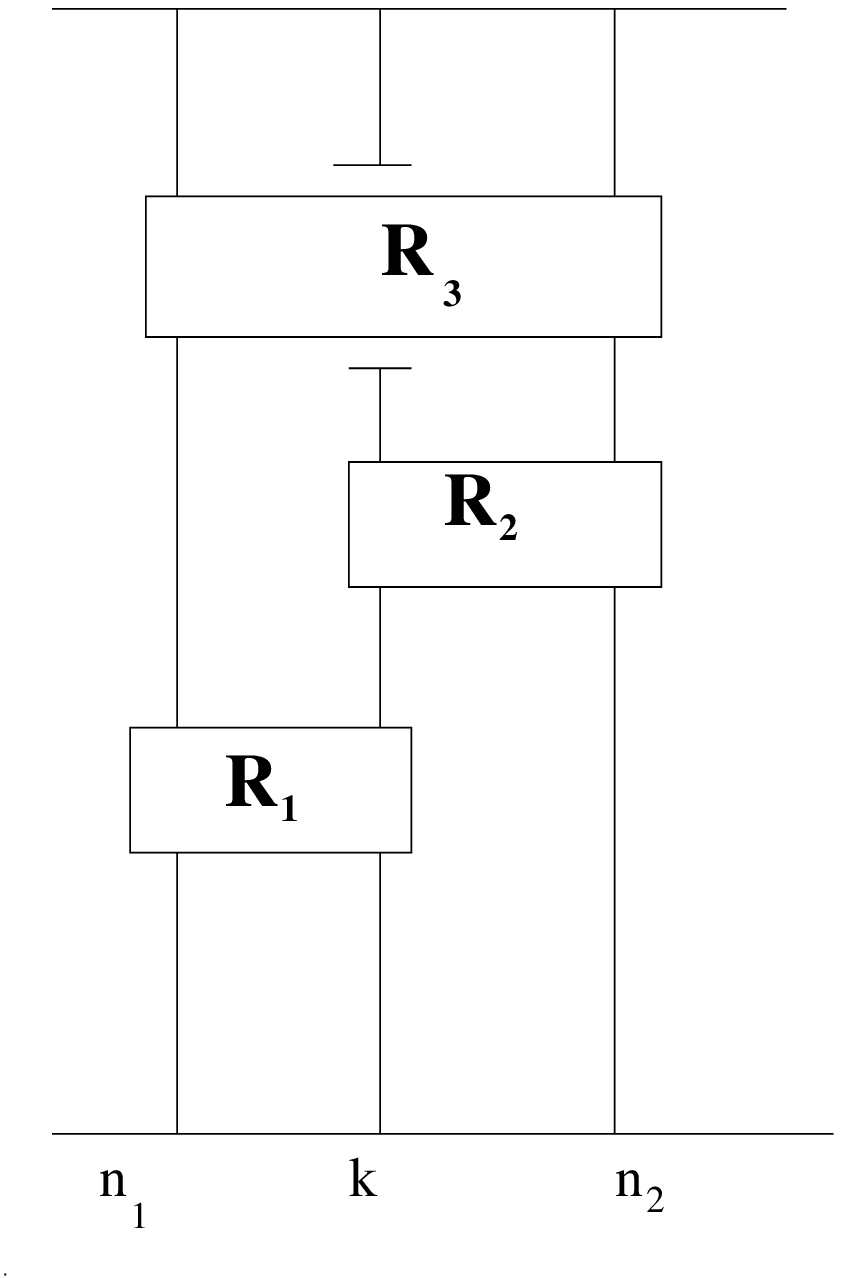} }

The sequence of projectors 
is shown in the \projdesc\ below and the derivation 
is explained in appendix 4. 
This allows us to write down a group integral
for the correlator. 
\eqn\gpint{\eqalign{&  
\langle E_{21}^k \chi_{R_1} ( \Phi_1 (x_1) )  Q_{12}^k  
\chi_{R_2} ( \Phi_1 (x_2) ) 
\chi_{R_3 } ( \Phi_1^{\dagger} (0) ) \rangle = 
{ (n_1+n_2)!  k ! (n_1+k)!(n_2+k)! \over n_1! n_2 !  } \cr
& \times {1 \over x_{1}^{2 n_1}} {1 \over x_{2}^{2 n_2}}
{1 \over (x_1 - x_2)^{2 k}}
  { Dim  ( R_1 )  \over d_{R_1}  }  { Dim  ( R_2 )  \over d_{R_2}  } 
    { Dim  ( R_3 )  \over d_{R_3}  }\cr 
& \int dU_1 dU_2 dV_1 
\chi_{R_1}(U_1^{\dagger}) ~ \chi_{R_2}(U_2^{\dagger})~
 \chi_{R_3 }(U_3^{\dagger}) ~ ( tr ( U_1 U_3 ))^{n_1}  ~ ( tr( U_2 U_3 )
)^{n_2} ~ ( tr (U_1 U_2) )^k .}}

It is useful to convert the traces in tensor 
spaces to traces in tensor products of irreps. of $U(N)$  
\eqn\gpinti{\eqalign{&  
\langle E_{21}^k \chi_{R_1} ( \Phi_1 (x_1) )  
Q_{12}^k  \chi_{R_2} ( \Phi_1 (x_2) ) 
\chi_{R_3 } ( \Phi_1^{\dagger} (0) ) 
= { (n_1+n_2)!  k ! (n_1+k)!(n_2+k)! \over n_1! n_2 !  } \cr
& \times {1 \over x_{1}^{2 n_1}} {1 \over x_{2}^{2 n_2}}
{1 \over (x_1 - x_2)^{2 k}}
  { Dim  ( R_1 )  \over d_{R_1}  }  { Dim  ( R_2 )  \over d_{R_2}  } 
    { Dim  ( R_3 )  \over d_{R_3}  }\cr 
&  \sum_{S_1, S_2 , S_3 } Dim ( S_1 )  ~ Dim ( S_2 ) ~ Dim (S_3) \cr 
&  \int dU_1 dU_2 dV_1 
\chi_{R_1}(U_1^{\dagger}) ~ \chi_{R_2}(U_2^{\dagger})~
 \chi_{R_3 }(U_3^{\dagger}) ~ \chi_{S_1}( U_1 U_3 )  ~ 
\chi_{S_2}( U_2 U_3 )
 ~  \chi_{S_3} ( U_1 U_2) \cr }}
$S_1$ runs over irreps of $U(N)$ with $n_1$ boves, 
$S_2$ has $n_2$ boxes and $S_3$ has $k$ boxes.

Now we have a sequence of projectors acting 
on a tensor product of irreps of $U(N)$
\eqn\trirrep{\eqalign{ &  
Dim  ( R_1 )  Dim  ( R_2 )  Dim  ( R_3 )\cr 
& \int dU_1 dU_2 dV_1 
\chi_{R_1}(U_1^{\dagger}) ~ \chi_{R_2}(U_2^{\dagger})~
 \chi_{R_3 }(U_3^{\dagger}) ~ \chi_{S_1}( U_1 U_3 )  ~ 
\chi_{S_2}( U_2 U_3 ) ~  \chi_{S_3} ( U_1 U_2) \cr 
&=  tr_{S_1 \otimes S_2 \otimes S_3 } ( P_{S_1, S_2}^{ R_1}
 P_{S_2,S_3}^{R_2} P_{S_1, S_3}^{R_3} )  \cr }} 
 
It is useful to draw a diagrammatic representation 
of this where a projector has been replaced by a 
sequence of two trivalent vertices. 
This corresponds, in formulae, to the expansion of projectors in terms
of a product of Clebsch-Gordan coefficients. 
\eqn\expcl{ 
P_{S_1 S_2}^{ R_1 } = \sum_{ \alpha, m_1, m_2, m }
 C_{S_1, m_1 ; S_2, m_2 }^{\alpha,  R_1, m }   
 C_{ \alpha,  R_1 , m }^{ S_1, m_1;  S_2, m_2 } } 
Here $\alpha $ runs over the $ g(S_1, S_2; R_1) $ occurences
of $R_1$ in the irrep. decomposition of $S_1 \otimes S_2 $. 
$m_1,m_2,m $ run over states in the irreps $ S_1,S_2$ and $R_1$ 
respectively. The diagrammatic presentation now makes 
no reference to tensor space and  is a graph with 
legs labelled by irreps. of $U(N)$ and vertices representing
Clebsch-Gordan coefficients.
 \ifig\unprojdone{ projector graph with U(N) irrep labels }
{\epsfxsize2.5in\epsfbox{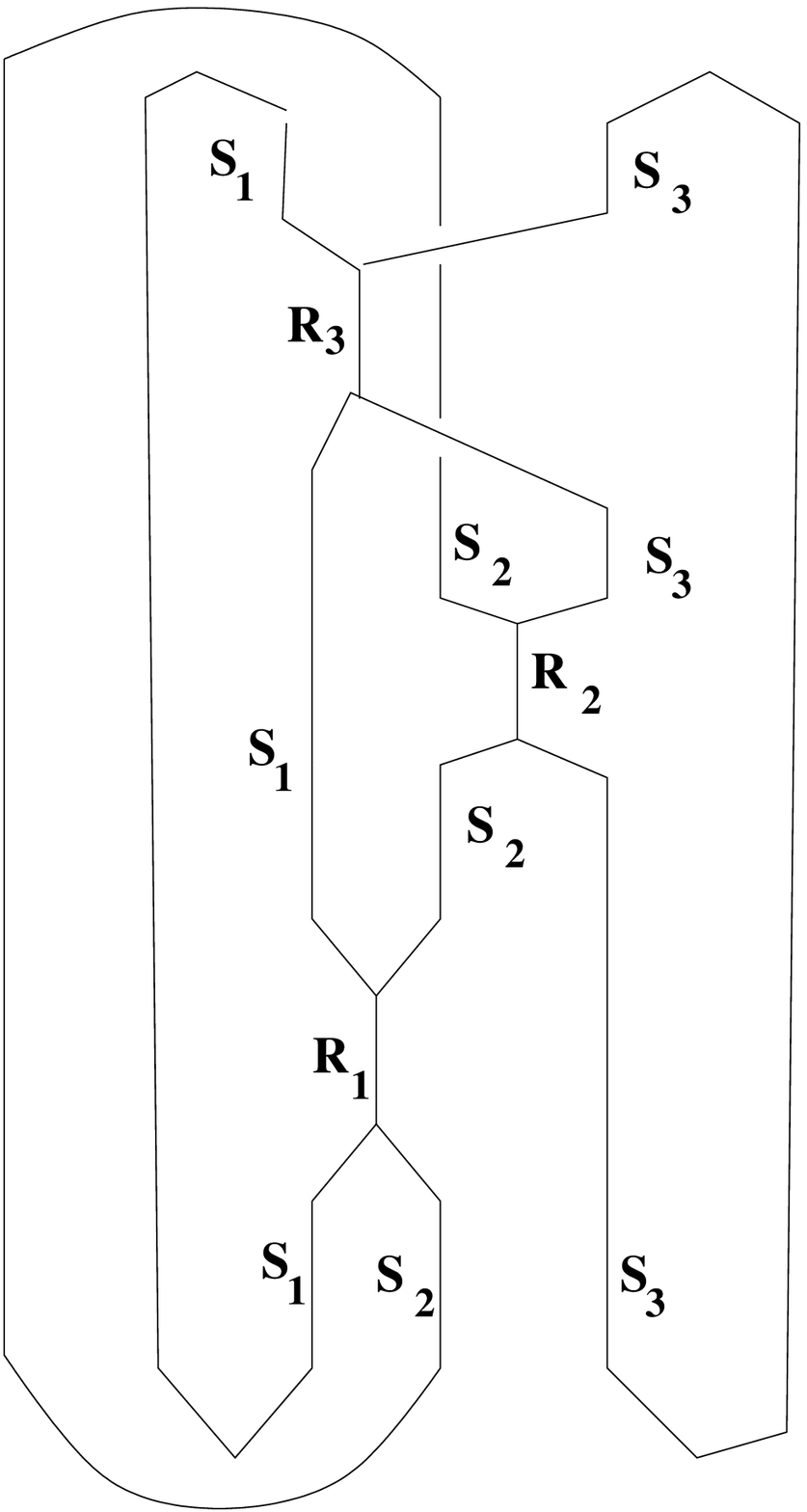} }
Converting  the projectors  in  \nexti\ to 
products of Clebsch's and representing the result diagrammatically
we arrive at \unprojdone. 
These { \bf projector graphs } are useful in the mapping 
of the SYM4 correlators to quantities in Chern-Simons theory as we will 
discuss in section 7.   

\subsec{ Correlators with general positions } 

\ifig\fourptfig{Projector diagram corresponding to the
four-point correlator with general positions.}
{\epsfxsize2.5in\epsfbox{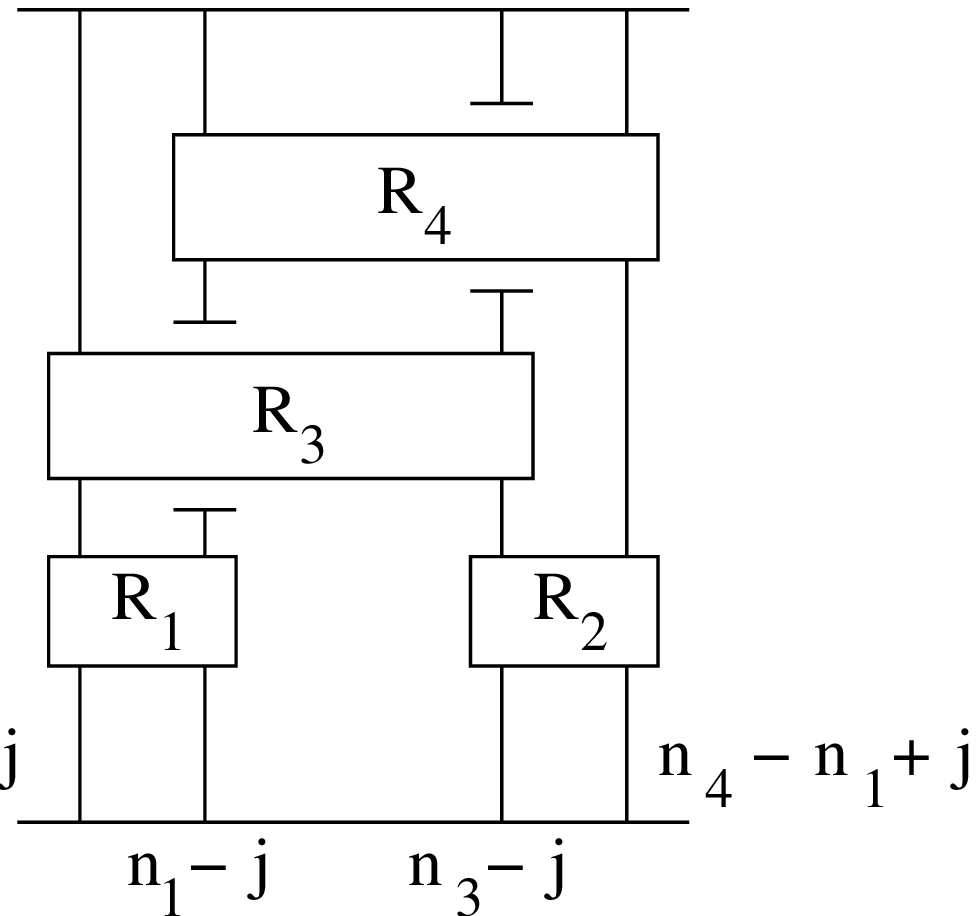} }

Another class of non-extremal correlators of interest are
generalizations of those considered in \cjr\ by
allowing arbitrary
position dependence.  Consider for example the four point
correlator $\langle \chi_{R_1} (\Phi (x_1)) \chi_{R_2} (\Phi (x_2))
\chi_{R_3} (\Phi^{\dagger} (x_3) 
\chi_{R_4} (\Phi^{\dagger} (x_4))  \rangle$ where now the
two $\Phi^{\dagger}$ operators are evaluated at different
locations.  
From the diagrammatic rules described in section 3 we see that
this correlator will involve a sum of traces of products
of projection operators corresponding to the diagram shown
in \fourptfig.

The sum can taken to be over the number $j$ of fields from
$\chi_{R_1}$ contracted with fields from $\chi_{R_3}$.  For
fixed $j$ this fixes the number of fields of $\chi_{R_1}$
contracted with fields from $\chi_{R_4}$ and moreover
the number of fields from $\chi_{R_2}$ contracted with
fields from $\chi_{R_3}$ and $\chi_{R_4}$ respectively.
In \fourptfig\ we have included a label on each
strand denoting the number of contractions occuring as 
described above.  Also we denote the number of boxes
in the Young diagram associated to representation $R_i$
by $n_i$.

Inserting the character expansions in terms
of unitary integrals and evaluating the contractions one
finds
\eqn\fourptgp{\eqalign{\langle \chi_{R_1} & 
(\Phi (x_1)) \chi_{R_2} (\Phi (x_2))
\chi_{R_3} (\Phi^{\dagger} (x_3)) 
\chi_{R_4} (\Phi^{\dagger} (x_4))  \rangle =
\left( \prod_{i=1}^{4} {Dim R_i \over d_{R_i}} \right) 
{1 \over (x_1 - x_4)^{2 n_1}} \cr
& \times
{1 \over (x_2 - x_4)^{2(n_4 - n_1)}}
{1 \over (x_2 - x_3)^{2 n_3}}
\sum_{j=sup(0,n_1 - n_4)}^{inf(n_1,n_3)} {n_1 ! n_2 ! n_3 ! n_4 !
\over j! (n_1 -j)! (n_4 - n_1 +j)! (n_3 -j)!} \cr
& \times
\left( {(x_1 - x_4)(x_2 - x_3) \over 
(x_1 - x_3)(x_2 - x_4)} \right)^{2 j}
\int (\prod_{i=1}^{4} dU_i \chi_{R_i}(U_i) )
(tr(U_{1}^{\dagger} U_{3}^{\dagger}))^j
(tr(U_{1}^{\dagger} U_{4}^{\dagger}))^{n_1 - j} \cr
& \times
(tr(U_{2}^{\dagger} U_{4}^{\dagger}))^{n_4 - n_1 + j}
(tr(U_{2}^{\dagger} U_{3}^{\dagger}))^{n_3 - j}.
}}
This can be simplified somewhat by expressing it in terms
of a trace of projection operators as
\eqn\fourptgpproj{\eqalign{\langle \chi_{R_1} & 
(\Phi (x_1)) \chi_{R_2} (\Phi (x_2))
\chi_{R_3} (\Phi^{\dagger} (x_3)) 
\chi_{R_4} (\Phi^{\dagger} (x_4))  \rangle =
\left( \prod_{i=1}^{4} {1 \over d_{R_i}} \right) 
{1 \over (x_1 - x_4)^{2 n_1}} \cr
& \times
{1 \over (x_2 - x_4)^{2(n_4 - n_1)}}
{1 \over (x_2 - x_3)^{2 n_3}}
\sum_{j=sup(0,n_1 - n_4)}^{inf(n_1,n_3)} {n_1 ! n_2 ! n_3 ! n_4 !
\over j! (n_1 -j)! (n_4 - n_1 +j)! (n_3 -j)!} \cr
& \times
\left( {(x_1 - x_4)(x_2 - x_3) \over 
(x_1 - x_3)(x_2 - x_4)} \right)^{2 j}
tr((P_{R_1} \otimes P_{R_2}) (P_{R_3} \otimes_{j} P_{R_4}))
}}
where the $j$ subscript on the tensor product
denotes
that the projection operator $P_{R_3}$ acts on the
first $j$ indices and the last $(n_3 + n_4 -j)$ indices
while $P_{R_4}$ acts on the remaining indices
as in \fourptfig.  
The main difference here
as compared to the earlier computation in \unmulti\
is that there is a nontrivial position dependence in 
the sum on $j$.  Were it not for this, the correlator
could be re-expressed as in \unmulti\ and evaluated
in terms of fusion coefficients and dimensions of 
representations.  
As it is however we cannot perform this simplification.
This can be generalized in a
fairly obvious way to higher point correlators.

\newsec{ Sum-rules for non-extremal correlators } 

We consider in this section sum rules for the
various non-extremal correlators evaluated in the previous
section.  
As in section 4, the sum rules follow easily when the
correlators are expressed in terms of traces of products
of projection operators, using $\sum_R P_R = I$.  

\ifig\threeptsumdiag{Diagram representing the trace appearing
in the sum rule for the non-extremal correlator \gpint.}  
{\epsfxsize3.0in\epsfbox{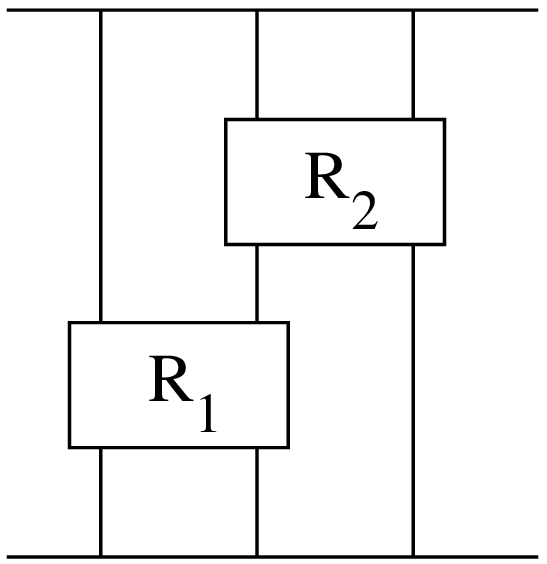}}

As an example consider the non-extremal three point correlator
\gpinti\ rewritten in terms of projection operators using
\trirrep.  If we multiply by $d_{R_3}$ and
sum over all the $R_3$ representations, then the
projector $P_{R_3}$ is replaced by the identity operator
resulting in 
\eqn\threeptsum{\eqalign{ 
\sum_{R_3} ~~
d_{R_3} \langle   E_{21}^k \chi_{R_1} ( \Phi_1 (x_1) )  &
Q_{12}^k  \chi_{R_2} ( \Phi_1 (x_2) ) 
\chi_{R_3} ( \Phi_1^{\dagger} (0) )\rangle
= {(n_1 +k) ! (n_2 +k) ! k! (n_1 + n_2)!
\over n_1 ! n_2 !} \cr 
& \times {1 \over d_{R_1} d_{R_2}}
{1 \over x_{1}^{2 n_1}} {1 \over x_{2}^{2 n_2}}
{1 \over (x_1 - x_2)^{2 k}}
tr((P_{R_1} \otimes 1_{n_2})
(1_{n_1} \otimes P_{R_2})).
}}
The trace appearing here can be evaluated in terms of
just fusion coefficients and dimensions of various representations.
Diagrammatically the trace is represented
in \threeptsumdiag.
  
\ifig\ftonesumfig{Projector diagram defining the
operator $F_{T_1}$ in terms of the projection operator
$P_{R_1}$.}
{\epsfxsize2.5in\epsfbox{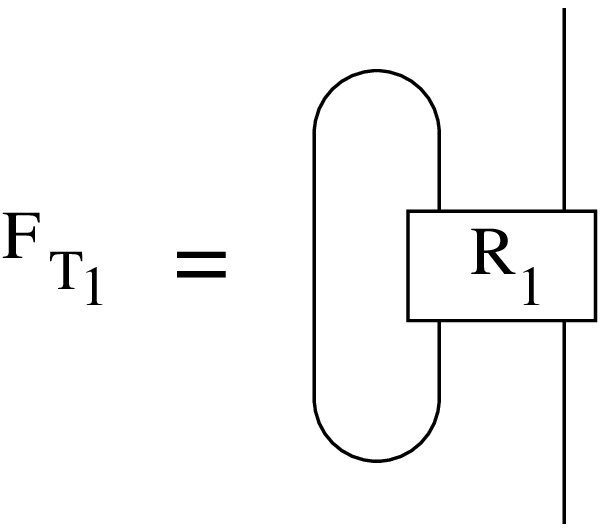} }

This sort
of diagram is of the general class discussed in appendix 3
that can be split, or factorized, into a pair of diagrams.  
The argument for the factorization identity is detailed
there, but the important point to note here is
that the operators $F_{T_1}$ and $F_{T_2}$ introduced in appendix 3
are given by
\eqn\ftonesum{\eqalign{F_{T_1} & = (tr_{n_1} \otimes 1_k) P_{R_1} \cr
F_{T_2} & = (1_k \otimes tr_{n_2}) P_{R_2}
}}
and where $F_{T_1}$ is represented diagrammatically in \ftonesumfig.

In terms of unitary group integrals $F_{T_1}$ is given by
\eqn\ftonesumunit{F_{T_1}
= Dim R_1 \int dU_1 \chi_{R_1}(U_1) (tr(U^{\dagger}_1))^{n_1}
\rho_k (U^{\dagger}_1)}
and similarly for $F_{T_2}$
\eqn\fttwosumunit{F_{T_2}
= Dim R_2 \int dU_2 \chi_{R_2}(U_2) (tr(U^{\dagger}_2))^{n_2}
\rho_k (U^{\dagger}_2).}
Plugging in the explicit
forms for $F_{T_1}$ and $F_{T_2}$ into the factorized form
(15.5) of the identity derived in appendix 3 leads to
\eqn\factsum{ tr((P_{R_1} \otimes 1_{n_2})
(1_{n_1} \otimes P_{R_2})) = \sum_{S_2} {1 \over d_{S_2} ~ Dim S_2}
tr(P_{R_1} (P_{S_2} \otimes 1_{n_1})) 
tr((1_{n_2} \otimes P_{S_2} ) P_{R_2}).
}
The latter traces are of the form evaluated in
\cjr\ and moreover follow easily from relation \multidecomp\ derived
in section 3.  Specifically one finds that 
\eqn\basethree{tr((P_{S_2} \otimes P_{S_1}) P_{R_1}) =
d_{S_2} d_{S_1} g(S_2, S_1;R_1) ~ Dim R_1.
}
To get exactly the traces on the right-hand-side of
\factsum\ one simply notes that $\sum_{S_1} P_{S_1} = I$.
This results in the sum rule
\eqn\threeptsumtwo{\eqalign{ & \sum_{R_3} 
d_{R_3} \langle  E_{21}^k \chi_{R_1} ( \Phi_1 (x_1) )  
 Q_{12}^k  \chi_{R_2} ( \Phi_1 (x_2) ) 
\chi_{R_3} ( \Phi_1^{\dagger} (0) )\rangle
=  
{(n_1 +k) ! (n_2 +k) ! k! (n_1 + n_2)!
\over n_1 ! n_2 !} \cr & \times 
{1 \over x_{1}^{2 n_1}} {1 \over x_{2}^{2 n_2}}
{1 \over (x_1 - x_2)^{2 k}}
{Dim R_{1} ~ Dim R_{2} \over d_{R_1} d_{R_2}}
\sum_{S_1,S_2,S_3} {d_{S_1} d_{S_2} d_{S_3} \over
Dim S_2} g(S_1,S_2;R_1) g(S_3,S_2;R_2).
}}

Finally we note that multiplying by $d_{R_1}$ or $d_{R_2}$
instead and summing would yield a correlator of a similar
form to that as in \threeptsum.  In particular the
same sort of trace would appear as is shown in the diagram
\threeptsumdiag.

\subsec{ Sum rules for Correlators with general positions } 

\ifig\fourptsumdiag{Diagram representing the trace appearing
in the sum rule for the correlator \fourptgpproj.}
{\epsfxsize3.0in\epsfbox{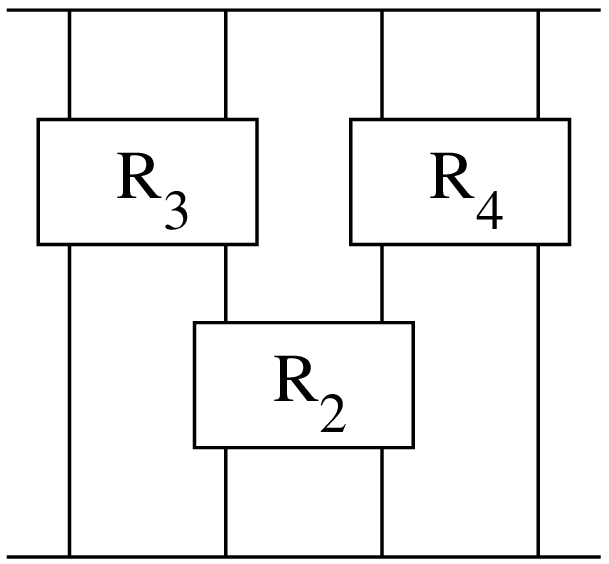}}

The sum rules apply also to correlators with general
position dependence.  As an example we consider the
four-point correlator evaluated in the last section.
Multiplying by $d_{R_1}$ say and summing over all
representations $R_1$ (as above one could instead
perform these same operations on any of the other
representations $R_2$, $R_3$, or $R_4$ and obtain
similar results) replaces the projection operator
$P_{R_1}$ in \fourptgpproj\ by the identity.  The
resulting trace that appears is represented diagrammatically
in \fourptsumdiag. 

\ifig\fttwosumfig{Projector diagram defining the
operator $F_{T_2}$ in terms of the projection operators
$P_{R_2}$ and $P_{R_4}$.} 
{\epsfxsize3.0in\epsfbox{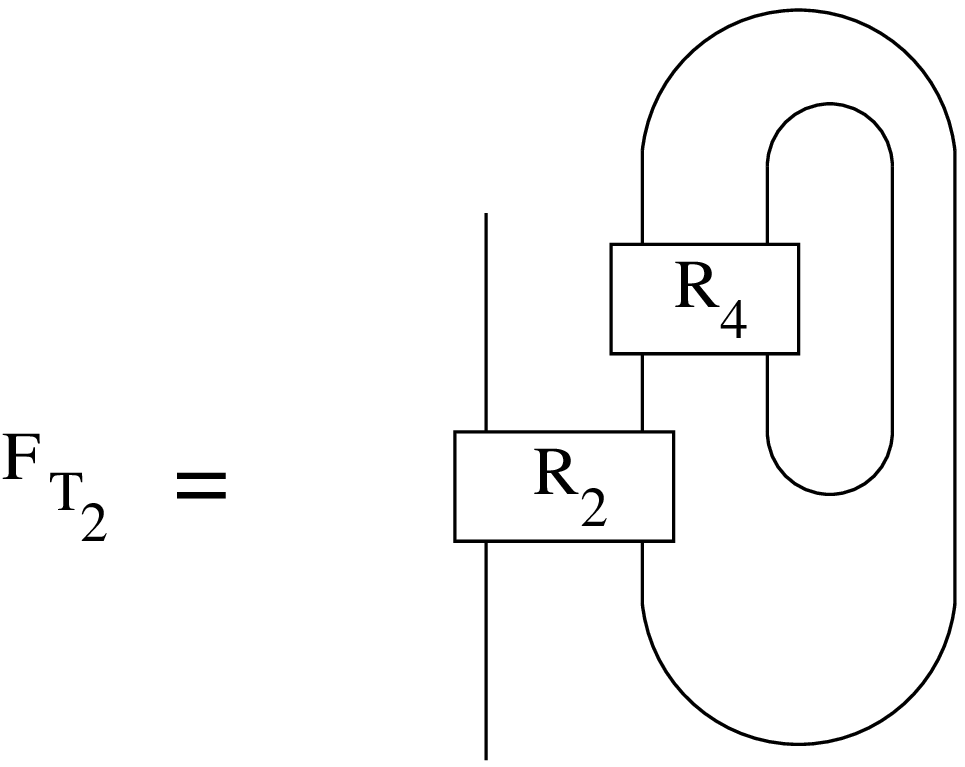}}

This diagram is also of the general
class analyzed in appendix 3 that can be factorized.
In this case one cuts the diagram in two places:  (1)
along the strand connecting $R_2$ to $R_3$ and (2) along
the strand connecting $R_2$ to $R_4$.  Considering
case (1) first, the operator $F_{T_1}$ defined in 
appendix 3 will be exactly as in figure \ftonesumfig,
or in terms of an equation \ftonesum, with $R_1$
replaced by $R_3$.  The other operator, $F_{T_2}$, defined
in appendix 3 will in this case be given by
\eqn\fttwofourpt{F_{T_2} = (1_{n_3 - j} \otimes tr_{n_4 - n_1 +j}
\otimes tr_{n_1 -j})(P_{R_2} \otimes 1_{n_1 -j})(1_{n_3 - j}
\otimes P_{R_4})
}
and is represented diagrammatically in \fttwosumfig.

Applying the factorization equation (15.5) from appendix 3 along
the strand described above in (1) produces
\eqn\twocutinter{\eqalign{tr((1_{n_1} \otimes P_{R_2}) 
(P_{R_3} \otimes_{j} P_{R_4})) & = \sum_{T_1}
{1 \over d_{T_1} ~ Dim T_1}
tr(P_{R_3} (1_{j} \otimes P_{T_1})) \cr
& \times
tr((P_{T_1} \otimes 1_{n_4})(P_{R_2} \otimes 1_{n_1 - j})
(1_{n_3 - j} \otimes P_{R_4})).
}}
Another application of the factorization equation along
the strand described in (2) above, or equivalently
along the strand connecting $R_2$ and $R_4$ in the
second trace appearing in the right-hand-side of \twocutinter,
yields
\eqn\twocut{\eqalign{tr((1_{n_1} \otimes P_{R_2}) 
(P_{R_3} \otimes_{j} P_{R_4})) & = \sum_{T_1,T_2}
{1 \over d_{T_1} ~ Dim T_1} {1 \over d_{T_2} ~ Dim T_2}
tr(P_{R_3} (1_{j} \otimes P_{T_1})) \cr
& \times
tr(P_{R_4} (1_{n_1 - j} \otimes P_{T_2}))
tr(P_{R_2}(P_{T_1} \otimes P_{T_2})).
}}
All traces appearing on the right-hand-side of \twocut\
can now be evaluated using \basethree, we find
\eqn\overlaptrace{\eqalign{tr((1_{n_1} \otimes P_{R_2}) 
& (P_{R_3} \otimes_{j} P_{R_4})) = Dim R_2 ~ Dim R_3 ~ Dim R_4
\sum_{T_1,T_2} \sum_{U_1,U_2} d_{U_1} d_{U_2} \cr
& \times {d_{T_1} d_{T_2}
\over Dim T_1 ~ Dim T_2} g(T_1,T_2;R_2)  g(U_1,T_1;R_3)
g(U_2,T_2;R_4).
}}
Substituting this into \fourptgpproj\ gives the final
form of the sum rule evaluated in terms of fusion
coefficients and dimensions of various representations.

For higher point correlators of the form
\eqn\highersum{\langle \chi_{R_1} 
(\Phi (x_1)) \cdots \chi_{R_n} (\Phi (x_n))
\chi_{S_1} (\Phi^{\dagger} (y_1)) \cdots 
\chi_{S_m} (\Phi^{\dagger} (y_m))  \rangle
}
where $n,m > 2$,
it is no longer true that multiplying by a single
dimension $d_{R_i}$ say and summing over representations
$R_i$ yields a correlator expressible just in terms
of fusion coefficients and dimensions of representations.
In such a case the trace of projection operators does not
simplify to a form that is factorizable.  To get such
traces one needs to multiply by $(n-1)$ of the $d_{R_i}$'s
and sum or by $(m-1)$ of the $d_{S_j}$'s and sum to
get a factorizable diagram that can be evaluated just
in terms of fusion coefficients.

\newsec{Comparison  to 3D Chern-Simons theory, 2D $G/G$ theory and 2D YM
theory } 

Highly supersymmetric correlators in super Yang-Mills
theories often capture a topological phase of the theory, 
as for example in \witsyfr. Often the connection 
to topology proceeds via instantons. However, the extremal 
correlators do not receive instanton corrections. So the standard
route to establishing deductively a connection between 
them and topological gauge theories is not available.  
Nevertheless  we 
remarked in previous sections that factorization equations, 
fusion relations and sum rules can be written down for the extremal 
correlators which have analogs in topological gauge theories. 
The group theoretic character of the 
correlators allows equalities between ratios of correlators 
and appropriate observables in topological gauge theories. 
By  exploiting known exact answers for topological theories
in lower dimensions, in particular  Chern-Simons theory in the $
k\rightarrow \infty $ limit 
and  two-dimensional 
Yang-Mills theory in the zero area limit, we describe these relations. 
We will be using results from the literature on 
three-dimensional Chern-Simons theory  and knot invariants
\refs{ \witjones, \witint, \mooreseib, \alvgaum, \schweig, 
\pasqsal}, as well as that on two-dimensional Yang-Mills 
\refs{ \mig, \rus ,  \witymii, \grotayii, \cmrrev  }.

For the extremal correlator we have 
\eqn\coneq{\eqalign{ 
& { \langle ~~  \chi_{R_1} \bigl( \Phi(x) \bigr)   ~ 
\chi_{R_2} \bigl( \Phi ( x ) \bigr)   ~ \chi_{S}
\bigl( \Phi^{\dagger}  ( 0 ) \bigr) ~~\rangle_{SYM4}  \over 
  { \langle ~~ \chi_{S} \bigl( \Phi(x) \bigr) 
 \chi_{S} \bigl( \Phi^{\dagger} ( 0 ) \bigr)
 ~~ \rangle_{SYM4}} }\cr 
& = ~ g(R_1, R_2 ; S ) \cr 
 & = ~ Z_{CS} \bigl( S^2 \times S^1 ; R_1, R_2, \bar S \bigr) \cr 
 & = ~ \langle ~~ \chi_{R_1} ( g )  \chi_{R_2} ( g ) 
\chi_{ \bar S} ( g ) ~~  \rangle_{G/G} \cr 
&   =  ~ Z_{2DYM} ~ ( R_1;  W_{R_2} ; \bar{S} )   \cr 
}}

\ifig\twdymthpt{ Diagram illustrating observable in 2DYM.  }
{\epsfxsize2.5in\epsfbox{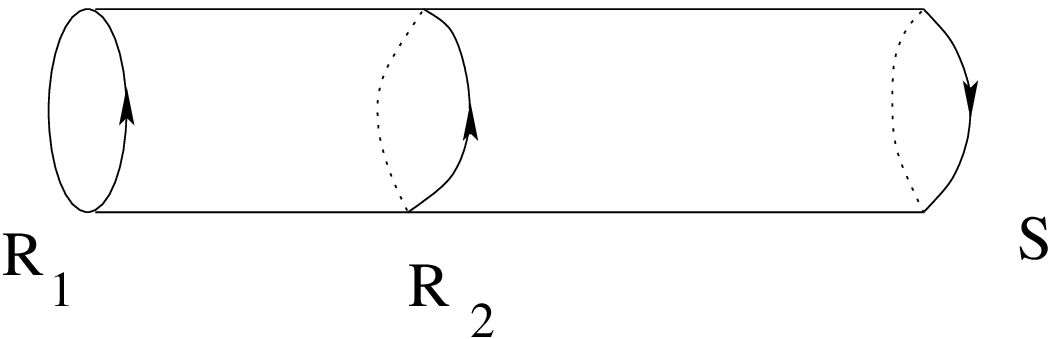} }

The first line is a special extremal 
 correlator in four-dimensional  $U(N)$ 
super-Yang-Mills theory, where the insertion points of
$R_1$ and $R_2$ are taken to coincide.
The second line is the Littlewood-Richardson coefficient, 
i.e the number of times the tensor product $ R_1 \otimes R_2$ 
contains $S$.   
 The third  line 
is a correlator of $ U(N)$ Chern-Simons theory on $S^2 \times S^1$
 with two Wilson lines carrying representations $R_1$ and $R_2$ 
winding in one direction along the $S^1$ and a third Wilson line 
in the $S$ representation winding in the opposite direction along the
 circle. By the standard relation between Chern-Simons 
theory in three dimensions with gauge group $G$
and $G/G$ topological field theory, this correlator is 
equal to a correlator for the $G/G$ model with insertions 
at three points on $S^2$  which can be viewed as the intersection of 
the three Wilson lines  in Chern-Simons
with an  $S^2$ at a fixed point of $S^1$ on 
$S^2 \times S^1$. The fifth line gives  a correspondence between the 
ratio of $4D$ correlators and  topological  two-dimensional Yang-Mills
theory with $U(N)$ gauge group on a cylinder. 
Topological two-dimensional Yang-Mills is the zero area limit 
of standard two-dimensional Yang-Mills. 
 Evaluating the 
path integral for two-dimensional Yang-Mills on the cylinder 
requires the specification of a boundary holonomy, or more generally 
the holonomies can be weighted by the character in a representation 
$R$. The relevant observable has boundary holonomies specified 
by $R_1$ on the left and $\bar S$ on the right, and there 
is a Wilson loop insertion in the irrep $R_2$ parallel to $R_1$. 
The diagram in \twdymthpt\ illustrates the relevant observable 
in two-dimensional Yang-Mills. 

This generalizes to
\eqn\genconeq{\eqalign{ 
&  \langle ~~ \prod_{i=1}^{k}  \chi_{R_i}\bigl( \Phi(x) \bigr)   
  \chi_{ S } \bigl( \Phi^{\dagger}  ( 0 ) \bigr) ~~\rangle_{SYM4}  \over 
 \langle~~ \chi_{S} \bigl( \Phi(0) \bigr)
  \chi_{S} \bigl( \Phi^{\dagger}  ( x) \bigr) ~ \rangle \cr 
& = Z_{CS} ( S^2 \times S^1 : R_1, R_2, \cdots R_n ; \bar S ) \cr  
& = \langle \prod_{i=1}^{k} 
\chi_{R_i} (g) \chi_{\bar S} ( g )  \rangle_{G/G }  \cr 
& =  Z_{2DYM} ( R_1;  W_{R_2}, W_{R_3} \cdots W_{R_k} , W_{\bar S_1}, 
\cdots W_{\bar S_{l-1}}  ; \bar S_l  ) \cr 
}}

It is also worth noting that for two point functions we have 
\eqn\twocor{\eqalign{  
& \langle ~~\chi_{R}( \Phi(x)  ) \chi_{ S }( \Phi^{\dagger} )   ( 0 )
~~ \rangle_{SYM4}   = x^{-2n } \delta_{RS} { n ! Dim R \over d_R  } \cr 
& = x^{-2n}   { n!\over d_R }  Z_{CS } (S^2 \times S^1  :  R, \bar S ) 
{ Z_{CS } (S^3 :  R ) \over Z_{CS } ( S^3 : \emptyset ) }  \cr 
&  = x^{-2n}  Z_{ 2DYM } ( S^1 \times I : R, \bar S ) { n ! Dim R \over d_R} \cr }} 
The relations in \twocor\ are useful in deriving the previous equations 
but, with the $x$ dependences still present,  
 do not convincingly exhibit the topological character 
of the correlators the way the earlier equations do.

\subsec{ Converting graphs to Wilson loops in $S^3$ } 

Equation \coneq\ uses only a special class of extremal 
correlators, those where all the holomorphic operators 
are at the same point and all the anti-holomorphic operators 
are also at a fixed point. The correlator is still extremal 
and non-renormalized if the anti-holomorphic operators 
are fixed at one point and holomorphic operators are at different
points. It is also possible to find ratios  involving such
correlators which map to correlators in topological gauge theories. 
\eqn\thrpt{\eqalign{  
& \langle ~~ 
\chi_{R_1} \bigl( \Phi(x_1) \bigr)   
\chi_{R_2} \bigl( \Phi ( x_2 ) \bigr)   \chi_{S}
\bigl( \Phi^{\dagger}  ( 0 )) ~~ \rangle_{SYM4 }  \over 
   \langle~ \chi_{R_1}( \Phi(0) )  \chi_{R_1} ( \Phi^{\dagger} ( x_1 )) 
   ~ \rangle_{SYM4}  ~~
  \langle~ \chi_{R_2}( \Phi(0) )  \chi_{R_2} ( \Phi^{\dagger} ( x_2 )) 
   ~ \rangle_{SYM4}       \cr 
& = { ( n_1+n_2 )  ! \over n_1! n_2! } ~ { d_{R_1} d_{R_2} \over d_S }~  
     g( R_1, R_2; S ) ~ { Dim S \over Dim ~ R_1 Dim ~ R_2} \cr   
& = { Z_{CS} ( S^3 :   W ( R_1, R_2, S) ) \over Z_{CS} ( S^3 :  R_1,
R_2 )  }   ~  {d_{R_1} d_{R_2} \over d_{S} } ~ {  n! \over n_1! n_2!  }
\cr  }}
\ifig\gphthrpt{ Diagram illustrating observable in Chern-Simons.  }
{\epsfxsize2.5in\epsfbox{ 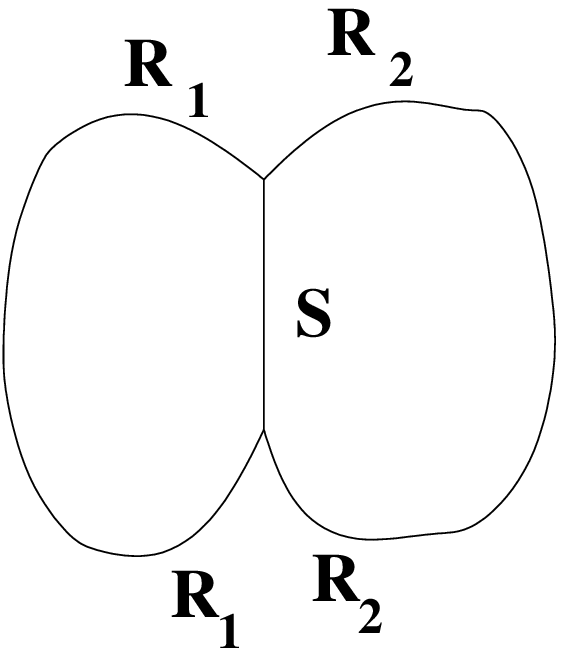} }
 In the last line of \thrpt\  we have taken advantage 
 of the fact that group theoretic graphs 
 with legs labelled by irreps and vertices by Clebsch Gordans 
  are large $k$ limits of the  normalized expectation 
values of the corresponding intersecting Wilson loops
  in Chern-Simons theory. $ W ( R_1, R_2, S )$ denotes 
 the Wilson loop shown in \gphthrpt.

\eqn\chsims{\eqalign{  
&{  Z_{CS} ( S^3 :   W ( R_1, R_2, S ) ) \over Z_{CS} ( S^3 :
\emptyset ) }
 = g(R_1, R_2 ; S ) Dim S \cr 
&{  Z_{CS} ( S^3 :   R_1, R_2 )  \over Z_{CS} ( S^3 : \emptyset) }
= Dim R_1 Dim R_2 \cr }}
As explained in \witint\ the vertices can be labelled 
by a choice of invariant tensor in $ R_1 \otimes R_2 \otimes \bar S
$. 
Here we are summing over all the choices with equal weight 
 to get the  fusion multiplicity $g(R_1, R_2 ; S )$ and $Dim  S$
appears because of the trace. The second equation just gives 
the expectation value for two disconnected Wilson loops.

\ifig\klobscs{ The $ k \rightarrow l $ observable in Chern-Simons.  }
{\epsfxsize2.5in\epsfbox{ 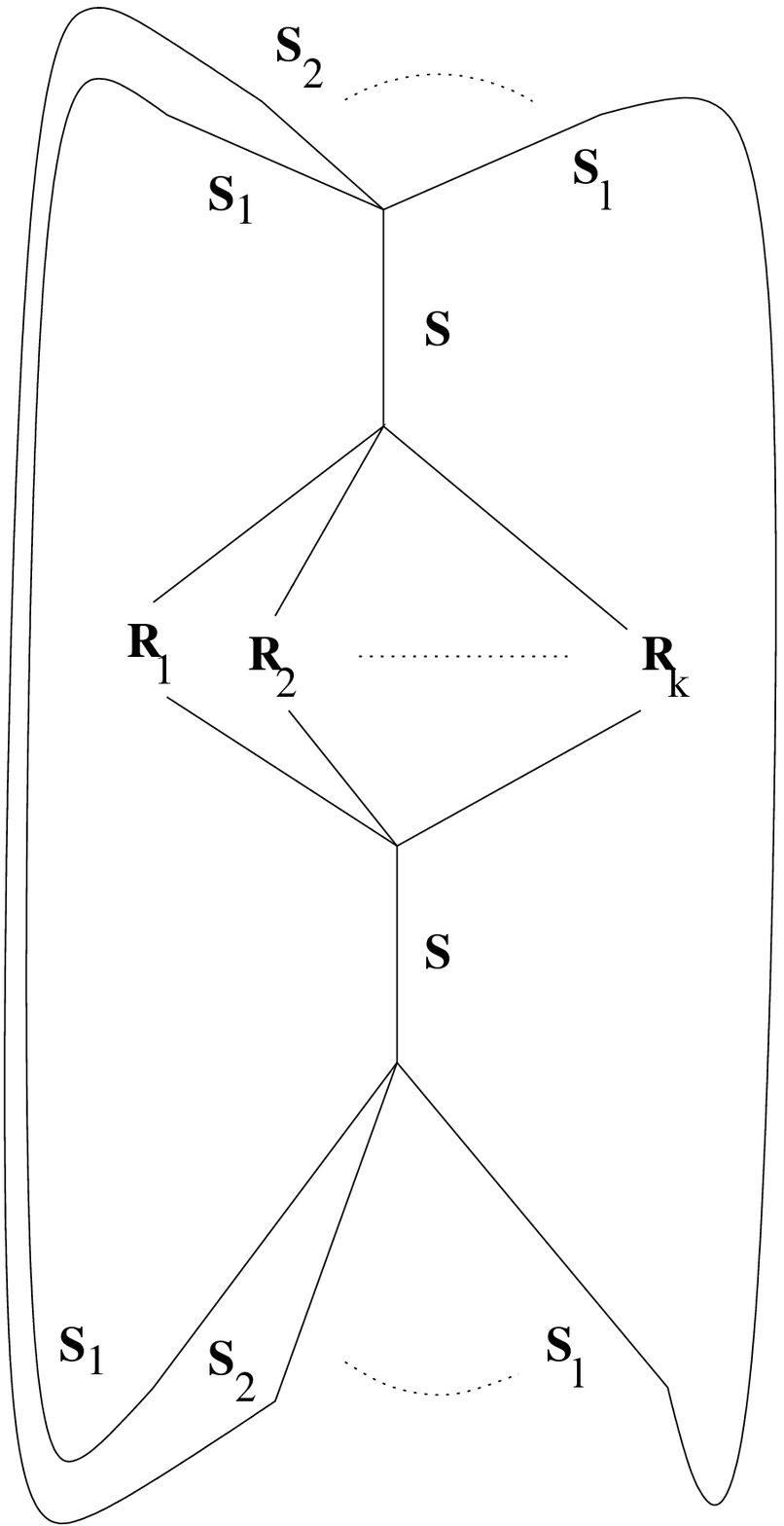} }

  For the more general case of the $k \rightarrow l $ 
 multipoint function \nrhtmx\ we can 
\eqn\multicorcs{\eqalign{  
&  {  \langle ~~ \chi_{ R_1 } \bigl( \Phi  (x_1) \bigr)
 \chi_{R_2}  \bigl( \Phi (x_2) \bigr) \cdots 
  \chi_{R_k}  \bigl( \Phi (x_k ) \bigr)   ~  
\chi_{S_1} \bigl( \Phi^{\dagger}   (0) \bigr) \cdots
\chi_{S_l} \bigl( \Phi^{\dagger } (0) \bigr)   ~~\rangle  
\over \langle ~ \chi_{R_1} \bigl( \Phi(x_1) \bigr) 
\chi_{R_1} \bigl( \Phi^{\dagger} ( 0 ) \bigr) ~\rangle \cdots 
\langle ~ \chi_{R_k} \bigl( \Phi(x_k) \bigr) 
\chi_{R_k} \bigl( \Phi^{\dagger} ( 0 ) \bigr) ~\rangle }  \cr 
& = {1 \over n (R_1) ! n (R_2) ! \cdots n(R_k)! } 
 { 1 \over Dim R_1 Dim R_2 \cdots Dim R_k } 
{ 1 \over d_{S_1} \cdots d_{S_l} } \cr 
& \times \sum_{S} \sum_{\gamma} 
tr \Bigl( ( P_{S_1} \otimes P_{S_2} \cdots P_{S_l} )  P_S 
\gamma^{-1} ( P_{R_1} \otimes
 P_{R_2}  \cdots P_{R_k} ) \gamma P_S \Bigr) \cr 
& = {n! \over n (R_1) ! n (R_2) ! \cdots n(R_k)! } 
{ 1 \over Dim R_1 Dim R_2 \cdots Dim R_k } \cr
& \times \sum_S tr_{S_1 \otimes \cdots \otimes S_l} ( P_S ( P_{R_1}
\otimes \cdots \otimes P_{R_k}) P_S) \cr
& = { n! \over n(R_1)! n(R_2)! \cdots n(R_k)! }   
\sum_S { Z_{CS}  ( S^3 ; W( R_1, R_2, \cdots R_k; S ; S_1, S_2 \cdots S_l ) ) 
 \over Z_{CS } ( S^3 ; R_i ) } \cr }}  
where $Z_{CS} ( S^3 ; R_i ) $ is the expectation value of 
 a Wilson loop in the $R_i$ representation in 
$k=\infty $ three dimensional $U(N)$ Chern-Simons theory, 
and   $Z_{CS}  ( S^3 ; W( R_1, R_2, \cdots R_k; S ; S_1, S_2 \cdots S_l
 )$ is the expectation value of the Wilson loop 
shown in \klobscs.

 The relation we used in \chsims\ for the relation between 
 Chern-Simons expectation values in $S^3$  and knotted group theoretic graphs 
 is quite general, and can also be applied to  
 quantities  like the one  in \unprojdone\  which appears 
 in non-extremal correlators. The extremal correlators 
 only involve the simple graphs in \gphthrpt\ and  \klobscs\ which
when evaluated group theoretically gives fusion coefficients and 
dimensions. The fusion coefficient is also the expectation value in 
 non-intersecting  Wilson loops in $S^2 \times S^1$ as we saw in
 \coneq. This is also true for the 
 extremal $SU(N)$ correlators as we will see later, which involves 
products of the simple graph. 
 The non-extremal correlators are of course renormalized 
 at finite coupling $g_{YM}^2$, so it appears that 
 correlators which are associated with complicated graphs 
 have non-trivial coupling dependence, while those 
 associated with simple graphs have trivial coupling dependence. 
 We will return to this point in section 9.

\subsec{ Sum Rules, 
 Relations between $S^2 \times S^1 $ and $S^3$ and Verlinde formula.  }

 In sections 4 and 6 we have written down sum rules 
 for both extremal and non-extremal correlators. 
 Here we will consider the extremal case and 
 show how they are equivalent to identities 
 we expect from the mapping to Chern-Simons theory. 
  The basic group theoretic identity \fusid\ leads to \smrlxti. 
  Using the map to Chern-Simons correlators 
 given in \coneq\ and \chsims\ this translates 
into 
\eqn\basgpcs{ \sum_{ S} Dim S ~ Z_{CS} ( S^2 \times S^1 ; R_1, R_2, \bar
 S ) 
 = { Z_{CS} ( S^3 : R_1, R_2 )  \over Z_{CS} ( S^3 : \emptyset ) }  }
 This is indeed the large $k$ limit of 
 the relation between expectation values in 
 $S^3$ and in $S^2 \times S^1$ which holds for any observable $ \cO$
 inserted on $S^3$ \witjones
\eqn\reltm{
 Z_{CS} ( S^3 : \cO ) =  \sum_{ S } S_{0 S } ~~Z_{CS} ( S^2 \times S^1 :
\cO , S ) } 
Here $S_{0S} $ is the matrix element of the modular $S$ matrix
 between the identity rep and $S$, and it obeys 
\eqn\sprop{\eqalign{  
&  S_{00} = Z_{CS} ( S^3 : \emptyset ) \cr 
& { S_{0 R } \over S_{00} } = Dim R \cr }  } 
At finite $k$ the right hand side of the second line 
is the quantum dimension, but at large  $k$ it is the ordinary 
dimension. Using \sprop\ it is clear that  
\basgpcs\ is  the same as \reltm.

It is also of interest to compare the 
identity \fusid\ underlying a class of  sum rules \smrlxti\ 
with the large $k$ limit of the Verlinde formula \verlinde. 
This is expressed in terms of $S_{R_1 R_2} $ 
which in the large $k$ limit is proportional to $ Dim ( R_1) Dim (R_2 )$. 
It can be expressed \witjones\ as : 
\eqn\vform{ 
{ S_{R_1 R_2} S_{R_1 R_3} \over S_{0 R_1} } = \sum_{ S} S_{R_1 S}
~~ g( R_2, R_3 ; S )  } 
Substituting the appropriate product of dimensions for each 
matrix element of the modular $S$-matrix, \vform\ implies
\eqn\vformimp{ 
Dim R_1  Dim R_2 Dim R_3  = \sum_{S } Dim ~R_1 Dim ~S
~~ g( R_2 , R_3 ; S  ) } 
which reduces to the \fusid. 

In section 3 we saw that there are two kinds of 
sum rules involving the fusion coefficient, 
as a result two kinds of sum rules for extremal three point 
functions. This arises from the fact that $g(R_1, R_2 ; R_3) $
is both a fusion number for unitary groups and a 
branching number for symmetric groups. The fusion number 
interpretation leads to \fusid\ which we related to Chern 
Simons in different ways.  The symmetric group 
interpretation leads to \bchid. 
It would be interesting to find the 
Chern-Simons interpretation of this latter sum rule. 
In section 6, it was observed  that the sum rules can also 
be applied to non-extremal correlators. 
In a sense then, these more general sum rules 
are generalizations of the large $k$ limit 
of the Verlinde formula. Given the rich geometry 
of the Verlinde formula in terms of dimensions 
of spaces of sections of holomorphic bundles over 
moduli spaces of complex structures in the context of two-dimensional 
rational CFT or equivalently in topological $G/G$ models,
the above remarks may be viewed as a hint that the correlators 
of four-dimensional  $N=4$ super-Yang-Mills theory might 
have an analogous geometrical meaning in four dimensions, although the precise 
geometric objects  relevant here are unknown. 
Higher dimensional versions of
the Verlinde formula  have  been discussed in \lomonesh\ and 
it would be interesting to explore any possible connections 
of \lomonesh\ with $N=4$ SYM.

\newsec{ Staggered Factorization  }

\ifig\contpat{ {Picture of contraction pattern  } }
{\epsfxsize3.0in\epsfbox{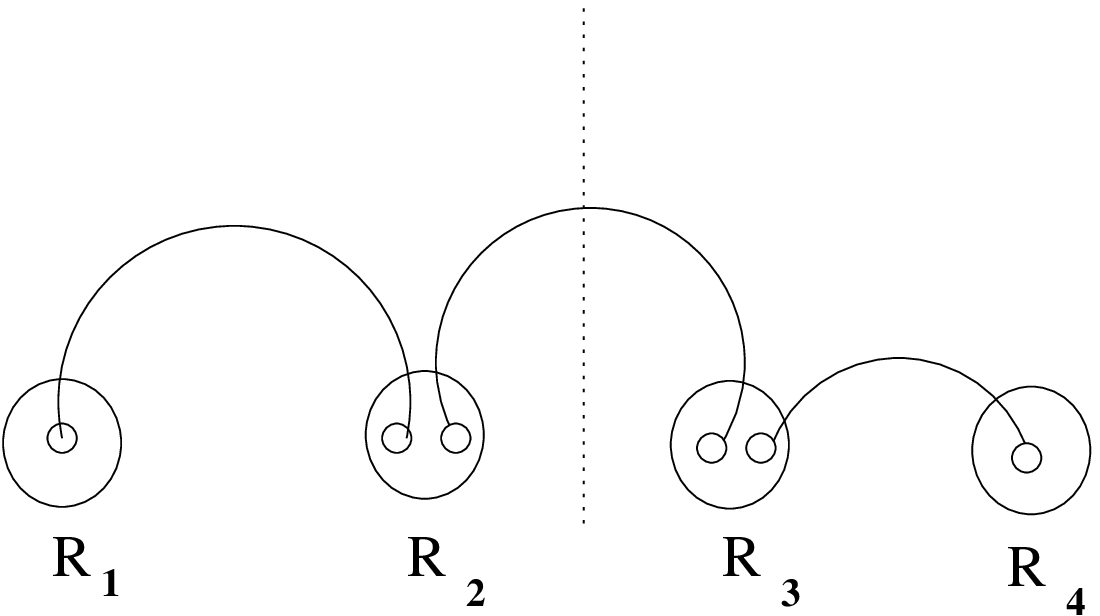} }

\ifig\contpatfac{ { Factorized contraction pattern  } }
{\epsfxsize4.0in\epsfbox{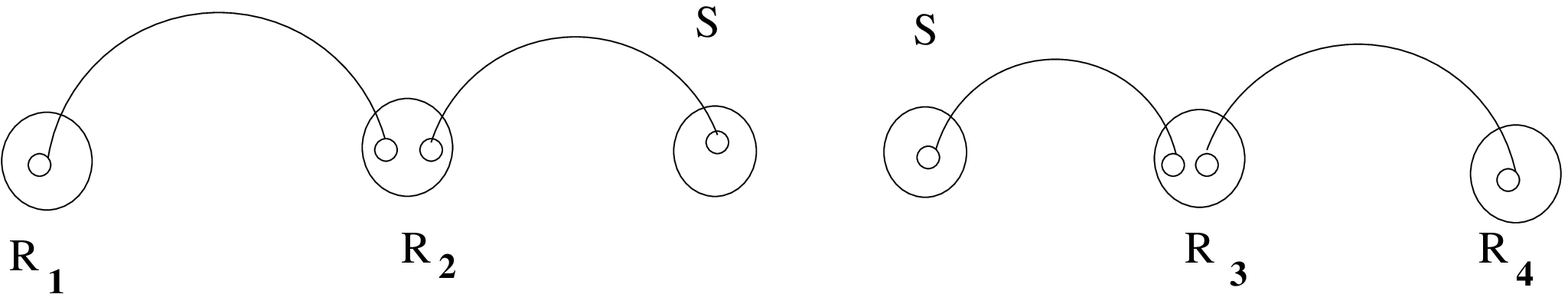} }

We now describe a class of factorization equations 
obeyed by  correlators which are obtained 
by a   simple patterns of contractions. 
To describe the contraction patterns, 
for each operator insertion we draw a 
circle. We draw a single line connecting 
two circles if there is a set of $\Phi$'s 
being contracted between them. If the contraction diagram 
can be disconnected by cutting a line, we 
can write a factorized expression. These contraction 
diagrams thus have to be ``one-particle reducible''
to be factorizable . We have used ordinary Feynman 
diagram language here, although each line 
describes a set of propagators rather than  a single propagator. 
A simple factorizable pattern of contractions is shown in \contpat. 
After factorization we get a sum over products of 
correlators of the form in \contpatfac. 
It is in fact simplest to describe the class of relevant 
correlators by the kind of projector diagrams 
that calculates their expectation value. This class of
diagrams is described in appendix 3, and given the form 
of the relevant projector diagrams we will call this 
factorization property ``staggered factorization'' 
which is a generalization of the basic factorization 
discussed in sections 2 and 3.

Consider  a  correlator  coresponding to \contpat\ 
\eqn\corfrm{ 
 \langle ~~ \chi_{R_1} \bigl( \Phi_3 (x_1) \bigr)   
  Q_{13}^{n_3} E_{21}^{n_2} \chi_{R_2} \bigl( \Phi_1 (x_2)  \bigr)   
    Q_{12}^{n_2} \chi_{R_3} \bigl( \Phi_1 (x_3)  \bigr) 
\chi_{R_4} \bigl( \Phi_1^{\dagger}  (x_4) \bigr) ~~\rangle  } 
Here $n(R_1) =n_3, n(R_2) = n_2 + n_3 , n(R_3) = n_1 + n_2 , and 
n(R_4) = n_1 $.  
   We would like to describe the  $SO(6) $ 
transformation properties of the operators involved. 
  One way to specify the content is to list 
  the  numbers $(n_1, n_2, n_3 ) $ which count the number 
  of $ \Phi_1, \Phi_2$, and $\Phi_3$ fields respectively. 
  We count $ \Phi_i^{\dagger } $ as contributing
  negatively to these three charges. 
Here, the charges are 
\eqn\chrgs{\eqalign{  
& \vec v_1 = ( 0,0, n_3 )  \cr 
& \vec v_2 = ( 0,n_2, -n_3 ) \cr 
& \vec v_3 = ( n_1, -n_2, 0 ) \cr 
& \vec v_4 = ( -n_1, 0,0 ) \cr }}

\ifig\stagfactone{Diagram corresponding to the trace
appearing in correlator }
{\epsfxsize1.6in\epsfbox{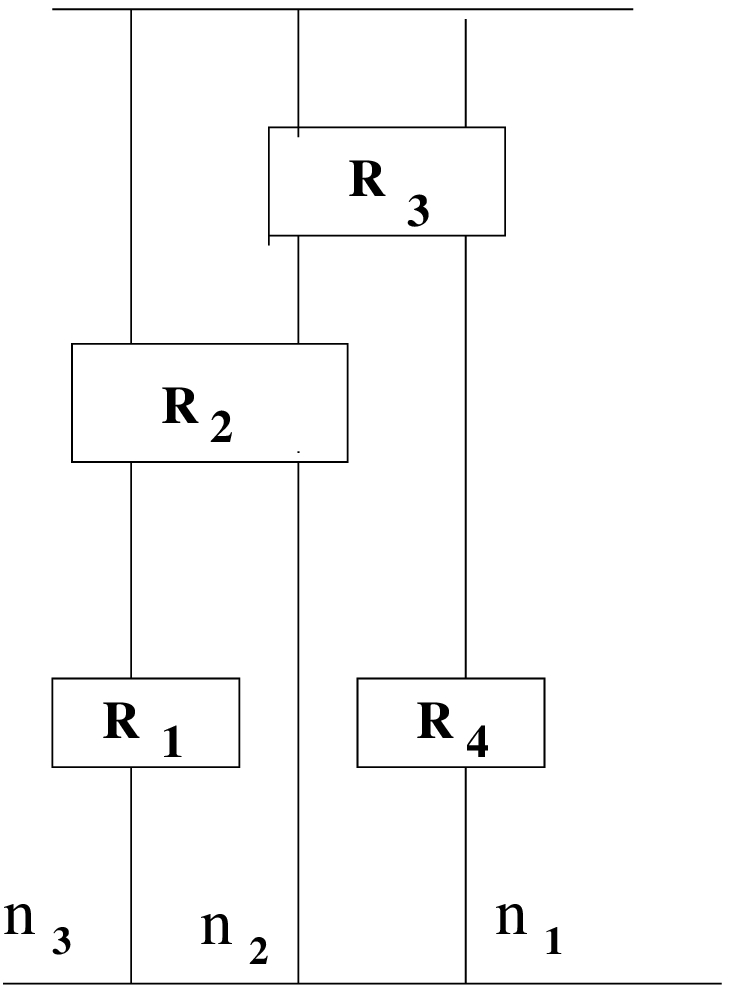} }

The correlator in \corfrm\ is
\eqn\corfrmi{\eqalign{& {n_2 !  n_3 ! (n_1+n_2)! (n_2 + n_3)!
\over d_{R_1} d_{R_2} d_{R_3} d_{R_4}} tr \left( (P_{R_1}
\otimes 1_{n_{R_3}}) (P_{R_2} \otimes 1_{n_{R_4}}) 
(1_{n_{R_1}} \otimes P_{R_3})(1_{n_{R_2}} \otimes P_{R_4}) \right) \cr
& ~~~ = {n_2 !  n_3 ! (n_1+n_2)! (n_2 + n_3)!
\over d_{R_1} d_{R_2} d_{R_3} d_{R_4}}  tr \left( (P_{R_1}
\otimes P_{R_3}) (P_{R_2} \otimes P_{R_4}) \right)
}}
where the trace is represented diagrammatically in 
\stagfactone.
Using the group integral version of projectors, we can 
express the trace as : 
\eqn\trgi{\eqalign{  
&  tr \left( (P_{R_1}\otimes P_{R_3}) (P_{R_2} \otimes P_{R_4})
\right) 
= Dim R_1 ~ Dim R_2 ~ Dim R_3 ~ Dim R_4 \cr  
& \int dU_1 dU_2 dU_3 dU_4 ~~ \chi_{R_1} (U_1^{\dagger} )
  \chi_{R_2} (U_2^{\dagger} )~~
 \chi_{R_3} (U_3^{\dagger} ) ~~ \chi_{R_4} (U_4^{\dagger} ) \cr 
&  tr_{n_3}( U_1 U_2 ) ~tr_{n_2} ( U_2 U_3 ) ~tr_{n_1} (U_4 U_3 ) \cr
& = \sum_{S} d_{R_1}  ~ Dim R_2 ~ d_S ~  Dim R_3 ~ d_{R_4} \cr 
&  \int dU_1 dU_2 
   \chi_{R_2} (U_1^{\dagger} )~~ \chi_{R_3} (U_2^{\dagger} )
\chi_{R_1} ( U_1 ) \chi_S ( U_1 U_2) \chi_{R_4} ( U_2 ) \cr }}

\ifig\facstfac{ Factorized product of traces  }
{\epsfxsize5.0in\epsfbox{ 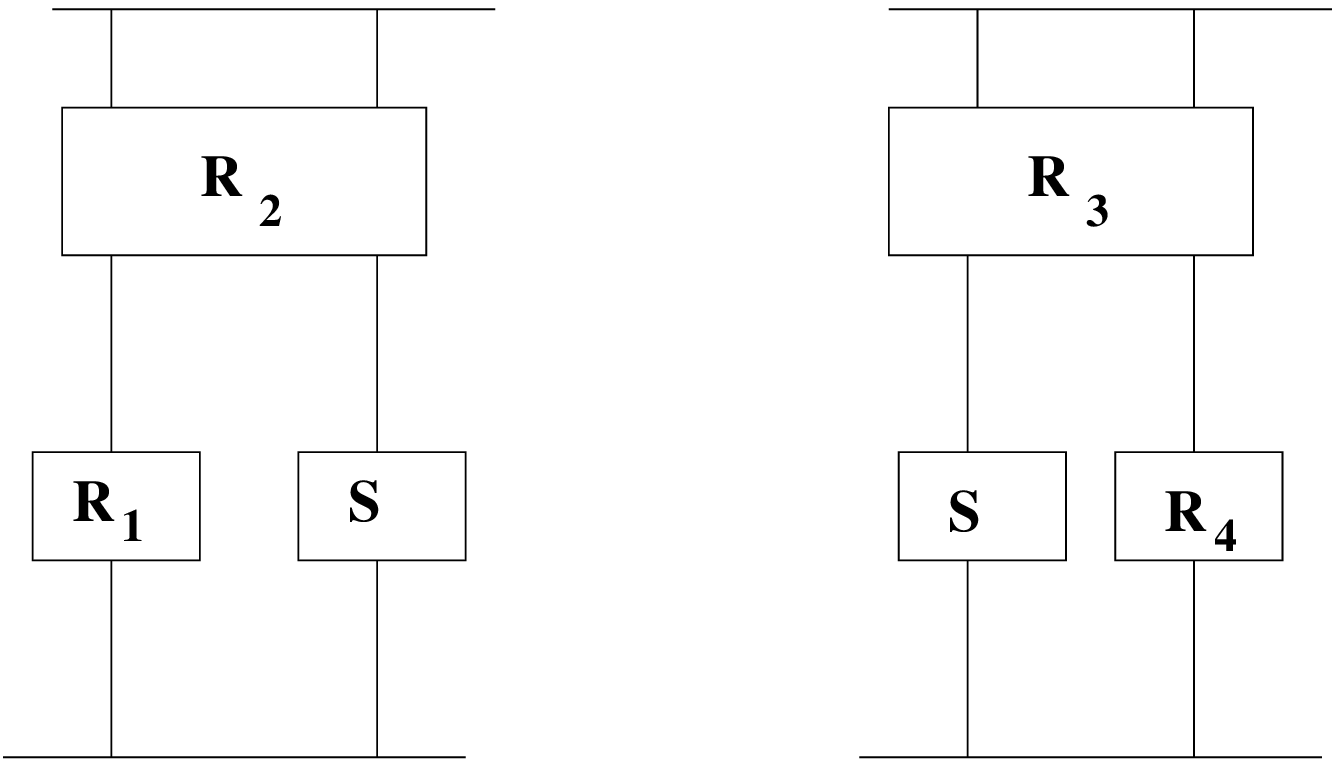 } }

The diagram in  \stagfactone\
is of the general form considered in appendix 3 that can be
factorized.  Specifically we can cut the diagram along the line
connecting $R_2$ and $R_3$ by inserting a complete set of 
irreducible projectors.  In detail, the operators $F_{T_1}$ and $F_{T_2} $
introduced in appendix 3 are  given by
\eqn\ftone{\eqalign{ 
& F_{T_1} = ( tr_{n_3}  \otimes {\bf 1 }_{n_2} ) ( P_{R_1}  \otimes
{\bf 1}_{n_2}  )
P_{R_2} \cr 
& F_{T_2} = ( {\bf 1}_{n_2} \otimes tr_{n_1}  ) ( {\bf 1}_{n_2}   \otimes
P_{R_4}) P_{R_3}   \cr }}
Using the factorization property 
\eqn\factprop{ tr_{n_2} (P_S F_{T_1} F_{T_2} ) = { 1 \over d_S Dim S }
tr_{n_2}  ( P_S F_{T_1}) tr_{n_2} ( P_S F_{T_2}) } 
proved in appendix 3
yields the relation
\eqn\stagfacteqn{\eqalign{ & tr \left( (P_{R_1}
\otimes P_{R_3}) (P_{R_2} \otimes P_{R_4}) \right) \cr
& ~~~~ = \sum_S  {1 \over d_S ~ dim S} tr \left( (P_{R_1} \otimes P_S)
P_{R_2} \right) tr \left( P_{R_3} (P_S \otimes P_{R_4}) \right) \cr 
& ~~~~ = \sum_S { 1 \over d_S Dim S } 
d_S d_{R_1} d_S d_{R_4} g(R_1, S ; R_2 ) g(S, R_4 ; R_3 )
~ dim R_2 ~ dim R_3 \cr 
}}
The factorized product of traces is shown 
diagrammatically in \facstfac.

These steps lead to the following result 
\eqn\anscorfrm{\eqalign{  
& \langle ~~ \chi_{R_1} \bigl( \Phi_3 (x_1) \bigr)   
  Q_{13}^{n_3} E_{21}^{n_2} \chi_{R_2} \bigl( \Phi_1 (x_2)  \bigr)   
    Q_{12}^{n_2} \chi_{R_3} \bigl( \Phi_1 (x_3)  \bigr) 
\chi_{R_4} \bigl( \Phi_1^{\dagger}  (x_4) \bigr) ~~\rangle \cr 
& = n_2! n_3! (n_2 +n_3) ! (n_1 +n_2 ) ! {1 \over (x_1 - x_2)^{2 n_2}}
{1 \over (x_2 - x_3)^{2 n_3}} {1 \over (x_3 - x_4)^{2 n_1}}
\cr
& \times \sum_S
{ Dim R_2 Dim R_3 \over Dim S } { d_S \over d_{R_2} d_{R_3} }
g(R_1, S; R_2 ) g( S,R_4 ; R_3 ) \cr }}
Since we now have fusion coefficients 
which are reminiscent of three-point functions, 
it is natural to convert \anscorfrm\ into a
 statement about correlators as indicated by \contpatfac. 
This is indeed possible, and when the RHS of \anscorfrm\ 
is expressed in terms of the appropriate three-point functions, 
all the dimensions and factorials cancel leaving a simple 
formula 
\eqn\anscorfrmi{\eqalign{ 
& \langle ~~ \chi_{R_1} \bigl( \Phi_3 (x_1) \bigr)   
  Q_{13}^{n_3} E_{21}^{n_2} \chi_{R_2} \bigl( \Phi_1 (x_2)  \bigr)   
    Q_{12}^{n_2} \chi_{R_3} \bigl( \Phi_1 (x_3)  \bigr) 
\chi_{R_4} \bigl( \Phi_1^{\dagger}  (x_4) \bigr) ~~\rangle \cr 
&= \sum_{S}  
\langle ~~ \chi_{R_1} \bigl( \Phi_3 (x_1) \bigr)   
  Q_{13}^{n_3} E_{21}^{n_2} \chi_{R_2} \bigl( \Phi_1 (x_2)  \bigr)
  \chi_{S}  \bigl( \Phi_2^{\dagger} (x_3) \bigr)  ~~\rangle \cr 
& \times \langle ~~ \chi_{S}  \bigl( \Phi_2  (x_2) \bigr) 
Q_{12}^{n_2} \chi_{R_3} \bigl( \Phi_1 (x_3)  \bigr) 
\chi_{R_4} \bigl( \Phi_1^{\dagger}  (x_4) \bigr) ~~\rangle 
{  1 \over    \langle \chi_{S}  \bigl( \Phi_2  (x_2) \bigr)
    \chi_{S}  \bigl( \Phi_2^{\dagger} (x_3) \bigr) \rangle }\cr }}

The key to this factorization equation is that the 
correlator is obtained from a simple pattern of contractions.
In the above, we fixed the pattern of contractions 
by choosing the $SO(6)$ quantum numbers of the 
operators. We can also fix the contraction patterns by working 
with just the highest weight or lowest weight states 
under the $SO(6)$ action, but picking out the desired contraction
patterns.  Consider 
as an example the
four-point correlator with general positions given by
\fourptgpproj.  Assume that $n_1 \leq n_3$ (and therefore
$n_4 \leq n_2$ since $n_1 + n_2 = n_3 + n_4$ for this
correlator to be non-zero) and extract 
the $j=n_1$ term from the sum over $j$ by a contour
integral, 
\eqn\intspc{ 
\int dy_1 dy_2 dy_3 ~~
y_1^{2 n_1-1 } y_2^{2(n_3 - n_1)-1} y_3^{2n_4-1} 
\langle \chi_{R_1} ( \Phi (x_1) ) \chi_{R_2} ( \Phi (x_2) ) 
\chi_{R_3} ( \Phi^{\dagger} (x_3) )  
\chi_{R_4} ( \Phi^{\dagger} (x_4) ) \rangle  } 
where $ y_1 = x_1 -x_3 , y_2 = x_2-x_3$, and $ y_3 = x_2 -x_4 $.
We find then the
exact same trace as appears on the left-hand-side
of \stagfacteqn\ (only with $R_2$ and $R_3$ exchanged).
Consequently we can factorize this contribution to 
the general position four-point correlator as above.


\ifig\bastagexi{ { A more general factorizable form  } }
{\epsfxsize3.0in\epsfbox{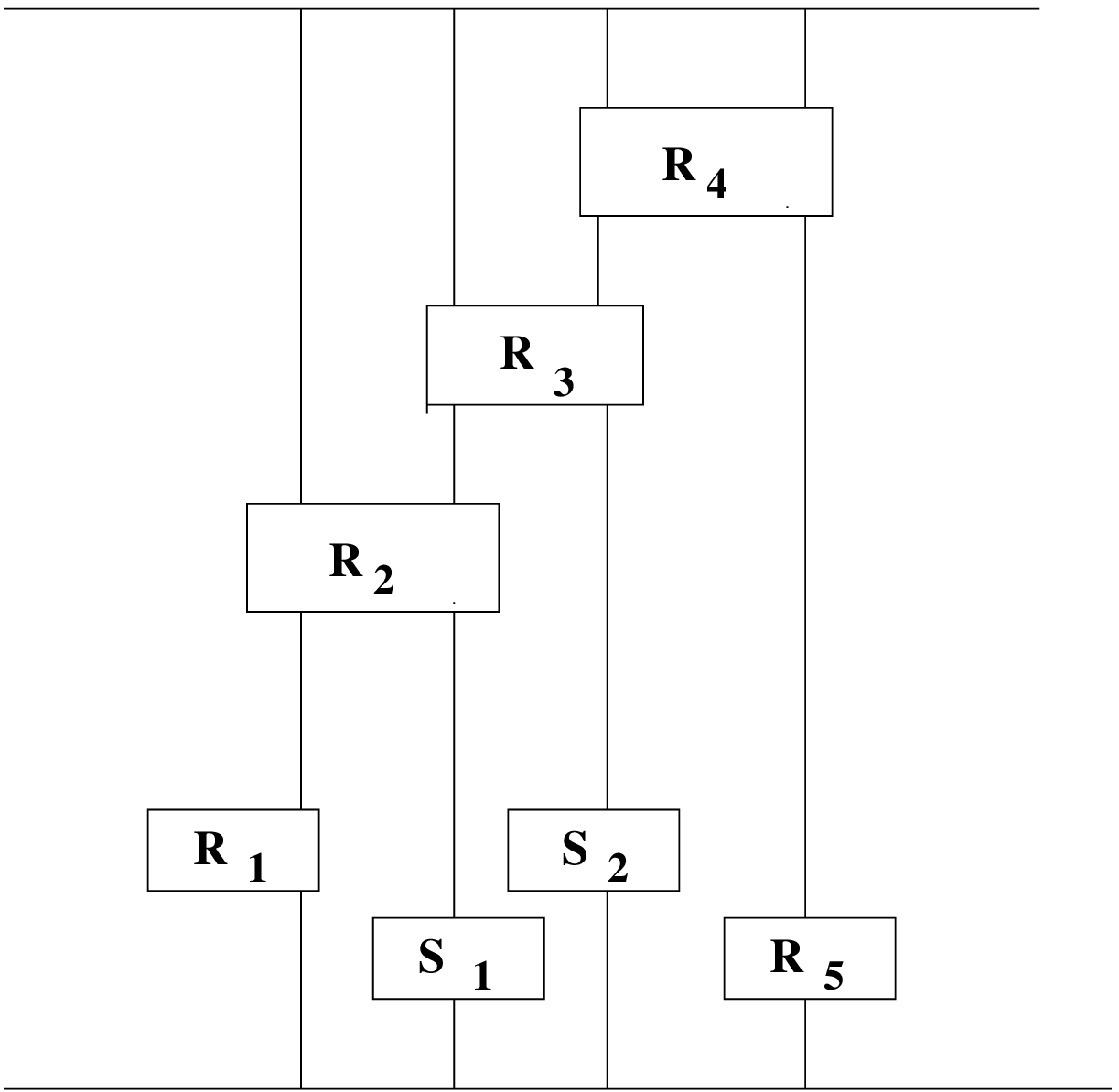} }

This approach to picking out 
simple contraction patterns
clearly generalizes to higher point functions as well. 
For example, consider the following contour
integral of  a correlator involving five insertions 
\eqn\fvptfac{\eqalign{  
&  \int dy_1 dy_2 dy_3 dy_4 ~~
y_1^{2 n_1-1 } y_2^{2n_2-1} y_3^{2n_3-1} y_4^{2n_4-1 } \cr  
& \langle ~~ 
\chi_{R_1} ( \Phi  ( x_1 )) \chi_{R_2} ( \Phi^{\dagger}  ( x_2 ))
\chi_{R_3} ( \Phi  ( x_3 )) \chi_{R_4} ( \Phi^{\dagger}  ( x_4 ))
\chi_{R_5} ( \Phi  ( x_5 )) ~~ \rangle}} 
where the projector diagram with the contractions
of interest is shown below. The 
$y$'s are differences of 
$x$ coordinates 
$y_1 =  x_1 -x_2 , y_2 = x_2-x_3 , y_3 = x_3 -x_4 , 
y_5 = x_4 -x_5 $ and the numbers of boxes $n(R_i)$ in the
Young diagrams for the representations $R_i$ are chosen to
satisfy  $n(R_1) = n_1 , n(R_2 )= n_1 + n_2 , n(R_3) =
n_2+n_3 , n(R_4) = n_3 + n_4,$ and $n(R_5)  = n_4 $
guaranteeing that a contraction pattern of the form
shown in \bastagexi\ exists.  This diagram is
clearly also factorizable using the results of appendix 3.  

It will be interesting to see if these integrated correlators
satisfy non-renormalization theorems.

\subsec{ Staggered factorization and tangential crossings in YM2 }

Staggered factorization has a simple meaning in two-dimensional Yang
Mills at zero area. The rules for constructing the partition function 
for manifolds with boundary and arbitrary Wilson loop insertions
are derived in \mig\rus\ and reviewed in \cmrrev. 

\ifig\twdymstfac{ Diagram illustrating a Wilson Loop observable in 2DYM  }
{\epsfxsize5.0in\epsfbox{ 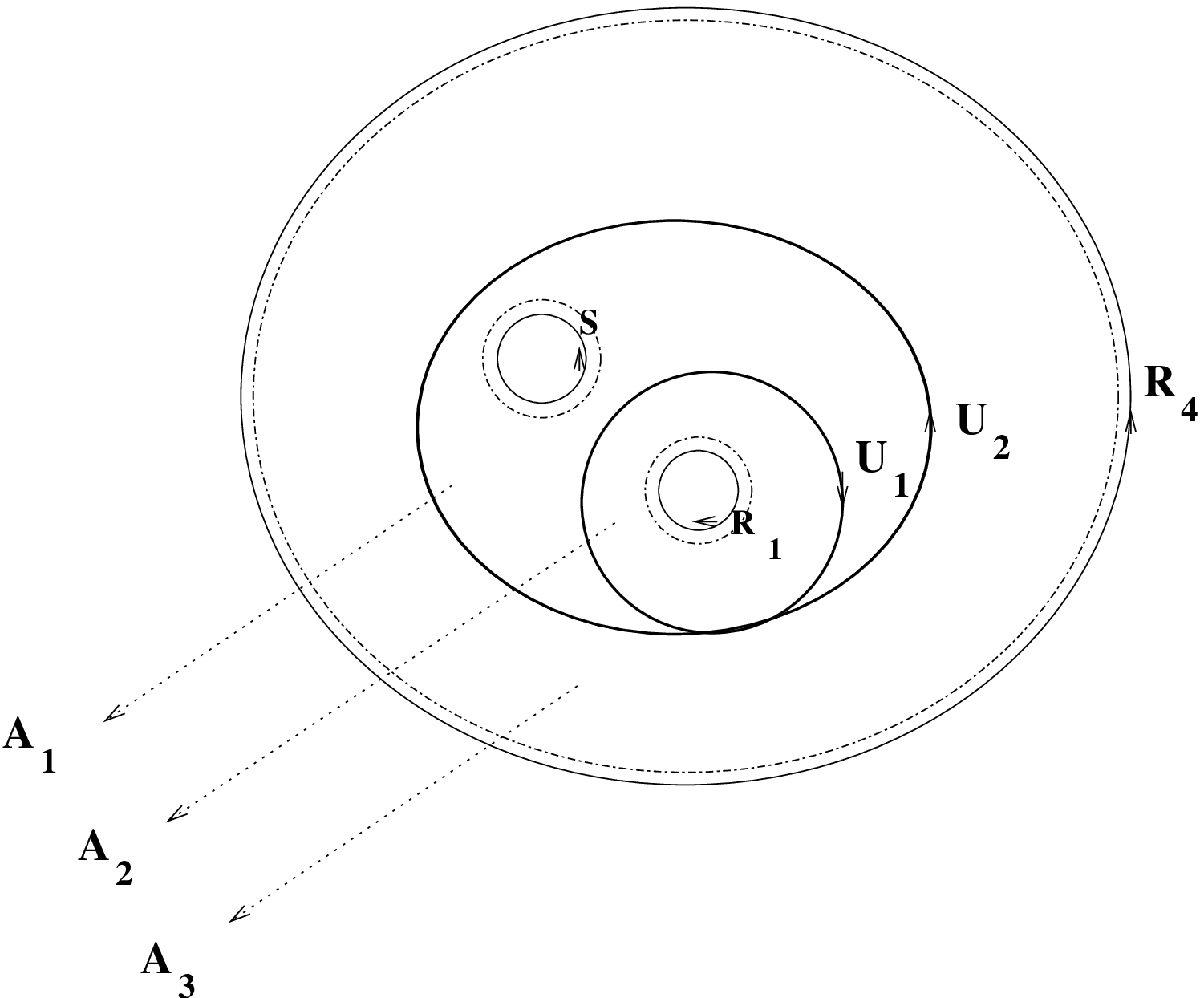 } }

Consider the partition function of topological 
two-dimensional Yang-Mills associated with the above 
diagram. Circles lined by dash-dotted lines are boundaries, 
with the dash-dots being in the interior of the two-dimensional
manifolds. 
The holonomies on the boundaries are weighted by 
a character in the irrep shown next to the boundary. 
The boundary irreps are $S, R_1, R_4$. In calculating 
the partition function we assign some weights 
for each region. The regions have been labelled 
$ A_1, A_2, A_3$. 
The appropriate factors are 
\eqn\regfcts{\eqalign{ 
& A_1 \rightarrow \chi_{S} ( U_1 U_2 ) \cr 
& A_2 \rightarrow  \chi_{R_1} ( U_1 )  \cr 
& A_3 \rightarrow  \chi_{R_4} ( U_2 ) \cr }}
There are also Wilson loop insertions
$ \chi_{R_2} ( U_1^{\dagger} ) $ and $ \chi_{R_3} ( U_2^{\dagger} )
$ along the contours in \twdymstfac\ labelled by $U_1$
and $U_2$ respectively. 

The expectation value of this observable 
is 
\eqn\expt{
Z =  \int dU_1 dU_2 
   \chi_{R_2} (U_1^{\dagger} )~~ \chi_{R_3} (U_2^{\dagger} ) 
   \chi_{R_1} ( U_1 ) \chi_S ( U_1 U_2) \chi_{R_4} ( U_2 ) }
This is precisely the expression that entered in 
\trgi. The staggered factorization argument 
leads to a simpler group integral. 
\eqn\nexpt{ 
Z =  \int dU_1 dU_2 
   \chi_{R_2} (U_1^{\dagger} )~~ \chi_{R_3} (U_2^{\dagger} ) 
   \chi_{R_1} ( U_1 ) { \chi_S ( U_1)  \chi_S (U_2) \over Dim S } 
           \chi_{R_4} ( U_2 ) } 
The latter expression is just the expression we would 
write, using the rules of \refs{\mig,\rus} 
for the Wilson loop expectation value where 
the two tangentially intersecting Wilson loops
are slid away so that they are not touching anymore. 
The factor $ {1 \over Dim S }$ arises because the 
region $A_2$ now has Euler character $-1$ as opposed
to $0$.
In a topological theory, we may naturally expect 
such a sliding move on tangential intersections to leave the 
expectation value invariant, and the rules
of  \mig\rus\
show that this is indeed  true in the zero area limit of 
two-dimensional Yang-Mills. To summarize then, staggered factorization 
of correlators in 4D SYM is mapped to deformations 
of Wilson Loops in 2DYM which disentangle tangential intersections.

\ifig\twdymstfaci{ Multiple tangential intersections
of Wilson loops in 2DYM  }
{\epsfxsize5.0in\epsfbox{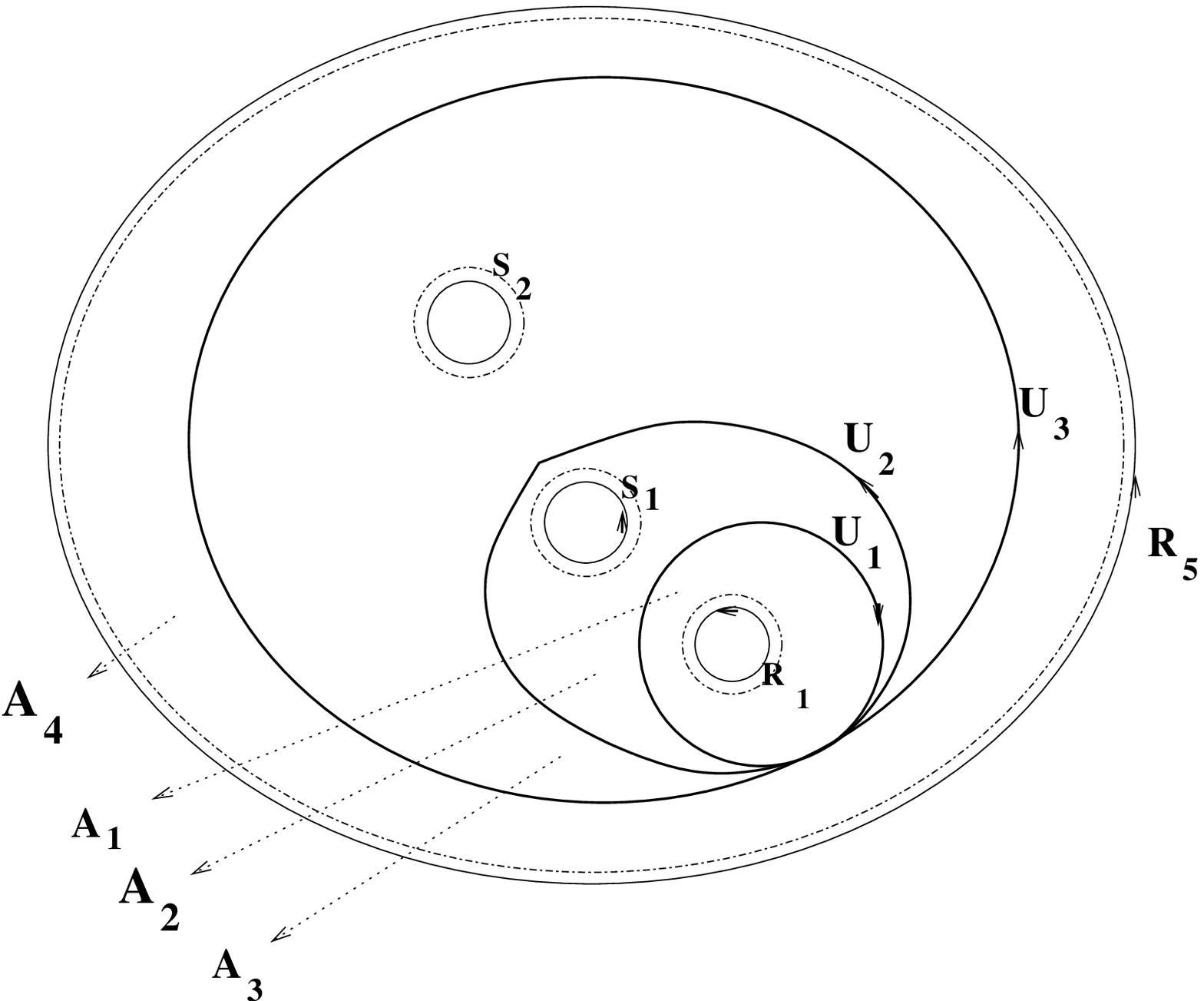}}

This remark is quite general. For example, 
the factorizable five-point correlator 
associated with the trace diagram in \bastagexi\ 
is related to the 2DYM observable with tangential 
crossings given in \twdymstfaci . In this case there 
are Wilson loop insertions 
$ \chi_{R_2} (U_1^{\dagger} )   \chi_{R_3} (U_2^{\dagger} )
  \chi_{R_4} (U_3^{\dagger} ) $.

\subsec{ Staggered factorization and connected sums of links in Chern
Simons }

\ifig\bastagfacexii{  Factorizable example using irreps  }
{\epsfxsize3.0in\epsfbox{ 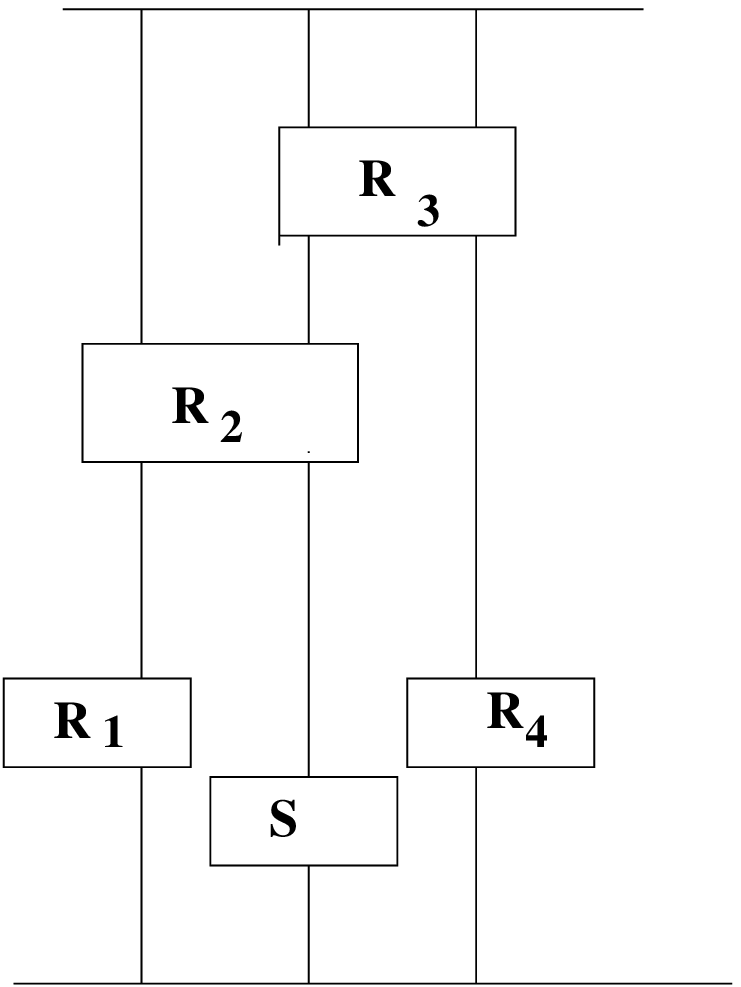 } }

Consider the projector diagram in \stagfactone\ 
which is a sequence of traces in tensor space. 
It is related to a sum over irreps $S$ when 
a complete set of irreducible projectors is 
inserted in $tr_{n_2} $. The diagram is now given 
in \twdymstfaci. 
Up to factors of $d_{R_1} d_S d_{R_4}$ it can be converted
to a trace in the tensor product of 
irreducibles of $U(N)$ given by 
$ R_1 \otimes S \otimes R_4 $. Replacing the 
projectors $R_2$ and $R_3$ by products 
of Clebsh-Gordan coefficients, it is given by a formula
which has a diagrammatic representation 
in terms of a graph with trivalent vertices as explained 
in section 5. As such it is related to 
the expectation value of a knotted graph
in $U(N)$ Chern-Simons. The same procedure 
applied to \facstfac\ gives a pair  of graphs. 
The geometrical operation which takes the pair of 
graphs to the single graph is to form  a connected 
sum of the pair of graphs.

These projector graphs have a q-deformation 
relevant to finite $k$ Chern-Simons and link invariants. 
The relation to standard link invariants is 
clearer if instead of the projectors $R_2, R_3$, 
we have $ \sigma R $
where $R$ is the universal R-matrix for $U_q( N )$ 
and $\sigma $ is a permutation. 
$\sigma R $ commutes with $U_q(N)$ ( see for example \pasqsal,
\alvgaum\ ) and hence is a related to a sum of projectors. 
In fact, in the example of \facstfac,  if we replace
$R_2$ and $R_3$ by $(\sigma R )^2$
we get exactly the example of a connected sum discussed in 
\witjones.  Since the derivation of
the staggered factorization in appendix 3
relied on properties of projectors acting on  tensor products, it  
generalizes to the quantum group case, and hence to finite 
$k$ Chern-Simons. 

To summarize, we have described a way to map 
correlators in SYM4 to expectation values of 
Wilson loops in CS3. The correlators in SYM4
can be simplified by a generalization of the basic 
factorization equation \facti\  when 
they map to Wilson loops which are connected sums of 
simpler Wilson loops.

\subsec{ Remarks on Staggered factorization and Non-renormalization Theorems } 

The extremal correlators have simple spacetime 
 dependence and a simple coupling dependence. 
The proof of the NR theorems indeed relies on 
relating spacetime and coupling dependences
\intril. 
The result of \cjr\ and the first few sections 
of this paper is that the extremal correlators
can be expressed in terms of simple group theoretic 
quantities like fusion coefficients and dimensions.
 The reason this is 
possible is given by the basic factorization and fusion 
equations we have discussed.  More general correlators
are given by more intricate group theory invariants
such as those described in section 5.

The basic factorization was then generalized to staggered 
factorization which allowed a more general class of correlators 
to be expressed in terms of products of dimensions
and fusion coefficients. 
So there is a more general class of correlators which 
have simple colour dependence.  It is natural to speculate that 
these more general correlators would also have 
simple spacetime and coupling dependence. 
Given that dimensions and fusion coefficients are related
to two and three-point correlators \cjr, this line
of reasoning is similar to \refs{\ErdmengerPZ,\HokerDM}.
There non-renormalization properties of correlators, and
in some cases pieces of correlators which are expressed
in terms of products of two and three-point functions
at zero coupling, were explored.  The next-to-extremal
correlators considered in \ErdmengerPZ\ can also
be evaluated in terms of dimensions and fusion
coefficients using the techniques of this paper.
We shall however defer the details to future work.

Some support for this line of thinking is given by 
the fact that the dependence on the representation content 
translates into a dependence on the wavefunctions 
of giant gravitons on the $S^5$ of the $ADS_5 \times S^5 $ 
dual. Now in a supersymmetric theory governed by a 
superalgebra $ SU(2,2|4)$ where the symmetries 
of the $S^5$ and the $ADS_5$ are unified in a single algebra, 
we might expect that simple depdence on the $S^5$ would be related 
to simple dependence on $ADS_5$. By the existing arguments
relating coupling and spacetime dependence, this would imply simple 
coupling dependence, and perhaps no coupling dependence 
at all.

It would be very interesting to develop this 
line of reasoning further and give a  proof 
of the generalized non-renormalization theorems 
suggested by the staggered factorization property.

\newsec{ General  derivation of Basic Factorization
 equation and Fusion equation  } 
 
Consider the four point function in \fact.
This is given by a 
path integral on $R^4$ with four copies of $B_4$ (four-ball)
removed and having four $S^3$ boundaries.
We can think of cutting 
along an $S^3$ which separates  the two chiral operators 
and the  two anti-chiral operators, and inserting a 
complete set of states. We now have two four-manifolds
with three boundaries each. One contains two chiral 
operators and we would like to argue that only chiral 
operators can flow through.

The leading term in the OPE of a field operator 
which creates a highest weight of the superalgebra
with another similar operator is 
also a highest weight.  We are taking advantage of the 
fact that a tensor product of two highest weight states 
of short reps of the superalgebra is itself a highest
weight state of a short rep.
Sub-leading terms in the OPE will, of course, contain descendants. 
If the highest weight operators are all coincident
then only the leading term is relevant. This leading 
highest weight is the only term which will 
contribute to correlators  when the coincident highest weight 
operators are in a correlator involving only 
lowest weight operators as the remaining operators. 
This leads us to expect, in general, an equation of the 
form \facti\ once the complete set of highest
weight states have been described, as we have done already for 
the case of $U(N)$.

 With this heuristic understanding of the
basic factorization equation we can expect 
that it will also hold true in the case of  
$SU(N)$ (and for any gauge group) 
where the explicit formulae for the correlators 
are more intricate. 
In the next section we will describe the 
complete set of highest weights in the case of 
$SU(N)$. We will also give a general formula 
for the extremal correlators. Explicit checks 
of the factorization equations should be 
possible, but do not look trivial since the Schur 
polynomial basis which diagonalized the two-point functions 
for $U(N)$ no longer does so for $SU(N)$.

\newsec{ Observables and  Correlators for $SU(N)$ } 

\subsec{ Classification of half-BPS operators } 

We start by making some remarks about the classification
of operators.
In the case of $U(N)$ we could convert from the
 conjugacy class to the Schur basis. This was very useful
 since the two-point correlators in the Schur basis were 
 orthogonal, and the three-point correlators were
 directly related to fusion coefficients. Also when 
 $n$ becomes comparable or bigger than $N$, the Schur basis 
 provides a useful way to characterize the independent gauge invariant 
 operators. 

 In the case of $SU(N)$ it is still true that we can change basis 
 to the Schur polynomials, but as we shall see they are not orthogonal. 
 Moreover they obey several relations. We can solve the relations 
 but the independent observables are most easily described 
 in terms of conjugacy classes, more precisely in terms of 
 those conjugacy classes corresponding to permutations 
 with no cycles of length $1$.  The triviality of the operator
$tr_n (\sigma \Phi)$ where $\sigma \in S_n$ and
contains a cycle (or cycles) of length 1 follows from the
$SU(N)$ condition that $tr(\Phi) = 0$.

 When $n$ is larger than or comparable to $N$ the 
 exact characterization of the invariants becomes
 more intricate. 
 We consider the Schur polynomial basis. 
 For small $n$ we imposed identities for each 
 conjugacy class with one or more cycles of length $1$. 
When $n \geq N$ we also have to impose the trace relations,
eg., for $n = N+1$ we have the relation
$tr(\Phi^{N+1}) = tr(\Phi) tr(\Phi^N) + \cdots$ and similarly
for larger $n$.
Once we have guaranteed the
 vanishing of conjugacy classes $[n_1, n_2 \cdots 1 , 1 \cdots ]$ 
 with $n_1 \ge n_2 \ge \cdots $ with 
 $n$'s reaching and including $N$, we don't need 
 to impose extra conditions. 
 For example,  the vanishing of $[N+1, 1 ]$ is guaranteed 
 by the vanishing of $[N, 1,1 ]$ and others,   because 
 $[N+1]$ can be expressed in terms of $[N,1]$ and other terms
 all involving no traces with powers of $\Phi $ greater than 
 $N$. 
 
  A complete set of observables could 
 alternatively be expressed in terms of 
 traces and their products. Just take polynomials in 
 $ tr ( \Phi^2 ) , tr ( \Phi^3 )  \cdots tr ( \Phi^N ) $.
 These correspond to conjugacy classes with no entry 
 of length $1$ or any entry of length greater than $N$.

\subsec{ Correlators } 

\ifig\masterdiagram{Diagram representing the multipoint correlation
function of $n$ scalars $\Phi$ and their conjugates }
{\epsfxsize6.0in\epsfbox{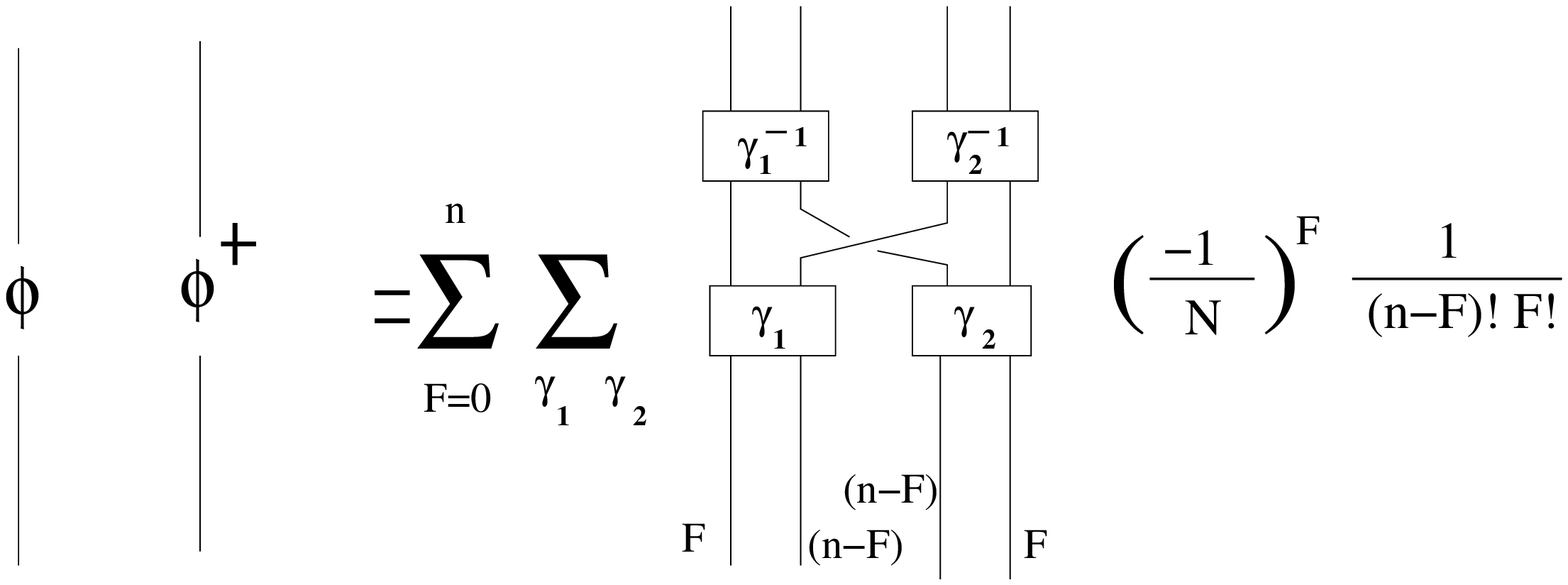} }

We now discuss the extremal correlators for the
gauge group $SU(N)$.
The story here is similar to the one developed for
the gauge group $U(N)$ in \cjr\ except that there
are now further $1/N$ corrections arising from the
modified two-point function, i.e., for $SU(N)$ one
has
\eqn\suntwo{ \langle \Phi^{i}_{j}(x) (\Phi^{\dagger})^{k}_{l}(y)
\rangle = {1 \over (x-y)^{2}} ( \delta^{i}_{l}
\delta^{k}_{j} - {1 \over N} \delta^{i}_{j} \delta^{k}_{l} ).
}
To evaluate the correlators of interest here we make use
of the multi-point correlator\foot{We drop the
position dependencies in writing this formula.
For the extremal correlators all $\Phi^{\dagger}$ (or $\Phi$)
operators will be evaluated at the same spacetime point, so
there is no loss in generality in dropping all position
dependencies for such correlators as it will only contribute
an overall factor and can be reinstated trivially.}
given diagrammatically in \masterdiagram.

Translating this into an explicit formula : 
\eqn\master{\eqalign{\langle \Phi^{I(n)}_{J(n)} 
(\Phi^{\dagger})^{K(n)}_{L(n)} \rangle & = \sum_{\gamma \in S_n}
\prod_{p=1}^{n}
(\delta^{i(p)}_{l_{\gamma(p)}} \delta^{k_{\gamma(p)}}_{j(p)}
- {1 \over N} (\delta^{i_p}_{j_p} 
\delta^{k_{\gamma(p)}}_{l_{\gamma(p)}} )) \cr
& = {1 \over n!} \sum_{F=0}^{n} (-{1 \over N})^F {n \choose F}
\sum_{\gamma, \gamma_1 \in S_n} 
\delta^{i_{\gamma_1(1)}}_{l_{\gamma \gamma_1(1)}}
\delta^{k_{\gamma \gamma_1(1)}}_{j_{\gamma_1(1)}} \cdots
\delta^{i_{\gamma_1(n-F)}}_{l_{\gamma \gamma_1(n-F)}}
\delta^{k_{\gamma \gamma_1(n-F)}}_{j_{\gamma_1(n-F)}} \cr
& ~ ~ \times
\delta^{i_{\gamma_1(n-F+1)}}_{j_{\gamma_1(n-F+1)}}
\delta^{k_{\gamma \gamma_1(n-F+1)}}_{l_{\gamma \gamma_1(n-F+1)}} 
\cdots
\delta^{i_{\gamma_1(n)}}_{j_{\gamma_1(n)}}
\delta^{k_{\gamma \gamma_1(n)}}_{l_{\gamma \gamma_1(n)}} \cr
& = {1 \over n!} \sum_{F=0}^{n} (-{1 \over N})^F {n \choose F}
\sum_{\gamma_1,\gamma_2 \in S_n} 
\delta^{i_{\gamma_1(1)}}_{l_{\gamma_2(1)}}
\delta^{k_{\gamma_2 (1)}}_{j_{\gamma_1(1)}} \cdots
\delta^{i_{\gamma_1(n-F)}}_{l_{\gamma_2(n-F)}}
\delta^{k_{\gamma_2 (n-F)}}_{j_{\gamma_1(n-F)}} \cr
& ~ ~ \times 
\delta^{i_{\gamma_1(n-F+1)}}_{j_{\gamma_1(n-F+1)}}
\delta^{k_{\gamma_2 (n-F+1)}}_{l_{\gamma_2 (n-F+1)}} 
\cdots
\delta^{i_{\gamma_1(n)}}_{j_{\gamma_1(n)}}
\delta^{k_{\gamma_2 (n)}}_{l_{\gamma_2 (n)}}.
}}
The first line of \master\ is obtained by carrying
out all contractions using the two-point function \suntwo.
The second line is obtained using the following reasoning.
We want to consider all terms of order $(1/N)^F$.  Consider
in particular the term
$\delta^{i_{1}}_{l_{\gamma (1)}}
\delta^{k_{\gamma (1)}}_{j_{1}} \cdots
\delta^{i_{(n-F)}}_{l_{\gamma (n-F)}}
\delta^{k_{\gamma (n-F)}}_{j_{(n-F)}}
\delta^{i_{(n-F+1)}}_{j_{(n-F+1)}}
\delta^{k_{\gamma (n-F+1)}}_{l_{\gamma (n-F+1)}} \cdots
\delta^{i_{n}}_{j_{n}}
\delta^{k_{\gamma (n)}}_{l_{\gamma (n)}}$
which appears at this order.  It is easy to see
that all other terms arising at this order can be
generated from this term by simply permuting
the subscripts $1,...,n$, i.e., by replacing
$i \rightarrow \gamma_1(i)$ and summing over all
permutations $\gamma_1 \in S_n$.  This however
overcounts by the factor $F! (n-F)!$, the number of
elements in the stabilizer subgroup
$S_{n-F} \times S_F \in S_n$
that preserve a given product of Kronecker
delta-functions, like the one given above.  Dividing
out by this factor we arrive at the second line.
The third line then follows by redefining the
summation index $\gamma_2 = \gamma \gamma_1$.

\ifig\suninter{Projection diagram arising in the
$SU(N)$ multi-point correlator.}
{\epsfxsize3.0in\epsfbox{ 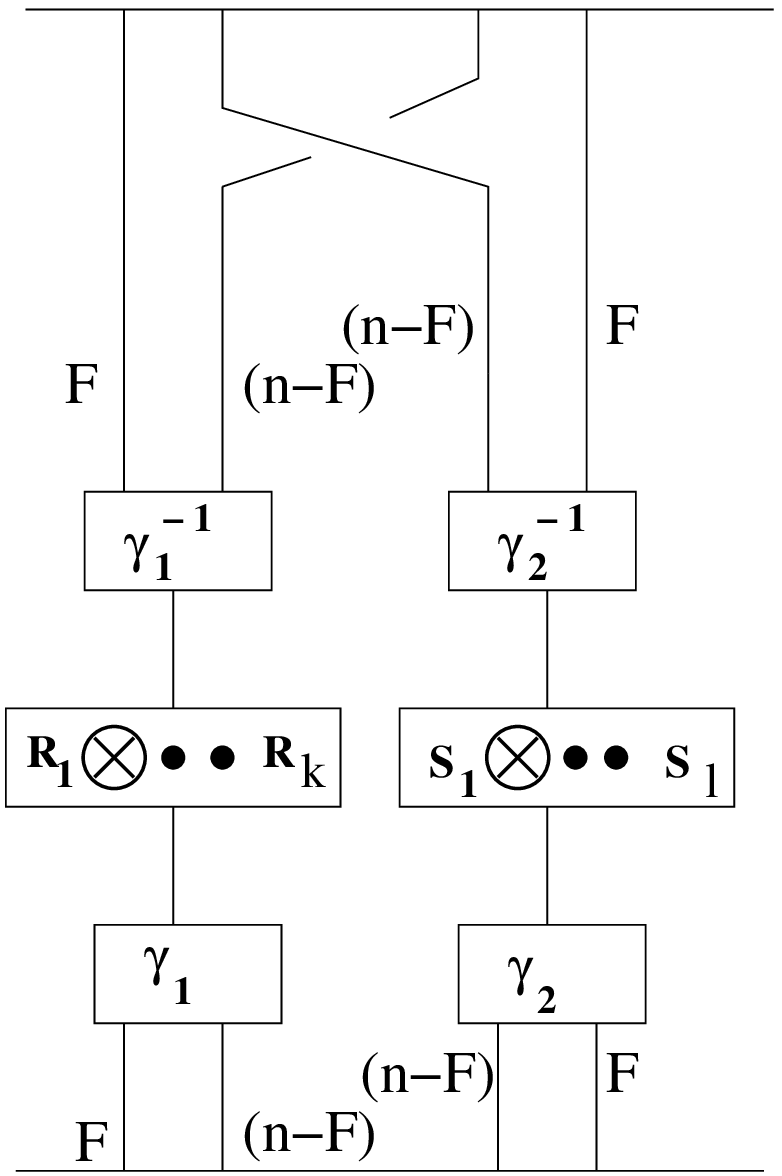} }

To compute the correlators of interest we project
the correlator \master\ onto the appropriate basis.
For example, for a correlator such as
\eqn\schur{\langle \chi_{R_1}(\Phi) \cdots
\chi_{R_k}(\Phi) \chi_{S_1}(\Phi^{\dagger})
\cdots \chi_{S_l}(\Phi^{\dagger}) \rangle}
we project \master\ onto the Schur polynomial
basis by contracting the free indices of \master\
with projection operators of the form given in
\proj\ for each operator $\chi_{R_i}(\Phi)$ and
$\chi_{S_j}(\Phi^{\dagger})$ appearing in the correlator.

\ifig\sunmultfig{Rearrangement of the diagram \suninter\ 
illustrating the multi-point $SU(N)$
correlator.}
{\epsfxsize3.0in\epsfbox{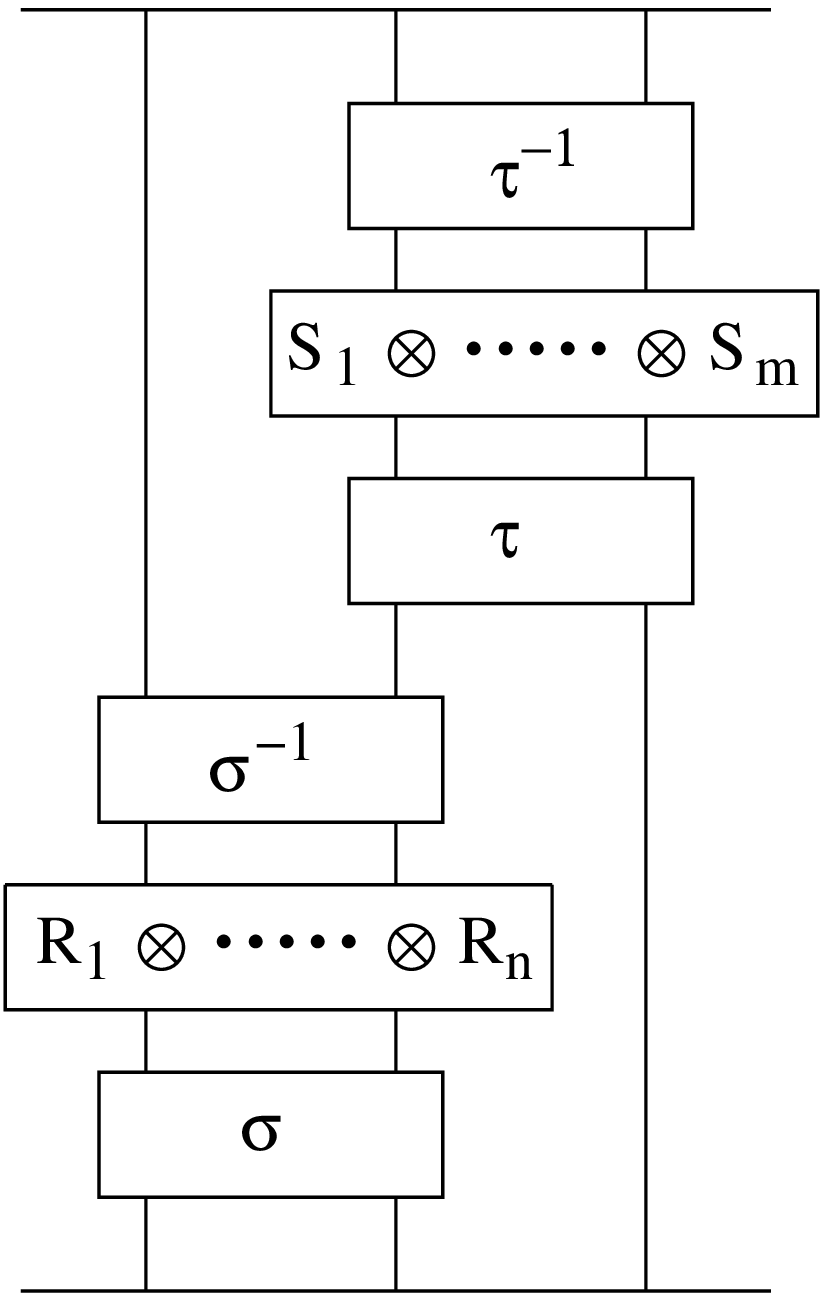}}

After contracting the $\Phi$'s using \masterdiagram\ we
get \suninter\ which can be rearranged to give \sunmultfig.
In terms of formulae we find
\eqn\sunmulti{\eqalign{\langle \chi_{R_1} & (\Phi) \cdots
\chi_{R_k}(\Phi) \chi_{S_1}(\Phi^{\dagger})
\cdots \chi_{S_l}(\Phi^{\dagger}) \rangle =
{1/n! \over d_{R_1} \cdots d_{R_k} d_{S_1} \cdots d_{S_l}}
\sum_{F=0}^{n} \sum_{\gamma_1,\gamma_2 \in S_n} (-{1 \over N})^F
\cr
& \times {n \choose F} tr \bigg( \Bigl( 1_{F} \otimes (\gamma_{1}^{-1}
(P_{R_1} \otimes \cdots \otimes P_{R_k}) \gamma_1 ) \Bigr)
\Bigl( ( \gamma_{2}^{-1}(P_{S_1} \otimes \cdots \otimes P_{S_l})
\gamma_2 )
\otimes 1_F \Bigr) \bigg).
}}

This can be simplified using the decomposition
\multidecomp\ derived earlier in the $U(N)$ case.
That is, the conjugated tensor product of projectors is replaced
by a sum of projectors onto irreducible representations to arrive at
\eqn\sunmultitwo{\eqalign{ \langle \chi_{R_1}(\Phi) & \cdots
\chi_{R_k}(\Phi) \chi_{S_1}(\Phi^{\dagger})
\cdots \chi_{S_l}(\Phi^{\dagger}) \rangle =
n! \sum_{F=0}^{n} (-{1 \over N})^F {n \choose F} \cr
& \times \sum_{T_1,T_2}
{g(R_1, \cdots, R_k; T_1) \over d_{T_1}}
{g(S_1, \cdots, S_l; T_2) \over d_{T_2}}
tr((1_F \otimes P_{T_1})(P_{T_2} \otimes 1_F)).
}}
Diagrammatically we have simply replaced the
boxes corresponding to the tensor product
of projection operators by projection operators
onto irreducible representations.
This diagram is in fact just \threeptsumdiag\
in section 6.  The factorization argument there 
applies word for word here and yields
\eqn\tracefact{tr((1_F \otimes P_{T_1})(P_{T_2} \otimes 1_F))
= \sum_{T} {1 \over d_T ~ Dim T}  tr \left( (1_F \otimes P_T) P_{T_1} \right)
tr \left( (1_F \otimes P_T) P_{T_2} \right).}
Finally the traces on the right-hand-side are evaluated
as in \basethree\ to arrive at the final expression
\eqn\sunmultifinal{\eqalign{\langle \chi_{R_1}(\Phi) & \cdots
\chi_{R_k}(\Phi) \chi_{S_1}(\Phi^{\dagger})
\cdots \chi_{S_l}(\Phi^{\dagger}) \rangle =
n! \sum_{F=0}^{n} (-{1 \over N})^F {n \choose F}
\sum_{T_1,T_2} \sum_{U_1,U_2} \sum_T
{d_{U_1} d_{U_2} d_T \over Dim T} \cr
& \times {Dim T_1 ~ Dim T_2 \over d_{T_1} ~ d_{T_2}} g(U_1, T; T_1)
g(U_2, T; T_2) g(R_1, \cdots, R_n; T_1) g(S_1, \cdots, S_m; T_2).
}}

Because of the $SU(N)$ condition $tr(\Phi)=0$, the correlator
\sunmultifinal\ must satisfy some non-trivial constraints.
We can trade in a character in the above correlator for
a trace by recalling the identity
\eqn\charsumident{\sum_R \chi_R (\sigma) \chi_R (\tau)
= \sum_{\gamma \in S_n} \delta(\sigma^{-1} \gamma \tau \gamma^{-1}).
}
Multiplying \sunmultifinal\ by $\sum_{R_i} \chi_{R_i} (\tau)$,
for example, for some fixed $i$, will replace $\chi_{R_i} (\Phi)$
by $\sum_{\gamma} tr(\gamma^{-1} \tau \gamma \Phi)$ after applying
\charsumident.  If $\tau$ has a cycle of length 1 then the
correlator must vanish.  Consider the case that $k=1$ and
multiply the correlator by $\sum_{R_1} d_{R_1}$ corresponding
to $\tau$ being the identity permutation.  On the right-hand-side
of \sunmultifinal\ the fusion coefficient $g(R_1;T_1) = 
\delta_{R_1,T_1}$ so that the sum over $T_1$ is trivial
and moreover the sum over $R_1$ just involves the terms
\eqn\intervanish{\sum_{R_1} Dim ~ R_1 ~ g(U_1,T;R_1) =
Dim U_1 ~ Dim T.}
The sum over $U_1$ now just becomes $\sum_{U_1} d_{U_1} Dim U_1 = N^F$
and the sums over $U_2$ and $T$ become
$\sum_{U_2,T} d_{U_2} d_T g(U_2,T;T_2) = d_{T_2}$.
This implies that the $T_2$ sum can be done producing
$\sum_{T_2} g(S_1,...,S_l;T_2) ~ Dim T_2 = Dim S_1 \cdots Dim S_l$.
The only $F$-dependent term produced in all these sums was
$N^F$, cancelling the $1/N^F$ in \sunmultifinal, yielding
for the $F$ sum 
\eqn\Fsum{
\sum_{F=0}^{n} {n \choose F} (-1)^F = 0.
}

More generally this correlator must vanish for any permutation
$\tau$ containing a 1 cycle and also for $k \geq 1$.
This condition gives rise to some non-trivial constraints
on the correlator as evidenced by the above argument
for the simplest case of $\tau=e$ and $k=1$.

\subsec{ Factorization and fusion identities for $SU(N)$ } 

The general arguments for factorization given in section 9
suggests that the $SU(N)$ correlators should satisfy
factorization equations similar to those satisfied by
the $U(N)$ correlators.  A natural guess for the four-point
function evaluated in the previous subsection would be
\eqn\facsu{\eqalign{ 
\langle \chi_{R_1} (\Phi) \chi_{R_2} (\Phi) &
\chi_{S_1} (\Phi^{\dagger}) \chi_{S_2} (\Phi^{\dagger}) \rangle \cr
& = \sum_{T_1,T_2} 
\langle \chi_{R_1} (\Phi) \chi_{R_2} (\Phi) T_1 (\Phi^{\dagger})
\rangle G^{T_1 T_2} \langle T_2 (\Phi)
\chi_{S_1} (\Phi^{\dagger}) \chi_{S_2} (\Phi^{\dagger}) \rangle.
} }
The operators $T_1 (\Phi)$ and $T_2 (\Phi)$ are either
polynomials in $tr(\Phi^2),...,tr(\Phi^{N})$ containing
exactly $n(R_1) + n(R_2) = n(S_1) + n(S_2)$ $\Phi$'s each,
or equivalently can be expressed in terms of Schur polynomials
subject to the constraints discussed earlier in the previous
subsection.  $G^{T_1 T_2}$ denotes the inverse of the two-point
function $\langle T_1 (\Phi) T_2 (\Phi^{\dagger}) \rangle$ (recall
that this is not diagonal in the Schur basis used in the
previous subsection) and the sum over $T_1,T_2$ runs over
a complete basis of such operators.
Using \sunmultifinal\ this factorization equation 
can be written in terms of products of fusion coefficients.   
We can therefore get 
some combinatoric identities which should be explicitly 
testable. 

\newsec{ Remarks on Giant gravitons }

\subsec{ A symmetry between giant gravitons 
and KK Modes } 

 Young diagrams with a small number $n \ll N$ of boxes
give rise to  
 operators with a small number of $ \Phi$'s 
 which are related to KK modes in the $AdS_5 \times S^5$
dual, \malda\GKP\WittenQJ. 
On the other hand Young diagrams which have a few very long columns, 
 of length close to $N$, are all giant graviton 
 states \refs{\bala,\cjr}.  Recall that the three-point function depends on 
the fusion coefficient \cjr\ as in
\eqn\thrpt{ 
\langle  \chi_{R_1} ( \Phi ) ( x) \chi_{R_2} ( \Phi ) ( x)  \chi_{S}
 ( \Phi^{\dagger}  ) ( 0)  \rangle 
= x^{-2n_1 -2n_2} g(R_1, R_2, S )  { n! Dim S  \over d_S}}
and appropriate ratios give exactly the fusion coefficient
\eqn\thrptrat{ 
{\langle  \chi_{R_1} ( \Phi ) ( x) \chi_{R_2} ( \Phi ) ( x)  \chi_{S}
 ( \Phi^{\dagger}  ) ( 0)  \rangle \over
\langle \chi_S (\Phi(x)) \chi_S 
(\Phi^{\dagger} (0)) \rangle}
= g(R_1, R_2, S ).} 
When $R_1$ and $R_2$ are small compared to $N$, 
$S$ must also be small. This just follows from 
the fact that $n(S) = n(R_1) + n(R_2)$. 

When $R_1$ and $R_2$ are both in the region 
of sphere giants, that is, they have a few columns of length 
comparable to $N$, the selection rule $n(S) = n(R_1) + n(R_2)$
would in principle allow $S$ to be a few rows of length comparable to $N$ 
for example, which by \cjr\ are related to $ADS$ giants 
\GrisaruZN\HashimotoZP\DasST.
The actual three point functions 
\thrpt\ for such $R_1$, $R_2$, and $S$
are zero because of the properties of $g(R_1,R_2; S)$. 
If $R_1$ and $R_2$ are in the sphere giant region specified 
above, their complex conjugates  $\bar R_1$ 
and $\bar R_2$ are small Young Diagrams. These fuse only 
into small Young Diagrams $\bar S$.  The fusion coefficients
satisfy the duality property 
\eqn\sym{ 
g( R_1, R_2 ; S ) = g( \bar R_1, \bar R_2 ; \bar S ).
} 
The conjugate of a small $\bar S$ is again in the region of 
sphere giants. This shows that the fusion rules 
for KK modes are identical to those for giant graviton modes. 
It would be very interesting to understand this symmetry 
from the point of view of three-brane world volumes
coupled to gravity. 

\subsec{ Saturating the factorization equation : Sphere giants  } 

Another consequence of this relation between 
fusions of sphere giants and KK modes is 
that when we consider an extremal  four point function
involving only sphere giants, the intermediate 
states which enter the factorization equation \facti\ 
are again sphere giants. 
 
Indeed in \facti\ if 
 $R_1,R_2,S_1,S_2 $ are all large 
 antisymmetric reps., then  
 a representation $S$ will only contribute 
 to the factorization sum if 
 $ g(R_1,R_2; S) g(S_1,S_2: S) $ is non-zero. 
 Now if $R_1, R_2,S_1,S_2$ are all large
 anti-symmetric reps then their conjugates 
 $ {\bar R_1} , { \bar R_2}, {\bar S_1} , { \bar S_2 } $ 
 are all small representations. For these 
 it is clear that  $g({ \bar R_1} ,{ \bar R_2} ; { \bar S} ) g ({ \bar
 S_1} ,{ \bar S_2} ; { \bar S } ) \ne 0 $ 
 requires that $\bar S$ be a small rep. This means
 that  $S$ is a large antisymmetric rep. 
This proves the claim that if $R_i, S_i$ are related
 to giant gravitons, then the factorization equations saturate
 on sphere giants. 
 
 Since sphere giants are related to moduli spaces
 of holomorphic maps \refs{ \bk , \mikh }, this means 
 that the factorization equations are saturated by 
 operators which are related to these moduli spaces. 
We expect the observables 
of a topological theory formulated on these 
moduli spaces to obey factorization equations which
would descend from 
the basic factorization equation of $N=4$ SYM.
   
 We expect some analogous results should hold 
 true for large symmetric irreps, i.e., Young
diagrams with a few rows of length order $N$, related to AdS 
 giants according to \cjr.  The arguments 
presented above for the sphere giants do not
immediately generalize to this case as the conjugate
of a large symmetric irrep consists of a large
number of boxes.  Nevertheless the Littlewood-Richardson
rule tells us which irreps occur when fusing a pair
of large symmetric irreps.  One finds that
the allowed Young diagrams again contain only
a few rows, at least one of which will have
order $N$ boxes, but some rows containing a
small number of boxes are also allowed.  Such
a Young diagram corresponds to a perturbation
of an $AdS$ giant.  Hence, it follows that
the factorization equations for $AdS$ giants
also saturates on $AdS$ giants and perturbations
thereof and do not contain, for example, sphere giants.  
It seems to be a general property, true for KK modes,
AdS giants and sphere giants, that the saturation
of factorization equations on a small set of operators
in the large N theory is related to the existence
of semiclassical objects admitting some form of
perturbative 1/N analysis.


\newsec{ Summary and Outlook } 

 We started with the general extremal 
 correlators derived in \cjr.  It was shown there 
 that there is a one-one correspondence between, on the one hand,  
 operators in $U(N)$ maximally supersymmetric gauge 
 theory in four dimensions transforming in 
 half-BPS representations of the superalgebra
 and, on the other hand,  irreducible representations of $U(N)$. 
 Operators belonging to half-BPS representations of 
 the superalgebra containing a highest weight state 
 of $R$-charge  $2n$ are mapped to representations 
 of $U(N)$ built from $n$ copies of the fundamental. 
 The extremal correlators were expressed in terms 
 of $U(N)$  and $S_n$ group theoretic data such as dimensions 
 of representations and fusion or branching coefficients.

 In this paper we generalized our results 
  and showed how to relate more general 
 correlators, including non-extremal ones, to 
 more general $U(N)$ group theoretic data. 
 Recalling that fusion coefficients of $U(N)$ 
 can be expressed in terms of an integral of 
 a product of characters, the more general 
 group theoretic data can be viewed as more general 
 group integrals involving, generically, many $U(N)$ 
 group variables. Such integrals occur for example in 
 lattice gauge theory or continuum two dimensional Yang-Mills 
 theory. The group theoretic quantities can also be described 
 in terms of sequences of projectors acting on tensor 
 spaces or on irreducible representations of $U(N)$. 
 These sequences are conveniently described in terms of 
 diagrams we called {\bf projector diagrams}. By using the 
 expression of projectors as products of Clebsch-Gordan 
 coefficients, the projector diagrams can be converted 
 to { \bf projector graphs}  where the vertices of the graphs 
 represent Clebsch-Gordan coefficients, and the projector graph 
 itself is a quantity analogous to $6J$ symbols. 
 
 It was found that extremal correlators 
 and some generalizations thereof  are associated 
 with simple graphs. When the graphs are simple, the correlator
 can be reduced entirely to fusion coefficients and dimensions. 
 The simple form of the extremal correlators 
 lead to some factorization equations and fusion identities 
 which were described in section 2. The relation to 
 projectors leads to sum rules. These sum rules 
 can quite generally be used to simplify 
 any projector diagram. The factorization and fusion identities, on 
 the other hand, only work for certain classes of diagrams. 

The simple group theoretic form of the extremal 
correlators, and in particular the factorization and fusion
identities, 
 suggest relations to topological gauge theories. 
We described concretely such relations in section 7. 
The topological gauge theories we discussed 
were three dimensional $U(N)$ Chern-Simons at large $k$ 
and two dimensional $U(N)$ Yang-Mills at zero area. A corollary of the 
Chern-Simons connection is that there are relations 
to $G/G$ models in two dimensions. 

Beyond the case of extremal correlators, 
since we still have the projector graphs 
associated to the correlator, we can use that 
to systematically map to Wilson Loops in Chern 
Simons. The projector graph, which begins as 
a device to summarize a group theoretic quantity,
emerges as a, possibly intersecting, Wilson loop for Chern 
Simons on $S^3$. There is 
also a connection to intersecting Wilson loops 
in two-dimensional Yang-Mills. The basics of these relations 
were described in Section 7. 
The basic factorization equation was found to have a generalization
which we called staggered factorization. Section 
8 described the physics of this generalization 
and shows that, under the maps to topological gauge theories
defined in section 7, the sum rules are related to the Verlinde formula, 
and staggered factorization is related to connected sums of 
Wilson loops in Chern-Simons and tangentially intersecting Wilson
loops in two dimensional Yang-Mills. 
Section 8 
includes some speculations on generalizations
of known non-renormalization theorems beyond the case 
of extremal correlators, which are motivated 
by staggered factorization.  Section  10 began a discussion 
of $SU(N)$ correlators. 
  In section 11 we remarked that there is 
 an intriguing symmetry between correlators 
 of giant gravitons and correlators of KK modes 
 in the $ADS_5 \times S^5$ dual to $N=4$ SYM. 
  We also suggested that the truncation of 
 the factorization sums is related to 
 the existence of semiclassical objects
described by the correlators of the $N=4$ SYM.

We now discuss some avenues for the future. 
Since we can write down the expressions for 
 correlators in terms of Clebsch-Gordan 
 coefficients,
  we can get a q-deformed version by converting 
 the Clebsch-Gordan to q-Clebsch Gordans.
 These q-deformed formulae will have 
 similar factorization properties. We saw 
 that the factorization equations were 
 related to unitarity and supersymmetry. 
 This leads one to suspect that there might 
 be a deformation of SYM preserving at least 
 some SUSY, which has correlators which are these 
 $q$-deformations. It would be interesting to look for 
 such a $q$-deformation, perhaps using recent results 
on deformations of $N=4$ SYM \ofer\oferi. 
 It is worth noting that, in general, the $q$-deformation
of the correlators  is not unique. 
 Whenever we have a tensor product of irreps 
 of $U(N)$,  we can project the product onto irreps by 
using  $\Delta$ or $\Delta^{\prime}$, two consistent 
ways for the quantum group to act  on the tensor product space,
related by a permutation.
 For all the factorizable correlators which can be reduced
to fusion coefficients, 
 there is no ambiguity, since fusion coefficients  are independent 
 of  whether we choose $ \Delta $ or $\Delta^{\prime}$. 
 For the cases where there is an ambiguity, one would
hope that it can be interpreted physically, for example 
in terms of a choice of regulator in defining the deformed 
theory.

 We expect these factorization equations 
 to be true on more general manifolds
 than $R^4$. Indeed if we consider 
 $ R^3 \times S^1$, for instance,  we can write down the two-point functions
 by summing over images. This will modify the 
 two point function. But the form of the answers 
 will remain the same. It might be interesting to make this 
 explicit.

 While the connections to Chern-Simons were established 
 by directly analyzing the correlators, a natural question 
 is to find a physical  rather than technical proof of 
 these connections. 
  One approach might be  to use field theory techniques, e.g., 
those of \wittqft\ relating 4D Donaldson to 3D
 Floer theory or
\blauthom\ to relate 3D Chern-Simons to 
 2D  $G/G$ models.  Here we would like to 
 relate the (quasi-) topological sector of a four dimensional gauge theory 
 to a topological three dimensional theory. 
 Correlation functions in a four dimensional 
theory are related to overlaps of wavefunctions 
which are functionals of three-dimensional fields living 
on boundaries of four dimensional manifolds. The wavefunctions are 
defined by path integrals on the four-manifold with boundary \wittqft. 
Making this wavefunction approach explicit in the case
of extremal correlators of $N=4$ SYM may be expected 
to lead to relations between the four dimensional theory and
three dimensional theories.  
 Another approach to a physical 
proof of the connections exhibited in section 7
may be to use stringy dualities,
 perhaps along the lines of \gopvaf\vafoog\vafmarin.

 A puzzle raised by this work is to explain 
 why there are relations between unitary group 
 integrals and correlation functions of Higgs fields.  
We have shown using 
Schur Duality that general extremal  correlation functions 
of half-BPS operators can be expressed in terms
of simple unitary group integrals. Non-extremal 
correlators involve more complicated group integrals. 
It would be interesting to find a physical 
explanation for the appearance of group integrals. 
Is there some way of mapping the
action involving the Higgs scalars into an action 
involving gauge fields, where the appearance of 
$U(N)$ group integrals would be more direct, 
the way it happens in lattice gauge theory for example ?

 We have found quantities in $N=4$ SYM which 
 are related to $G/G$ field theories on two dimensional 
 closed manifolds. The system of equations 
 satisfied by the observables of $G/G$  have a generalization 
 to the case where two-dimensional manifolds with boundary 
 are included. This open-closed topological theory is related to 
 $K$ theory \moseg. It would be of interest to find 
 observables related to $N=4$ SYM and its $ADS \times S$ dual 
 which would map to this open-closed topological set-up.

\bigskip

\noindent{ \bf Acknowledgements: }
 We wish to thank for  instructive  discussions Costas Bachas, 
 David Berenstein,  Stefan Cordes,  Robert Dijkgraaf, Louise
Dolan, Antal Jevicki, 
 David Lowe, 
 Gregory Moore,  Savdeep Sethi, and Chung-I-Tan. 
 SR would like to thank the Newton Institute,
 Cambridge  for hospitality  while part of this work was done. 
 This research was supported  by DOE grant  DE-FG02/19ER40688-(Task A).

\newsec{ Appendix 1 :  Schur Duality and group integrals,  $P_R
= P_r $ } 
 
 Consider n-fold tensor space, $ V^{\otimes n } $ where $V$ is the fundamental
 representation of $U(N)$. This space has an $S_n$ action 
 permuting the vectors in different factors of the tensor product.  
 The theorem described in 
 \decomp\ has some powerful implications for 
 relations between unitary and symmetric group actions on 
 tensor space. These relations are useful at various points in this
 paper. 

Consider  a fixed Young Diagram. There is an irrep. $R$  
of $U(N)$ and an irrep  $r$ of $S_n$  associated with it. 
For an irrep $r$ consider the following element 
of the group algebra of $S_n$ : 
\eqn\projsr{ P_r = d_r { 1 \over n! } \sum_{\sigma} \chi_r ( \sigma ) \sigma
} 
In this sum $ \sigma $ is in $S_n$. Instead of 
$\chi_r ( \sigma )$ we could equally well have written 
$ \chi_r ( \sigma^{-1} )$ since a permutation and its inverse
are in the same conjugacy class of $S_n$. 
Using character identities, reviewed for example in \cjr, 
one can prove that 
\eqn\prjpf{ P_r P_s  = \delta_{rs} P_r } 
This means that $P_r$ is a projector.  One further checks
that $tr(P_r) = d_r ~ Dim R$. 
Similarly 
\eqn\projsR{ P_R = Dim R \int dU \chi_R ( U^{\dagger} ) U } 
is a projector for $U(N)$, using the standard normalization of the 
measure where 
$$ \int dU \chi_R ( U )  \chi_S ( U^{\dagger } ) =
\delta_{RS}. $$
Again one can check that $tr(P_R) = d_r ~ Dim R$.
Inserting either projector in tensor space 
gives zero on any state which is not in the  $ R \otimes r $ 
subspace and $1$ on any state in $ R \otimes r $. This is 
essentially the content of \decomp. 

\ifig\ungpint{ Basic $U(N)$ group integral } 
{\epsfxsize5.0in\epsfbox{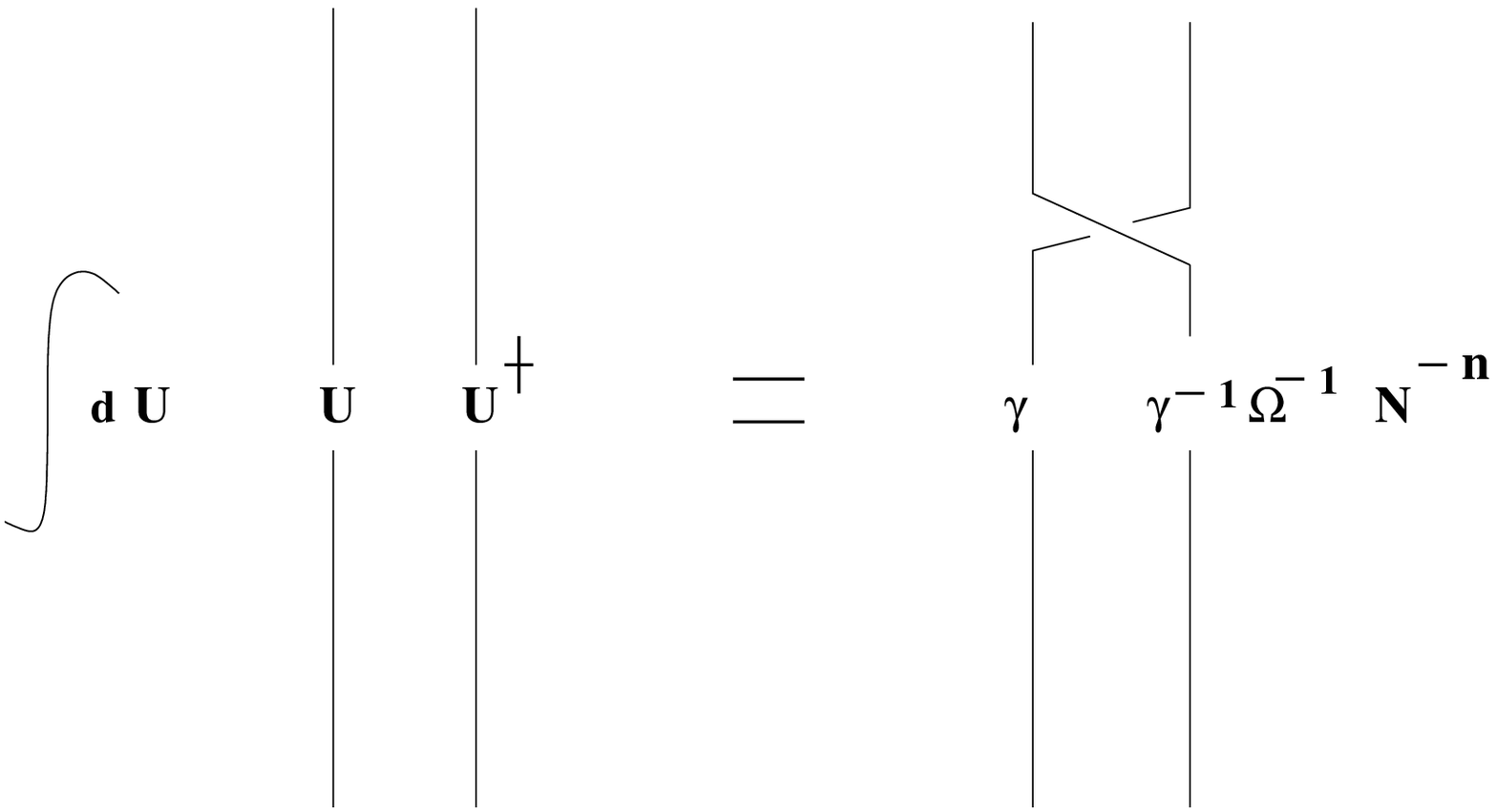} }

It is instructive to give a derivation of this
in physics language by using $U(N)$ group integrals 
familiar from lattice gauge theory and two-dimensional 
Yang-Mills literature \samuel\obzub\grotayii. 
The result we will need can be expressed in diagrammatic notation, 
as for example in \eulwil. The basic unitary group integral
is expressed diagrammatically in \ungpint\ where a sum 
over the element $ \gamma$ in $S_n$ 
is understood.

 In this formula $ \Omega^{-1} $ is an element of the 
 group algebra of $S_n$ with coefficients which are functions of $N$
 and is defined by the equations 
\eqn\omegps{\eqalign{ 
& Dim ~ R = { N^n \over n! }\chi_r ( \Omega ) \cr 
& \Omega ~ \Omega^{-1}   = 1 \cr 
& \chi_r ( \Omega^{-1}  ) = { d_r^2 \over \chi_r ( \Omega ) } \cr }}
 It is a useful fact that $ \Omega $ is actually in the center 
of the group algebra of $S_n$. Further details 
and uses of this object can be found in \grotayii. 

\ifig\pppfne{ Diagrammatic manipulation for projector relations } 
{\epsfxsize6.0in\epsfbox{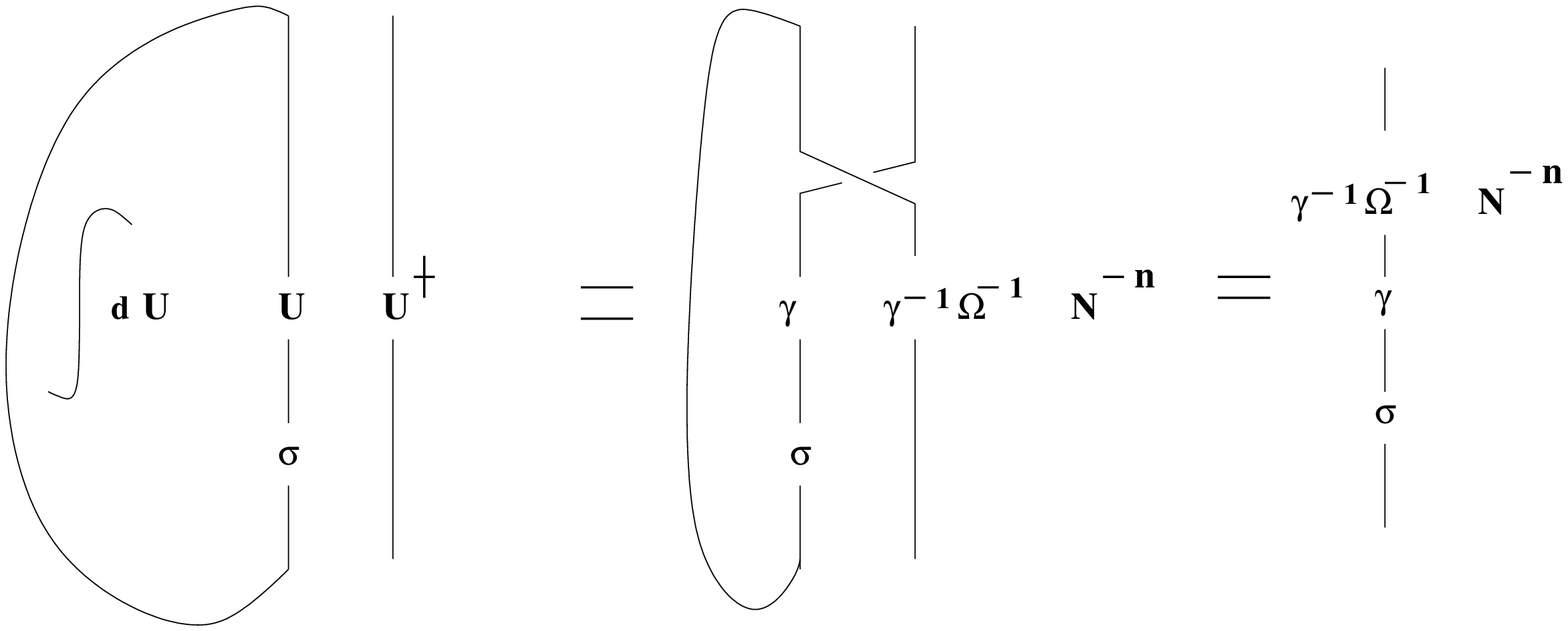} }

Consider $ \rho_n ( P_R ) $, i.e the operator in $ End ( V^{\otimes n})$
 given by acting with $P_R$. 
\eqn\maniprhopR{\eqalign{  
& \rho_n ( P_R ) = Dim R \int dU \chi_R ( U)  \rho_n ( U^{\dagger} ) \cr 
& = Dim ~ R  \sum_{\sigma} { \chi_r ( \sigma ) \over n! } 
 \int dU ~ tr_n ( \sigma U ) ~~ \rho_n ( U^{\dagger} ) \cr 
& = Dim R \sum_{\sigma}  { \chi_r ( \sigma ) \over n! }
  \int dU  ~ ( tr_n \otimes 1 ) ~  \rho_n (\sigma  U ) \otimes \rho_n
 ( U^{\dagger} ) \cr 
&  = Dim R ~~ \sum_{\sigma}  { \chi_r ( \sigma ) \over n! }  ~~ X \cr }}
In the second line we used the  expansion of
$ \chi_R ( U ) $ in terms of traces. In the last line we defined 
a quantity $X$, which is conveniently manipulated diagrammatically
as illustrated in \pppfne.

The diagrammatic steps 
show that 
\eqn\watx{\eqalign{  
& X \equiv ( tr_n \otimes 1 ) \int dU  ~ \rho_n (\sigma  U ) \otimes \rho_n
 ( U^{\dagger} ) \cr 
& = N^{-n} ~n! ~ \rho_n ( \Omega^{-1} \sigma )  \cr }}
The $n! $ comes from doing the sum over $ \gamma$ in $S_n$. 
Collecting terms in \maniprhopR\ we write 
\eqn\colpr{\eqalign{  
 \rho_n ( P_R ) & = {Dim  ~ R \over N^n} 
\sum_{\sigma } \chi_r ( \sigma^{-1}  )
 ~  \rho_n ( \Omega^{-1} \sigma ) 
\cr 
& = {Dim ~ R \over N^n} ~\sum_{\sigma }  \chi_r ( \sigma^{-1} \Omega^{-1}   )  
~~ \rho_n (\sigma ) \cr 
& = {Dim ~ R \over d_R} {1 \over N^n} \chi_R (\Omega^{-1})  
\sum_{\sigma} \chi_r (\sigma^{-1})  \rho_n (\sigma) \cr
& = { d_R \over n! }  ~ \sum_{\sigma } 
 \chi_r ( \sigma ) ~ \rho_n (\sigma )}}
where we have used \omegps\ to obtain the final answer.

The upshot is the simple equation 
\eqn\upsh{ \rho_n ( P_R ) = \rho_n ( P_r ) } 
which  also follows from \decomp. 
Taking advantage of this relation 
we easily convert from the unitary group 
form of the projector to the symmetric group form. 
We sometimes use a noncommital notation 
$R$ for both projectors, and correspondingly 
$ \chi_R $ for both symmetric group characters 
and unitary group characters or their extension 
to complex matrices. The result \upsh\ holds 
for all values of $n$ and $N$. 
When $n \ge N$, and the Young Diagram considered 
is not an admissible one as an irrep of $U(N)$, 
i.e., it has columns of length larger than $N$, 
then $P_R$ is zero, and therefore $ \rho_n ( P_R )$
is zero. In that case  $P_r$ is not zero, but $ \rho_n ( P_r )$
is still zero.

\newsec{ Appendix 2 : Schur Duality and group integrals --
Fusion coefficients and Branching coefficients } 

The basic group integral in \ungpint\ can also be used 
to derive another useful relation between unitary and symmetric
groups. The Littlewood-Richardson coefficient $ g(R_1, R_2 ; R_3) $ 
is the fusion coefficient for $U(N)$, i.e the number of times 
the representation corresponding to the Young Diagram 
 $R_3$ appears in the tensor
product $ R_1 \otimes R_2$. It is also the number of times 
the representation $ R_1 \otimes R_2$ of $S_{n_1} \times S_{n_2} $
appears when the representation $R_3$ of $S_{n}$
is decomposed into irreps of the subgroup $S_{n_1} \times S_{n_2} $,
i.e., it is a branching coefficient.
 Here $R_3$ has $n$ boxes and $R_1, R_2$ have $n_1,n_2$ respectively
with $n= n_1 +n_2$. 
These facts, proved for example 
in \fulhar,  were used in \cjr\ to  derive the relation between 
three-point functions and  $ g(R_1, R_2 ; R_3) $. 
In terms of characters, the above facts are stated as
\eqn\fusbran{\eqalign{ 
&   g( R_1, R_2 ; R_3) = \int dU \chi_{R_1} ( U)
 \chi_{R_2} ( U) \chi_{R_3} ( U^{\dagger}  ) \cr 
& = \sum_{\sigma_1, \sigma_2} \chi_{R_3} ( \sigma_1. \sigma_2 ) 
{ \chi_{R_1} ( \sigma_1 ) \over n_1! } 
{ \chi_{R_2} ( \sigma_2 ) \over n_2! } \cr }} 

It is instructive to derive \fusbran\ using the 
basic group integral in \ungpint.  Using the expansion 
of the Schur Polynomials in terms of traces
\eqn\fusbrander{\eqalign{ 
&   \int dU \chi_{R_1} ( U)
 \chi_{R_2} ( U) \chi_{R_3} ( U^{\dagger}  ) \cr 
& = \sum_{\sigma_1, \sigma_2 } 
{ \chi_{R_1} ( \sigma_1) \over n_1! } 
{ \chi_{R_2} ( \sigma_2) \over n_2! } { \chi_{R_3} ( \sigma_3) \over n_3! }
tr_{n_1 +n_2 }( \sigma_1 \circ \sigma_2  ~~ U ) tr_n ( \sigma_3
U^{\dagger} )  \cr 
&=  \sum_{\sigma_1, \sigma_2 } { \chi_{R_1} ( \sigma_1) \over n_1! }
{ \chi_{R_2} ( \sigma_2) \over n_2! } 
{ 1 \over d_{R_3} }  ~ tr_{n }( \sigma_1 \circ \sigma_2  ~~ U )
  ~ tr_{n} ( P_{R_3}  U^{\dagger} ) \cr  
& = \sum_{\sigma_1, \sigma_2 } { \chi_{R_1} ( \sigma_1) \over n_1! }
{ \chi_{R_2} ( \sigma_2) \over n_2! } 
{ 1 \over d_{R_3} } ~~ Y.} } 
The last line is a definition of $Y$, an object which 
we will write and manipulate diagrammatically using the rules 
in section 3. 

\ifig\fusbranc{ Diagrammatic manipulation for projector relations } 
{\epsfxsize6.0in\epsfbox{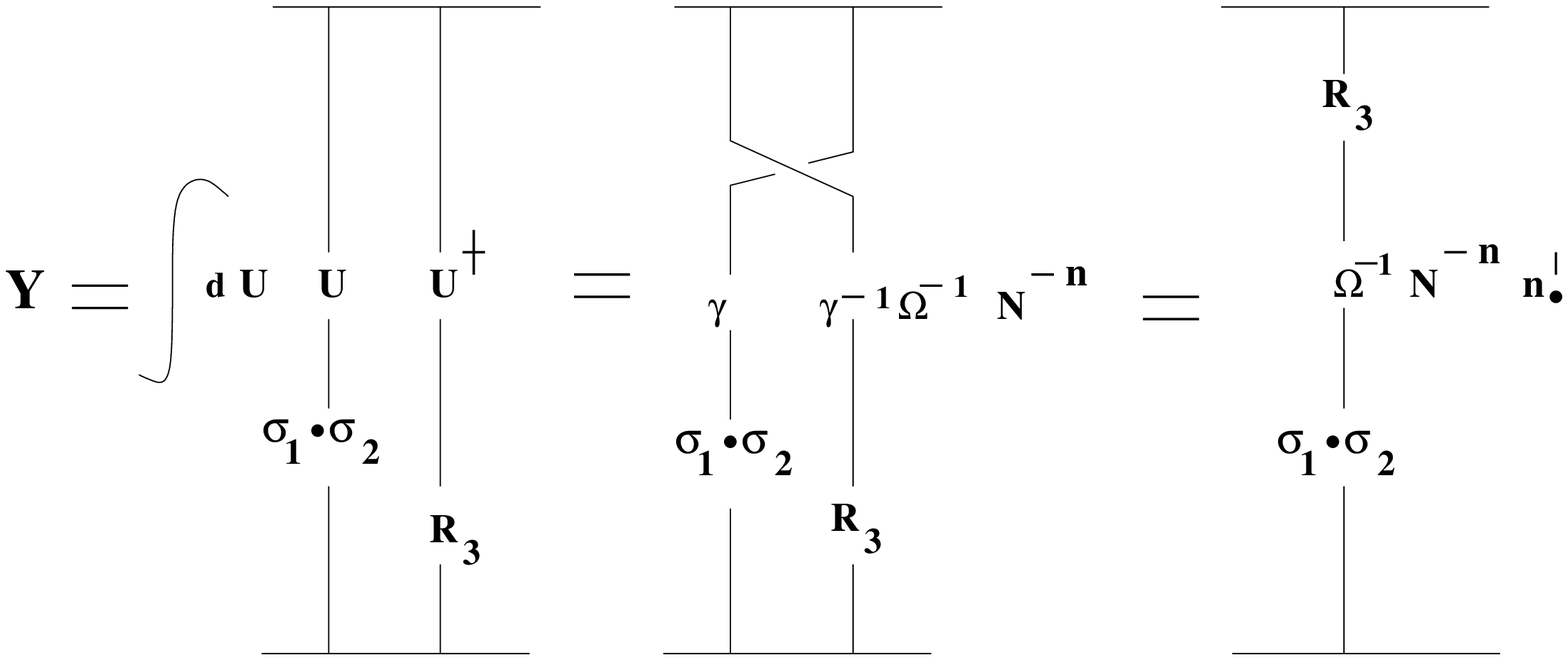} }

The diagrammatic manipulations show
\eqn\resdiy{ 
Y = n! N^{-n} tr_n  (  P_{R_3} (\sigma_1\circ \sigma_2) \Omega^{-1} )  
} 
where in the diagram we have written $R_3$ 
for the projector $P_{R_3}$ and have used the fact that
$\gamma$ commutes with action of $P_{R_3}$ to perform the
sum over $\gamma$ and obtain the factor of $n!$. 
Since the irrep $R_3$ of $S_n$ occurs 
in tensor space with a multiplicity $Dim R_3$, using \decomp, 
we have further
\eqn\resdiyi{\eqalign{  
& Y = n! N^{-n} Dim R_3 ~ \chi_{R_3}  ( (\sigma_1\circ \sigma_2)
\Omega^{-1} )  \cr 
& = n!  N^{-n} Dim R_3 ~ \chi_{R_3}  ( \sigma_1\circ \sigma_2) 
{ 1\over d_{R_3} }   \chi_{R_3} ( \Omega^{-1} ) \cr 
&= d_{R_3} \chi_{R_3}  ( \sigma_1\circ \sigma_2) \cr }}
where we have used \omegps. 
Combining this expression 
for $Y$ with \fusbrander\ we have the proof
that the integral in \fusbran\ is the same as the 
sum in \fusbran, so that $g$ is indeed both a
fusion coefficient for unitary groups and a
branching coefficient for Symmetric groups.

\newsec{Appendix 3:  Derivation of Staggered Factorization}

\ifig\tonettwo{A factorizable diagram along the
strand shown with arbitrary operator insertions denoted
by $T_1$ and $T_2$.}
{\epsfxsize3.0in\epsfbox{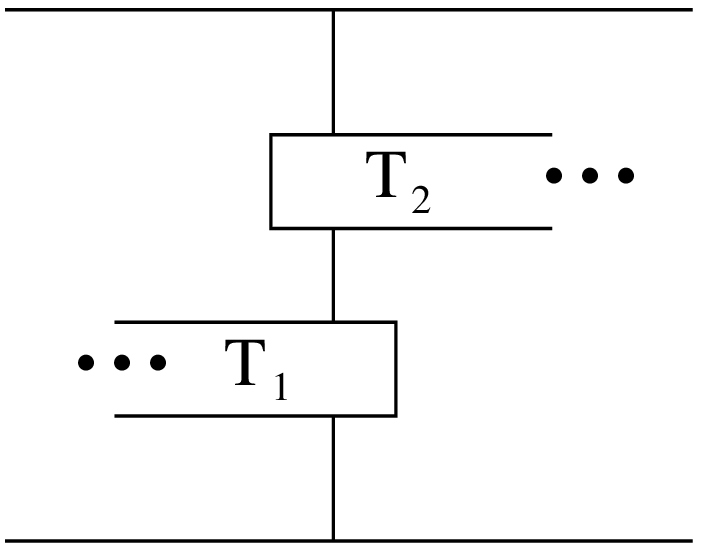} }

In this appendix we discuss the factorization of a diagram
of the form shown in \tonettwo.
The dots in the figure denote that the diagram can be as
complicated as one likes to the left and right of the
strand shown, but that there is one strand that flows
through two operators denoted $T_1$ and $T_2$ and is
then contracted with itself, and moreover there are no
free strands anywhere else in the diagram.  In terms of a formula
the operators $T_1$ and $T_2$ are expressed in terms of
unitary integrals as 
\eqn\ttwoop{\eqalign{
F_{T_1} & := \int dU_1 G_{T_1} (U_1,U_{1}^{\dagger}) 
\rho_{n}(U_{1}^{\dagger}) \cr
F_{T_2} & := \int dU_2 G_{T_2} (U_2,U_{2}^{\dagger}) 
\rho_{n}(U_{2}^{\dagger}).
}}
The diagram \tonettwo\ then corresponds to the expression
\eqn\basefact{\eqalign{ tr(F_{T_1} F_{T_2}) =
\int dU_1 dU_2 
G_{T_1} (U_1, U_{1}^{\dagger}) G_{T_2}(U_2,U_{2}^{\dagger})
(tr(U_{1}^{\dagger} U_{2}^{\dagger}))^{n}
}}
where the number of lines in the strand is taken to
be $n$.
In expressing the diagram in terms of $U(N)$
group integrals with $U_{1(2)}$ corresponding to 
the $T_{1(2)}$ operator insertion, the factors $G_{T_1}$ and 
$G_{T_2}$ contain all the other dependence of the
graph from operator insertions and from tracing over other strands 
on the left and right respectively.  The only constraint
is that both factors are $U(N)$ invariant.

\ifig\commute{Diagram illustrating that the projector $P_{R}$
commutes with the operator $T_2$. }
{\epsfxsize3.0in\epsfbox{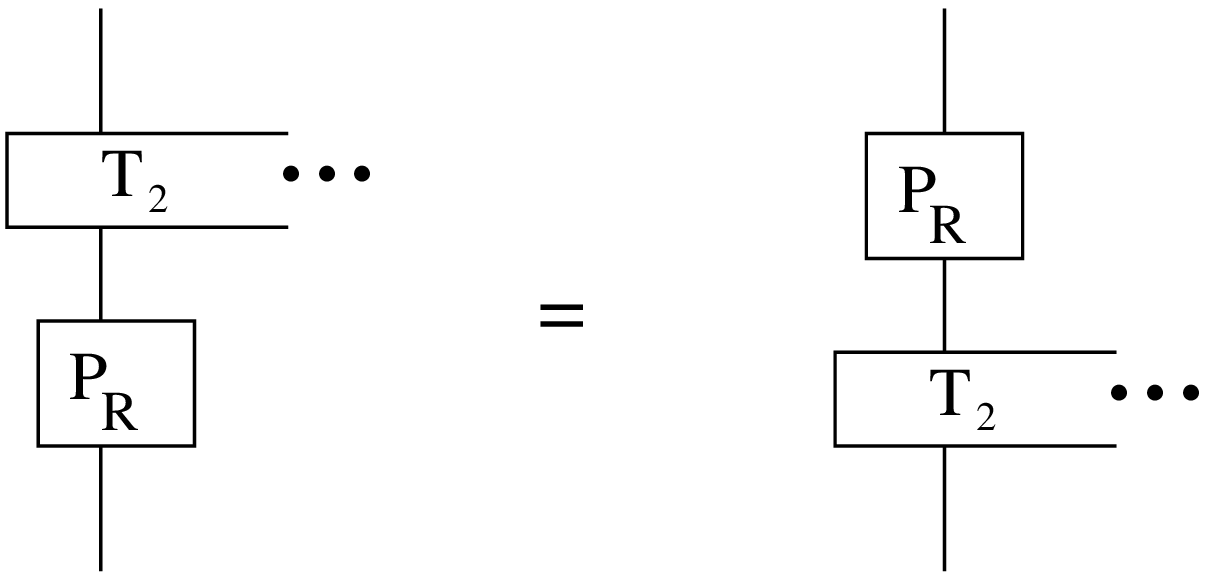} }

To factorize the diagram we insert the unit operator
$I = \sum_R P_R $
along the strand between the $T_1$ and $T_2$ operators, i.e.,
we write
\eqn\baseinter{tr(F_{T_1} F_{T_2}) = \sum_R tr(P_R F_{T_1} F_{T_2}).}
We now claim that the projection operator $P_{R}$
commutes with
$F_{T_1}$ and $F_{T_2}$.  This is expressed diagrammatically
in \commute for $T_2$.  

\ifig\factfrmi{The diagram corresponding to the factorized
form of  \tonettwo. }
{\epsfxsize3.0in\epsfbox{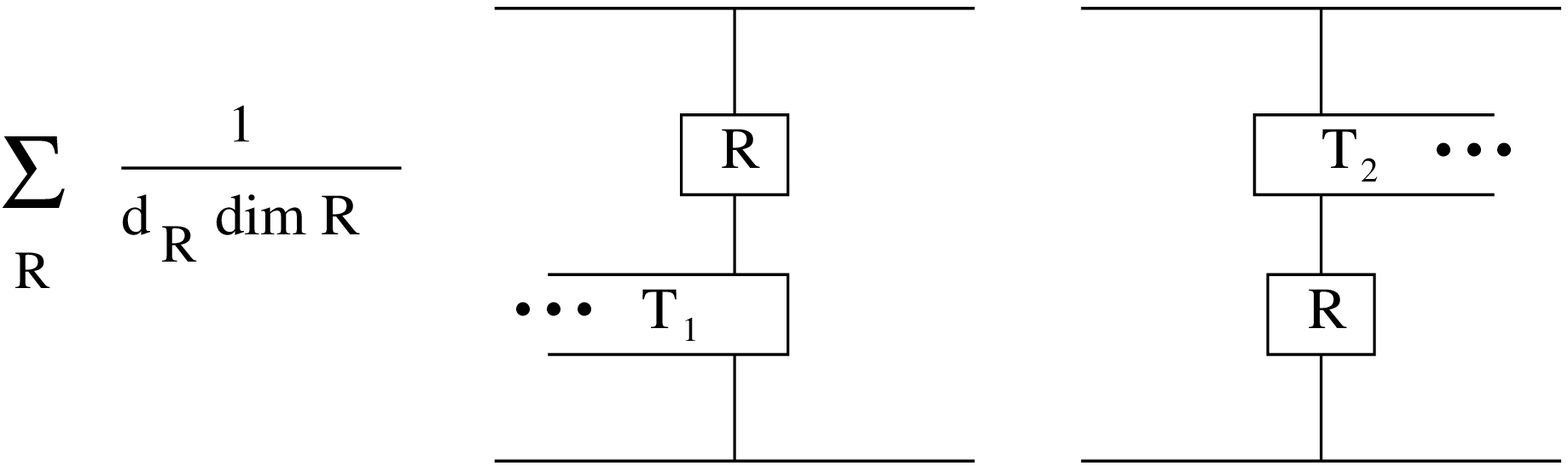} } 

This statement follows since the symmetric
group and unitary group actions on the vector space $V^{\otimes n}$
commute, and $P_{R}$ is  a sum of symmetric group
elements acting on  $V^{\otimes n}$ while $F_{T_1}$ and
$F_{T_2}$ have been expressed in terms of unitary group
integrals.  Because $P_R$ projects onto a single
irreducible representation of the product group
$U(N) \times S_n$, we can now apply Schur's lemma
to factor the trace appearing on the right-hand-side
of \baseinter,
\eqn\baseintertwo{tr(P_R F_{T_1} F_{T_2}) = {1 \over d_R ~ DimR}
tr(P_R F_{T_1}) tr(P_R F_{T_2}),}
where the factor of $d_R ~ DimR$ arises from the fact
that $tr(P_R) = d_R ~ DimR$.  The final factorization
equation is then simply
\eqn\basefacttwo{\eqalign{tr(F_{T_1} F_{T_2}) =
\sum_R {1 \over d_R ~ DimR}
tr(P_R F_{T_1}) tr(P_R F_{T_2}).}}
The diagrammatic interpretation is that the original
diagram in \tonettwo\ has been split into a sum of products
of two diagrams as illustrated in \factfrmi.

Alternatively, one may expand the factor of 
$tr(U_{1}^{\dagger} U_{2}^{\dagger}))^{n}$ appearing in the unitary
integral \basefact\ in terms of unitary group characters to obtain
\eqn\basefactalt{\eqalign{
\sum_{R} &  d_R \int dU_1 dU_2 
\chi_{R} (U_{1}^{\dagger} U_{2}^{\dagger}) G_{T_1} (U_1, U_{1}^{\dagger}) 
G_{T_2}(U_2,U_{2}^{\dagger}) \cr
& = \sum_{R} {d_R \over Dim R}  
(\int dU_1 \chi_{R} (U_{1}^{\dagger}) G_{T_1} (U_1, U_{1}^{\dagger}))
(\int dU_2 \chi_{R} (U_{2}^{\dagger}) \
G_{T_2}(U_2,U_{2}^{\dagger})) \cr
& = \sum_R {1 \over d_R ~ Dim R} tr(P_R F_{T_1}) tr(P_R F_{T_2}).
}}
In the second line we have used the fact that $F_{T_1}$ and
$F_{T_2}$ commute with the action of the unitary group.
To see this let, for example, $F_{T_1}$ act on $\rho_n (U)$.
By redefining the integration variable 
$U_1 \rightarrow U U_1 U^{\dagger}$, the commutativity
property follows since the measure and $G_{T_1}$ are
both invariant.  Taking the unitary group operator
$F_{T_1}$ inside the character $\chi_R$ as
$\chi_R (F_{T_1} U_{2}^{\dagger})$ and applying Schur's
lemma now to the group $U(N)$ yields the second line
in \basefactalt.  This time the factor of $Dim R$ arises
because a single copy of the $U(N)$ irrep $R$ has
this dimension.  Finally the last line follows by
rewriting the integrals as traces in $V^{\otimes n}$.

\newsec{ Appendix 4: Derivation of non-extremal correlators - Descendants  }   

 Here we present the diagrammatic derivation of the 
 non-extremal correlator of Schur polynomials considered
in section 5. 
Let $E_{21}$ be a generator of $SO(6)$ which converts 
 $\Phi_1$ to $\Phi_2$ and let $Q_{12}$ be a generator 
 which converts $\Phi_1$ to $\Phi_2^{\dagger}$, see
appendix 5 for explicit formulae for Lie algebra elements. 
 Let us consider the correlator 
\eqn\nxtcor{ \langle E_{21}^k \chi_{R_1} ( \Phi_1 )  Q_{12}^k 
 \chi_{R_2} ( \Phi_1 ) \chi_{S_1} ( \Phi_1^{\dagger} ) \rangle.  }
 $R_1$ is an irrep associated with a Young Diagram with 
 $(n_1 + k )$ boxes, $R_2$ an irrep. associated with a Young Diagram 
 having $(n_2 + k) $ boxes, and $S_1$ an irrep associated with
a Young diagram with $(n_1 + n_2)$ boxes 
 After acting with the $k$ powers of 
 $E_{21}$  and $ Q_{12} $,  we  get $k$ copies of 
 $\Phi_2$ from the first set of $\Phi_1$'s and
 $k$ copies of $\Phi_2^{\dagger}$ from the 
 second set. Irrespective of the  position of the $\Phi_2$ 
 and $\Phi_2^{\dagger} $, the correlator 
 to be computed can be put in the form shown in the diagram below. 
 There is a combinatoric factor of $ {  ( n_1 + k ) !  \over n_1 !   } 
 { ( n_2 + k )! \over n_2!     }$ which has to multiply this  
 diagram. 

\ifig\lhnxt{ { Diagram of correlator} }
{\epsfxsize2.8in\epsfbox{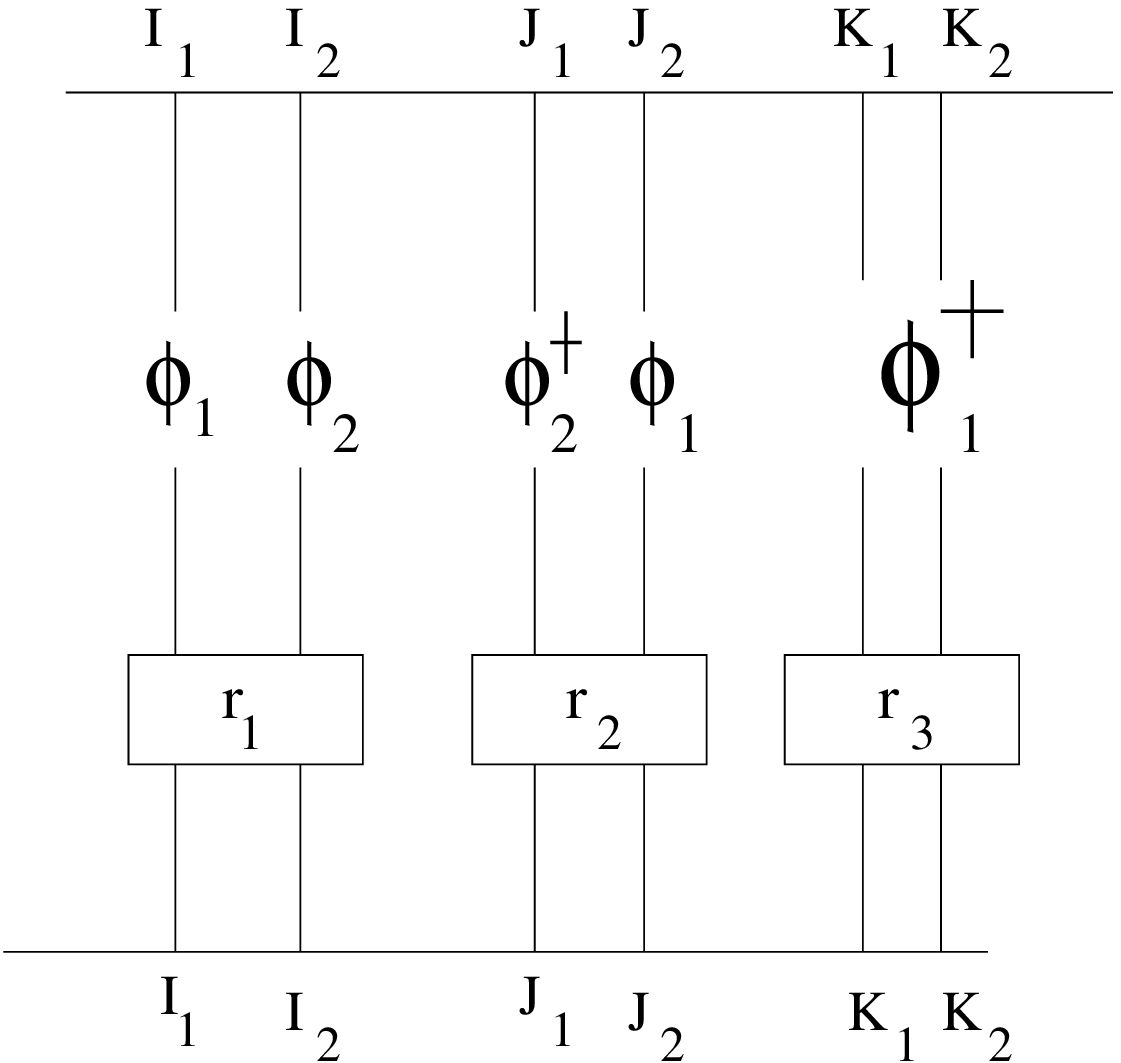} }

 Now we evaluate the correlator using fig. 3. 
 The $\Phi_2$ contractions involve a 
 permutation $\gamma_1$ in $S_{k}$. The $\Phi_1$ contractions involve a
 permutation $\gamma_2$ in $S_{n_1+n_2}$. 

\ifig\rhnxt{ {  Projector diagram obtained after contractions } }
{\epsfxsize2.4in\epsfbox{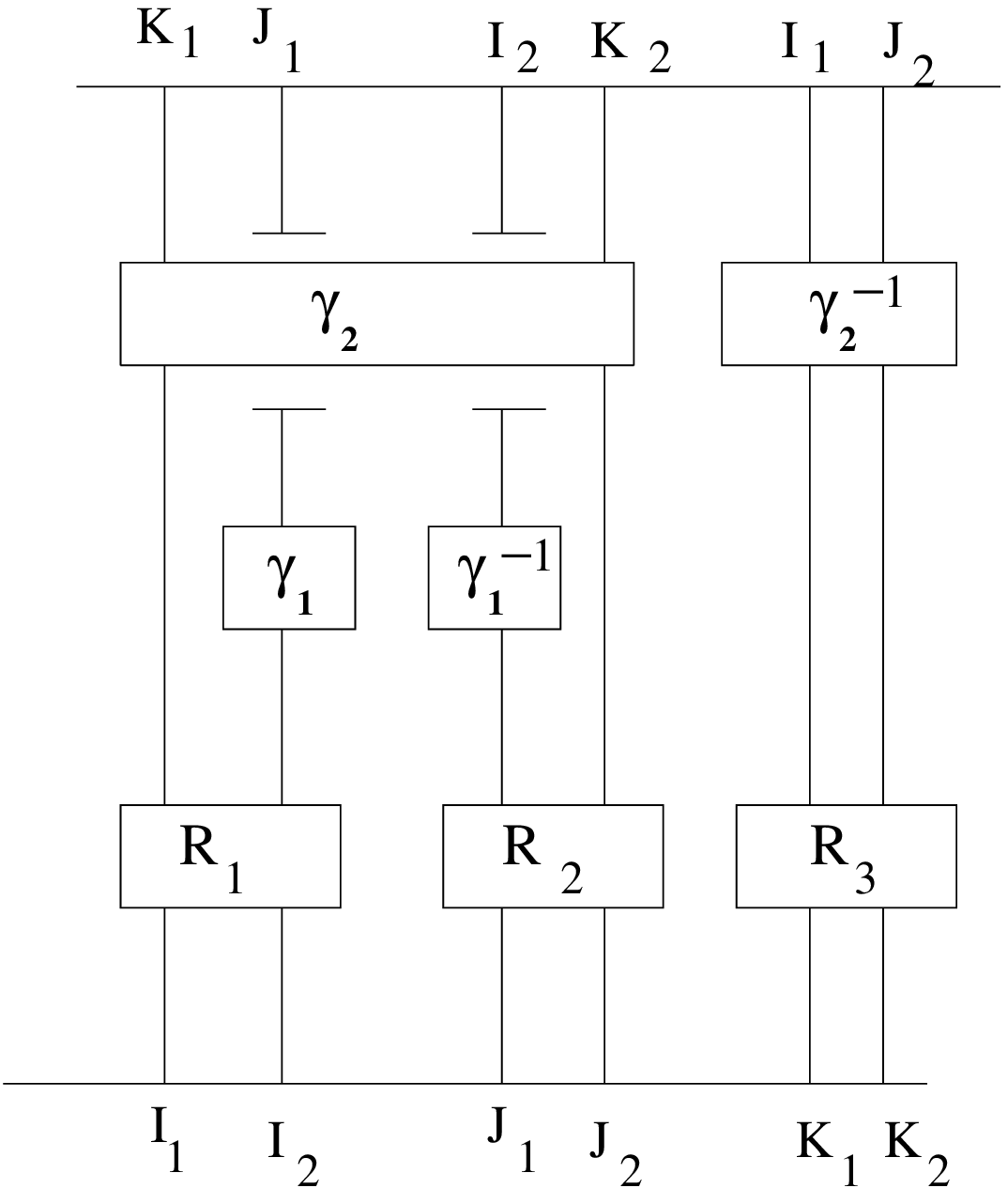} }

Now inspection of this diagram shows that it 
 can be simplified to a three strand diagram of 
 the form given in \nexti.

\newsec{ Appendix 5 : $SO(6)$ algebra } 

We need to describe with some care the 
$SO(6) $ algebra in order to show that 
the operators satisfying simple identities
are indeed directly related to half-BPS 
operators. 

Take the standard action of $SO(6)$ on 
$ x_1, \cdots x_6 $. Let us form 
combinations
\eqn\combs{\eqalign{ 
&  z_1 = x_1 + i x_4   \cr 
& z_2 = x_2 + i x_5 \cr 
& z_3 = x_3 + i x_6 \cr }}

The Cartan subalgebra  is spanned by 
\eqn\cart{\eqalign{ 
&  H_1  = z_1 {\di \over \diz{1} }   - \zb{1}{ \di \over \dizb{1}}  \cr  
&   H_2  = z_2 {\di \over \diz{2} }   - \zb{2}{ \di \over \dizb{2 } }  \cr 
&  H_3  = z_3 {\di \over \diz{3} }   - \zb{3}{   \di \over \dizb{3} } \cr }}

Additional  generators of the $SO(6)$ Lie algebra 
 are, for $i \ne j $ running from $1$ to $3$ :  
\eqn\lieal{
  E_{ij} = z_i { \di \over \diz{j}   } -
 \zb{j} { \di \over \dizb{i} } } 
We also take, for $ i < j $, 
\eqn\ilej{\eqalign{  
&  P_{ij} =   z_i { \di \over \dizb{j}   } - z_j { \di \over \dizb{i}
 } \cr 
& Q_{ij} = 
 - \zb{i} { \di \over \diz{j}   } + \zb{j} { \di \over \diz{i} }
\cr 
}}

It is easy to check that the above
operators preserve $ z_1 \zb{1} +  z_2 \zb{2} +  z_3 \zb{3} $. 

A monomial $z_1^n $ has weights  $ H_i ( z_1^n  ) = ( n, 0, 0 )$. 
We can convert $z_1$'s to $\zb{3}$'s by acting with $Q_{13}$
to obtain
\eqn\getzb{\eqalign{ 
& Q_{13} ( z_1 ) = \zb{ 3 } \cr 
& Q_{13} ( \zb{3} ) = 0 \cr 
& Q_{13}^{k} (   z_1^n  ) = {n ! \over (n-k)!} \zb{ 3 }^{k} z_{1}^{n-k}.}}
Similarly 
we can convert $z_1^n $ to a combination 
of $z_2$ and $z_1$ by observing : 
  \eqn\getzth{\eqalign{ 
& E_{21} ( z_1 ) = z_{ 2 } \cr 
& E_{21} ( z_{2} ) = 0 \cr 
& E_{21}^{k} (   z_1^n  ) = { n ! \over (n-k)!} 
z_{ 2 }^{k} z_{1}^{n-k}.}}
Combining these operations and noting that $E_{21}$ and
$Q_{13}$ commute, then we find that
\eqn\ztwozthreebar{Q_{13}^{k_1} E_{21}^{k_2} (   z_1^n  ) =
{n! \over (n-k_1 - k_2)!} \zb{ 3 }^{k_1 }
z_{2}^{k_2} z_{1}^{n-k_1-k_2},
}
a fact which is used in section 8.

\newsec{Appendix 6:  Extremal Correlators of traces- A Change of Basis } 

In this appendix we consider various extremal correlators
of traces of operators (as opposed to correlators
of operators evaluated in  the Schur
basis that we have been using).  As this just
involves a change of basis we can use the
previous results of \cjr\ (which were recalled 
in section 3)  for the extremal
correlators of operators evaluated in the Schur basis.
Since traces and multi-traces  can be used to give 
an alternative basis to the Schurs in the space 
of gauge invariant polynomials built from $\Phi$.
Our results on factorization and fusion equations and  
sum rules can be restated in this alternative basis. 
Correlators of single traces have been of interest recently in
regard to the pp-wave limit of the $AdS$/CFT correspondence.
The two- and three-point correlators of traces have
also been evaluated exactly in the recent papers
of \refs{\cpss, \harvmit   }.

As a first example we consider the two-point function
$\langle Tr(c_1 \Phi) Tr(c_2 \Phid) \rangle$.
Using the identity
\eqn\identsum{\sum_R \chi_R (\tau) \chi_R (\sigma)
= \sum_{\gamma} \delta(\sigma^{-1}
\gamma \tau \gamma^{-1})}
one may rewrite this correlator in terms of the
Schur basis as
\eqn\twotraceint{\langle Tr(c_1 \Phi) Tr(c_2 \Phid) \rangle
 = \sum_{R,S} \chi_R (c_1) \chi_S (c_2) \langle \chi_R (\Phi)
\chi_S (\Phid) \rangle.}
Plugging in the previous result for the two-point function \untwopoint\
yields the expression
\eqn\twotrace{
\langle Tr(c_1 \Phi) Tr(c_2 \Phid) \rangle
=\sum_{R} {n!  (Dim_N R) \over d_R}
\chi_R (c_1) \chi_R (c_2).}

This can now be generalized in the obvious way to 
higher point correlators of traces of the form
$\langle (\prod_{k=1}^{K} Tr(c_k \Phi))
(\prod_{l=1}^{L} Tr(d_l \Phid)) \rangle $
where now the cycles $c_k$ are of length
$n_k$ and the cycles $d_l$ are of length
$m_l$ where $n=n_1 + \cdots + n_K = m_1 + \cdots + m_L$.
Specifically applying the identity \identsum\
one finds
\eqn\higher{\eqalign{\langle (\prod_{k=1}^{K} Tr(c_k \Phi))
& (\prod_{l=1}^{L} Tr(d_l \Phid)) \rangle  =
\langle Tr(\prod_{k=1}^{K} (c_k \Phi))
Tr(\prod_{l=1}^{L} (d_l \Phid)) \rangle \cr
& = \langle Tr  \bigl((c_1 \circ \cdots \circ c_K) \Phi \bigr)
Tr \bigl((d_1 \circ \cdots \circ d_L) \Phid \bigr) \rangle \cr
& = {1 \over (n!)^2} \sum_{\gamma,\rho \in S_n} 
\langle Tr \bigl((\gamma^{-1} (c_1 \circ \cdots \circ c_K) \gamma \Phi
\bigr)
Tr \bigl(\rho^{-1} (d_1 \circ \cdots \circ d_L) \rho \Phid \bigr) 
\rangle \cr
& = \sum_R {n!  (Dim_N R) \over d_R} \chi_R (c_1 \circ \cdots \circ
c_K) \chi_R (d_1 \circ \cdots \circ d_L)}}
analogously to the two-point function.

We have so far dropped the position dependence of the
scalars.  Providing that all $\Phi$ (or $\Phid$) operators are
evaluated at the same position, then it is a trivial matter
to put the coordinate dependence back in, resulting only
in an overall coordinate dependent factor.  If however 
the correlator has a more general
coordinate dependence, then the correlators will no longer
be extremal and we will get more complicated
projector diagrams as in section 5.

The sums in the two-point function \twotrace\ and 
the three-point function following from \higher\ can actually
be done explicitly.  To see this we need a few facts from
the theory of symmetric groups, see eg. \fulhar.  
The first fact is that the
character $\chi_{R}(c_1)$ for $c_1 \in S_n$ an $n$-cycle
is simply $(-1)^s$ for a representation $R$
corresponding
to a Young diagram with partition $(n-s,1,...,1)$ and
zero otherwise.  We will refer to such Young diagrams 
as ``hooks''.  Combined with the branching formula
(see also appendix 2 where this is related to
unitary group characters)
\eqn\branching{
\chi_{R}(c_1 \circ c_2) = \sum_{R_1,R_2}
g(R_1, R_2; R) \chi_{R_1}(c_1) \chi_{R_2}(c_2)
}
we can also compute the character $\chi_{R}(c_1 \circ c_2)$
for representations $R$ corresponding to hook Young diagrams.  
This follows from the Littlewood-Richardson
rule which tells us that hook representations $R$ 
are only contained in tensor products $R_1 \otimes R_2$
provided that $R_1$ and $R_2$ also correspond to
hooks.
Another result that we use is that the 
factor $n! (Dim_{N} R)/d_R$ appearing in both
\twotrace\ and \higher\ is given by
\eqn\factor{{n! (Dim_{N} R) \over d_R} = {(N+n-s-1)! \over
(N-s-1)!}}
for the hook representation $R$ described above.  
Using this information along with
the following sum \prud,
\eqn\factsum{\sum_{k=m}^{n} {a \choose k} 
{b \choose k}^{-1} = {b+1 \over b-a+1}
\left[ {a \choose m} {b+1 \choose m}^{-1}
- {a \choose n+1} {b+1 \choose n+1}^{-1}
\right],}
we find for the two-point function \twotrace\ of traces
\eqn\twoexact{\langle Tr(c_1 \Phi) Tr(c_1 \Phid) \rangle =
{N \over n+1} \left( {(N+n)! \over N!} - {(N-1)! \over (N-n-1)!}
\right)}
and for the three-point function following from \higher\ 
\eqn\threeexact{\eqalign{\langle Tr(c_1 \Phi) Tr(c_2 \Phi)
Tr(c_3 \Phid) \rangle & = {1 \over n+1} \bigg( {(N+n)! \over (N-1)!}
- {(N+n_2)! \over (N- n_1 -1)!} - {(N+n_1)! \over (N- n_2 -1)!}
\cr 
& + {N! \over (N-n-1)!} \bigg),}}
in agreement with \refs{\cpss,\harvmit } where these expressions were
derived from a complex matrix model.

More generally the same kinds of manipulations can be applied
to correlators of traces involving not just $\Phi_1$ and
its conjugate but also the fields $\Phi_2$ and $\Phi_3$ and
their conjugates.  For example, consider the two-point function
\eqn\twoptmixed{\langle E_{31}^{k_1} E_{21}^{k_2} tr(c_1 \Phi_1)
E_{13}^{k_1} E_{12}^{k_2} tr(c_2 \Phi_{1}^{\dagger}) \rangle}
where the permutations $c_1$ and $c_2$ are $n$-cycles in 
$S_n$ and the $E_{ij}$ operators are defined in appendix 5. 
Their purpose here is simply to convert $\Phi_1$'s to
$\Phi_2$'s and $\Phi_3$'s and similarly for the conjugate
fields.  This two-point function differs from the one
without the $E_{ij}$ operators only by an overall factor.
The derivatives bring down a factor of $(n! / (n-k_1 - k_2)!)^2$.
The contractions give factors
of $k_1 !$, $k_2 !$, and $(n-k_1-k_2)!$ from the
$\Phi_3$, $\Phi_2$, and $\Phi_1$ contractions
respectively.  Without the $E_{ij}$ operators one
instead would just get an $n!$ from the $\Phi_1$
contractions.  Up to this difference in the overall
factor however, the resulting projector diagrams 
for the two-point function \twoptmixed\ and \twotraceint\
are identical.  As a result one finds
\eqn\mixed{\langle E_{31}^{k_1} E_{21}^{k_2} tr(c_1 \Phi_1)
E_{13}^{k_1} E_{12}^{k_2} tr(c_2 \Phi_{1}^{\dagger}) \rangle =
{n! k_1 ! k_2 ! \over (n-k_1 -k_2)!}   
\langle  tr(c_1 \Phi_1) tr(c_2 \Phi_{1}^{\dagger}) \rangle.}

The same argument holds
for the more general multi-point correlator derived
in \higher\ and its generalization to include
$\Phi_2$ and $\Phi_3$ fields, although we shall not
attempt to give a general formula.

\subsec{Normalizations}

There are two different natural ways to normalize
the correlators discussed above: (1) as a multi-point
correlator where one divides by the norm of each
single trace operator, and (2) as an overlap
of states where one divides by the norms of the
complete $\Phi$ and $\Phid$ operators.
The latter normalization leads to a correlator
whose magnitude is bounded above by one, a fact
which follows from the Schwarz inequality.

To be more explicit, consider the normalized multi-point
correlator
\eqn\normmulti{{\langle (\prod_{k=1}^{K} Tr(c_k \Phi))
(\prod_{l=1}^{L} Tr(d_l \Phid)) \rangle \over
\parallel \prod_{k=1}^{K} Tr(c_k \Phi) \parallel
\parallel \prod_{l=1}^{L} Tr(d_l \Phi) \parallel}}
where the norms in the denominator are defined as
$\parallel \prod_{k=1}^{K} Tr(c_k \Phi) \parallel
= \langle \prod_{k=1}^{K} Tr(c_k \Phi)
\prod_{k=1}^{K} Tr(c_k \Phi^{\dagger}) \rangle$.
The numerator is given in \higher.  The denominator
also follows from \higher\ by replacing the $d_l$'s
by $c_k$'s or vice versa.  That is
\eqn\cnorm{\parallel \prod_{k=1}^{K} Tr(c_k \Phi) \parallel^2
= \sum_R {n!  (Dim_N R) \over d_R} \chi_R (c_1 \circ \cdots \circ
c_K)^2
}
with a similar expression for 
$\parallel \prod_{l=1}^{L} Tr(d_l \Phi) \parallel$.
The Schwarz inequality bounds the inner product $(u,v)$
for two vectors $u$ and $v$ by
\eqn\schwarz{|(u,v)| \leq \sqrt{\parallel u \parallel
+ \parallel v \parallel.}}
Applied to the normalized correlator \normmulti\ we
see that it is bounded above by one.

\listrefs

\end